\documentstyle[preprint,eqsecnum,aps]{revtex}
\tighten
\begin{document}
\preprint{\parbox{1.2in}{UCONN 96-07}   }
\draft
\title{Gauss's law and gauge-invariant operators and states in QCD}
\author{Lusheng Chen\thanks{ e-mail address: chen@main.phys.uconn.edu},
Mario Belloni\thanks{e-mail address: mario@main.phys.uconn.edu},  and
Kurt Haller\thanks{ e-mail address: KHALLER@UCONNVM.UCONN.EDU}}
\address{Department of Physics, University of Connecticut, Storrs,
Connecticut
06269}
\date{\today}
\maketitle
\begin{abstract}
In this work, we prove a previously published conjecture that a
prescription we gave for constructing states that implement
Gauss's law for `pure glue' QCD is correct. We also construct
a unitary transformation that extends this prescription so that
it produces additional states that implement Gauss's law for
QCD with quarks as well as gluons.  Furthermore, we use the
mathematical apparatus developed in the course of this work
to construct gauge-invariant spinor (quark) and gauge (gluon) field
operators.  We adapt this $SU(3)$ construction for the $SU(2)$
Yang-Mills case, and we consider the dynamical implications
of these developments.
\end{abstract}
\bigskip
\bigskip



\section{Introduction}

The need to implement Gauss's law in QCD and Yang-Mills theory,
and the technical problems that complicate the implementation
of Gauss's law in non-Abelian theories have been discussed by a
number of authors\cite{goldjack,jackiw,khymtemp}. Strategies
for implementing Gauss's law have also been developed\cite{lands}.
In earlier work\cite{bellchenhall}, we
constructed states that implement
Gauss's law for Yang-Mills theory
and QCD --- in fact, for any `pure glue' gauge theory,
in a temporal gauge formulation that has
a non-Abelian $SU (N)$ gauge symmetry.
In that work, a state vector ${\Psi}\,|{\phi}\rangle$
was defined for which
\begin{equation}
\{\,b_{Q}^{a}({\bf{k}}) + J_{0}^{a}({\bf{k}})\,\}
{\Psi}\,|{\phi}\rangle =0\;,
\label{eq:subcon}
\end{equation}
where $b_{Q}^{a}({\bf{k}})$ and $J_{0}^{a}({\bf{k}})$
are the Fourier transforms of
$\partial_{i}{\Pi}^a_{i}({\bf r})$ (${\Pi}^a_{i}({\bf r})$ is
the momentum conjugate
to the gauge field) and of the gluon color charge density
\begin{equation}
J_{0}^{a}({\bf{r}})=g\,f^{abc}A_{i}^{b}({\bf{r}})\,\Pi_{i}^{c}({\bf{r}})
\label{eq:charge}
\end{equation}
 respectively. Since the chromoelectric field
$E^a_{i}({\bf{r}})=-{\Pi}^a_{i}({\bf{r}})$,  Eq.~(\ref{eq:subcon})
expresses
the momentum space representation of the non-Abelian `pure glue' Gauss's
law, and $\{\,b_{Q}^{a}({\bf{k}}) + J_{0}^{a}({\bf{k}})\,\}$ is
referred to as the `Gauss's law operator' for the `pure glue' case;
 $|{\phi}\rangle$ is a perturbative state annihilated by
$\partial_{i}\Pi^a_{i}({\bf{r}})$. In Ref.~\cite{bellchenhall},
we exhibited an explicit form for the operator ${\Psi}$, namely
\begin{equation}
{\Psi}={\|}\,\exp({\cal{A}})\,{\|}\;,
\label {eq:Apsi}
\end{equation}
where bracketing between double bars denotes a normal
order in which all gauge fields
and functionals of gauge fields appear
to the left of all momenta conjugate to gauge fields.  ${\cal{A}}$
was exhibited as an operator-valued series in
Ref.~\cite{bellchenhall}.
Its form was conjectured to all orders, and verified
for the first six orders.
\bigskip

In the work presented here we will
extend our previously published results in the
following ways:  we will prove our earlier conjecture that the
state ${\Psi}|{\phi}\rangle$ implements the
`pure glue' form of Gauss's Law;  we will extend our work
from the `pure glue' form of the theory to
include quarks as well as gluons; we will
construct gauge-invariant operator-valued
spinor (quark) and gauge (gluon) fields;
and we will adapt the QCD formulation to apply to the
 $SU(2)$ Yang-Mills theory.

\section{Implementing the `pure glue' form of Gauss's law}
\label{sec-Implementing}
Our construction of ${\Psi}$ in Ref.~\cite{bellchenhall} was
informed by the realization that the operator ${\Psi}$ had
to implement $\{\,b_{Q}^{a}({\bf{k}}) + J_{0}^{a}({\bf{k}})\,\}\,
{\Psi}\,|{\phi}\rangle=
{\Psi}\,b_{Q}^{a}({\bf{k}})\,|{\phi}\rangle\,,$ or equivalently that
\begin{equation}
[\,b_{Q}^{a}({\bf{k}}),\,{\Psi}\,]=-J_{0}^{a}({\bf{k}})\,
{\Psi}\,+\,B_Q^{a}({\bf{k}})\;,
\label{eq:psicomm}
\end{equation}
where $B_{Q}^{a}({\bf{k}})$ is an operator product that has
$\partial_{i}\Pi^a_{i}({\bf{r}})$ on its extreme right and therefore
annihilates the same states as
$b_{Q}^{a}({\bf{k}})$, so that
$B_{Q}^{a}({\bf{k}})\,|{\phi}\rangle=0$ as well as
$b_{Q}^{a}({\bf{k}})\,|{\phi}\rangle=0$.
 To facilitate the discussion of the structure of ${\Psi}$,
the following definitions are useful:
\begin{equation}
a_{i}^{\alpha} ({\bf{r}}) = A_{Ti}^{\alpha}({\bf{r}})\;
\label{eq:bookai}
\end{equation}
denotes the transverse part of the gauge field, and
\begin{equation}
x_i^\alpha ({\bf{r}}) = A_{Li}^{\alpha}({\bf{r}})\;
\label{eq:bookxi}
\end{equation}
denotes the longitudinal part, so that
$[\,a_i^\alpha({\bf{r}})+x_i^\alpha({\bf{r}})\,]=
A_{i}^{\alpha}({\bf{r}})$. We also will make use of the combinations
\begin{equation}
{\cal{X}}^\alpha({\bf{r}}) =
[\,{\textstyle\frac{\partial_i}{\partial^2}}A_i^\alpha({\bf{r}})\,]\;,
\end{equation}
and
\begin{equation}
{\cal{Q}}_{(\eta)i}^{\beta}({\bf{r}}) =
[\,a_i^\beta ({\bf{r}})+
{\textstyle\frac{\eta}{\eta+1}}x_i^\beta({\bf{r}})\,]\;,
\label{eq:bookaiQ}
\end{equation}
where $\eta$ is an integer-valued index.
\bigskip

We will furthermore refer to the composite operators
\begin{equation}
\psi^{\gamma}_{(\eta)i}({\bf{r}})= \,(-1)^{\eta-1}\,
f^{\vec{\alpha}\beta\gamma}_{(\eta)}\,
{\cal{R}}^{\vec{\alpha}}_{(\eta)}({\bf{r}})\;
{\cal{Q}}_{(\eta)i}^{\beta}({\bf{r}})\;,
\label{eq:psindef2}
\end{equation}
in which ${\cal{R}}^{\vec{\alpha}}_{(\eta)}({\bf{r}})$ is given by
\begin{equation}
{\cal{R}}^{\vec{\alpha}}_{(\eta)}({\bf{r}})=
\prod_{m=1}^\eta{\cal{X}}^{\alpha[m]}({\bf{r}})\;,
\label{eq:XproductN}
\end{equation}
and $f^{\vec{\alpha}\beta\gamma}_{(\eta)}$ is
the chain of $SU(3)$ structure functions
\begin{equation}
f^{\vec{\alpha}\beta\gamma}_{(\eta)}=f^{\alpha[1]\beta b[1]}\,
\,f^{b[1]\alpha[2]b[2]}\,f^{b[2]\alpha[3]b[3]}\,\cdots\,
\,f^{b[\eta-2]\alpha[\eta-1]b[\eta-1]}f^{b[\eta-
1]\alpha[\eta]\gamma}\;,
\label{eq:fproductN}
\end{equation}
where repeated indices are to be summed. For $\eta =1$,
the chain reduces to
$f^{\vec{\alpha}\beta\gamma}_{(1)}\equiv f^{\alpha\beta\gamma}$;
and for $\eta =0$,
$f^{\vec{\alpha}\beta\gamma}_{(0)}\equiv -\delta_{\beta ,\gamma}$.
Since the only properties of the structure functions that we will use
is their antisymmetry and the Jacobi identity, the
formalism we develop will be applicable to
$SU(2)$ as well as to
other models with an $SU(N)$ gauge symmetry.
\bigskip

The composite operators introduced so far
can help us to understand qualitatively how ${\Psi}$
can implement Eq.~(\ref{eq:subcon}).  We observe, for example,
the product
\begin{equation}
\psi^{\gamma}_{(1)i}({\bf{r}})= \,f^{\alpha\beta\gamma}\,
{\cal{X}}^\alpha({\bf{r}})\;{\cal{Q}}_{(1)i}^{\beta}({\bf{r}})
=\,f^{\alpha\beta\gamma}\,
{\cal{X}}^\alpha({\bf{r}})\,[ a_i^\beta({\bf{r}}) +
{\textstyle\frac{1}{2}}x_i^\beta({\bf{r}}) ]\;,
\end{equation}
which as part of the expression
\begin{equation}
{\cal{A}}_1=ig{\int}d{\bf{r}}\,\psi^{\gamma}_{(1)i}({\bf{r}})\,
\Pi^{\gamma}_{i}({\bf{r}})\;,
\end{equation}
has the property that its commutator with $b_{Q}^{a}({\bf{k}})$,
\begin{eqnarray}
[\,b_Q^a({\bf{k}}),\,ig{\int}d{\bf{r}}\,\psi^{\gamma}_{(1)i}({\bf{r}})\,
\Pi^{\gamma}_{i}({\bf{r}})\,]\,=&&
-g\,f^{a\beta\gamma}{\int}d{\bf{r}}\;e^{-i{\bf{k\cdot r}}}\;
 A_i^\beta({\bf{r}})\;\Pi_i^\gamma({\bf{r}})
\nonumber\\
&&-{\textstyle\frac{g}{2}}\,f^{a\beta\gamma}\,{\int}d{\bf{r}}\,
e^{-i{\bf{k\cdot r}}}\,
{\cal{X}}^\beta\;[\,\partial_i\Pi_i^\gamma({\bf{r}})\,]\;,
\label{eq:thetabqcom}
\end{eqnarray}
generates $-J_{0}^{a}({\bf{k}})$ when it acts on a state annihilated
by $b_Q^a({\bf{k}})\,.$
The expression $\exp({\cal{A}}_1)$
would therefore have been an appropriate choice for $\Psi$,
were it not for the fact that the commutator
$[\,b_Q^a({\bf{k}}),\,{\cal{A}}_1\,]$ fails
to commute with ${\cal{A}}_1$.
When Eq.~(\ref{eq:subcon}) is applied to a candidate
${\Psi}_{cand}=\exp({\cal{A}}_1)$, the commutator
$[\,b_Q^a({\bf{k}}),\,{\cal{A}}_1\,]$
is often produced within a polynomial consisting of ${\cal{A}}_1$
factors --- for example
${\cal{A}}_1^{(n-s)}\,[\,b_Q^a({\bf{k}}),\,
{\cal{A}}_1\,]\,{\cal{A}}_1^s\,$.
$[\,b_Q^a({\bf{k}}),\,{\cal{A}}_1\,]$ does not commute with
${\cal{A}}_{1},$ and can not move freely to annihilate the
state at the right of ${\Psi}_{cand}\,$, thereby excluding
$\exp({\cal{A}}_1)$ as a viable choice for $\Psi$.
\bigskip

The normal ordering denoted by bracketing between
double bars eliminates this problem, but only at the expense
of introducing another problem in its place ---
one that is more benign, but that nevertheless must be addressed.
When normal ordering is imposed,
the result of commuting  $\exp({\cal{A}}_1)$ with $b_Q^a({\bf{k}})$
is not the formation of $J_{0}^{a}({\bf{k}})$
to the left of ${\Psi}_{cand}$, but the formation of only
$f^{a\beta\gamma}\,{\int}\,d{\bf{r}}\,e^{-i{\bf{k\cdot r}}}\,
 A_i^\beta({\bf{r}})$ to the {\em left} of it, and of
$\Pi_i^\gamma({\bf{r}})$
to the extreme
{\em right} of all the gauge fields in the series
representation of the exponential.
Unwanted terms will be generated as
$\Pi_i^\gamma({\bf{r}})$ is commuted,
term by term, from the extreme right of
 ${\Psi}_{cand}$ to the extreme left to form the
desired $J_{0}^{a}({\bf{k}})$.  To compensate for
these further terms, we modify
${\Psi}_{cand}$ by adding additional expressions to ${\cal{A}}_1$ to
eliminate the unwanted commutators
generated as $\Pi_i^\gamma({\bf{r}})$
is commuted from the right to the left hand sides of
operator-valued polynomials.  The question
naturally arises whether the process of adding terms
to remove the unwanted contributions from
earlier ones, comes to closure --- whether
an operator-valued series ${\cal A}$,
that leads to a $\Psi$ for which Eq.~(\ref{eq:subcon})
is satisfied, can be specified to all orders.
In Ref.~\cite{bellchenhall}
we conjectured that this question can be answered affirmatively, by
formulating a recursive equation for ${\cal A}$, which we
verified to sixth order.
\bigskip

In Ref.~\cite{bellchenhall} we represented ${\cal A}$ as
the series ${\cal A}=\sum_{\,n=1}^\infty{\cal A}_{n}$;
we also showed that the requirement that ${\cal A}$ must satisfy
to implement Eq.~(\ref{eq:subcon}),
can be formulated as
\begin{equation}
{\|}\,[\,b_{Q}^{a}({\bf{k}}),\,
\sum_{n=2}^\infty{\cal A}_n\,]\exp({\cal A})\,{\|}\,
-\,{\|}\,g\,f^{a\beta\gamma}\int d{\bf{r}}\,e^{-i{\bf{k\cdot r}}}
A^{\beta}_{i}({\bf{r}})\,
[\,\exp({\cal A}),\,\Pi_{i}^{\gamma}({\bf r})\,]\,{\|}\approx 0\;,
\label{eq:psicom1f}
\end{equation}
where $\approx$ indicates a `soft' equality,
that only holds when the equation acts on a
state $|{\phi}\rangle$ annihilated by $b_Q^a({\bf k})$.
The commutator
$[\,\exp({\cal A}),\,\Pi_{i}^{\gamma}({\bf r})\,]$ in
Eq.~(\ref{eq:psicom1f}) reflects the fact that when
the gluonic `color' charge density is assembled to
the left of the candidate $\Psi$, the momentum conjugate
to the gauge field must be moved from the extreme
right to the extreme left of ${\|}\,\exp({\cal{A}})\,{\|}$.
Since ${\cal A}$ is a complicated multi-linear
functional of the gauge fields, but has a simple linear
dependence on $\Pi_i^{\gamma}({\bf{r}})$,
it is useful to represent it as
\begin{equation}
{\cal{A}}=
i{\int}d{\bf{r}}\;
\overline{{\cal{A}}_{i}^{\gamma}}({\bf{r}})\;
\Pi_i^{\gamma}({\bf{r}})\;,
\label{eq:Awhole}
\end{equation}
where
\begin{equation}
\overline{{\cal{A}}_{i}^{\gamma}}({\bf{r}})=
\sum_{n=1}^\infty g^n
{\cal{A}}_{(n)i}^{\gamma}({\bf{r}})\;,
\label{eq:calAbar}
\end{equation}
and the ${\cal{A}}_{(n)i}^{\gamma}({\bf{r}})$ are
elements in a series whose initial term is
${\cal{A}}^{\gamma}_{(1)i}({\bf{r}})=\psi^{\gamma}_{(1)i}({\bf{r}})$.
All the ${\cal{A}}_{(n)i}^{\gamma}({\bf{r}})$ consist of gauge fields
and functionals of gauge fields only; there are no
conjugate momenta, $\Pi_i^{\gamma}({\bf{r}})$, in
any of the ${\cal{A}}_{(n)i}^{\gamma}({\bf{r}})$.
We also showed in Ref.~\cite{bellchenhall}, that
Eq.~(\ref{eq:psicom1f})
is equivalent to
\begin{equation}
[\,b_Q^a({\bf{k}}),\,{\cal{A}}_n\,]\approx g\,f^{a\beta\gamma}
{\int}d{\bf r}\,e^{-i{\bf{k\cdot r}}}\,
A^\beta_i({\bf{r}})\,
[\,{\cal{A}}_{n-1},\,
\Pi^\gamma_i({\bf{r}})\,]\;,
\label{eq:recrel}
\end{equation}
for ${\cal{A}}_{n}$ with $n>1\,,$ where the ${\cal A}_{n}$
form the series ${\cal A}=\sum_{\,n=1}^\infty{\cal A}_{n},$
and each ${\cal A}_{n}$ can be represented as
\begin{equation}
{\cal A}_{n}=ig^n{\int}d{\bf r}\;{\cal{A}}_{(n)i}^{\gamma}({\bf{r}})
\Pi_{i}^{\gamma}({\bf{r}})\;.
\label{eq:asubn}
\end{equation}
 If ${\cal A}_{n}$ satisfies
Eq.~(\ref{eq:recrel}), then the ${\Psi}$ defined in
Eq.~(\ref{eq:Apsi}) will also necessarily satisfy
Eq.~(\ref{eq:subcon}), and the state ${\Psi}\,|{\phi}\rangle$
will implement the non-Abelian `pure glue' Gauss's law.
\bigskip

In Ref.~\cite{bellchenhall} we gave the form of
${\cal A}$ as a functional of the auxiliary operator-valued
constituents
\begin{equation}
{\cal{M}}_{(\eta)}^{\vec{\alpha}}({\bf{r}})
=\prod_{m=1}^\eta{\textstyle \frac{\partial_{j}}{\partial^{2}}
\overline{{\cal A}_{j}^{\alpha [m]}}({\bf r})}=\prod_{m=1}^\eta
\overline{{\cal Y}^{\alpha[m]}}({\bf{r}})
=\overline{{\cal Y}^{\alpha[1]}}({\bf{r}})\,
\overline{{\cal Y}^{\alpha[2]}}({\bf{r}})\,\cdots
\overline{{\cal Y}^{\alpha[\eta]}}({\bf{r}})\;,
\label{eq:defM}
\end{equation}
and
\begin{equation}
\overline{{\cal B}_{(\eta) i}^{\beta}}({\bf r})=
a_i^{\beta}({\bf r})+\,
(\,\delta_{ij}-{\textstyle\frac{\eta}{\eta+1}}
{\textstyle\frac{\partial_{i}\partial_{j}}{\partial^{2}}}\,)
\overline{{\cal A}_{i}^{\beta}}({\bf r})\;,
\label{eq:calB1b}
\end{equation}
where
\begin{equation}
\overline{{\cal Y}^{\alpha}}({\bf r})=
{\textstyle \frac{\partial_{j}}{\partial^{2}}
\overline{{\cal A}_{j}^{\alpha}}({\bf r})}\;\;\;
\mbox{\small and}\;\;\;
{\cal Y}^{\alpha}_{(s)}({\bf r})=
{\textstyle \frac{\partial_{j}}{\partial^{2}}
{\cal A}_{(s)j}^{\alpha}({\bf r})}\;.
\label{eq:defY}
\end{equation}
The defining equation for ${\cal A}$ is the recursive
\begin{equation}
{\cal{A}}=\sum_{\eta=1}^\infty
{\textstyle\frac{ig^\eta}{\eta!}}{\int}d{\bf r}\;
\{\,\psi^{\gamma}_{(\eta)i}({\bf{r}})\,+
\,f^{\vec{\alpha}\beta\gamma}_{(\eta)}\,
{\cal{M}}_{(\eta)}^{\vec{\alpha}}({\bf{r}})\,
\overline{{\cal{B}}_{(\eta) i}^{\beta}}({\bf{r}})\,\}\;
\Pi^\gamma_i({\bf{r}})\;.
\label{eq:inteq2}
\end{equation}
In Ref.~\cite{bellchenhall}, we presented
this form as a conjecture that we had verified to sixth order only.
In this work, we will prove that ${\Psi}\,|{\phi}\rangle$
satisfies the `pure glue' Gauss's law by showing
that the ${\cal A}$ given in Eq.~(\ref{eq:inteq2}) satisfies
Eq.~(\ref{eq:recrel}).
\bigskip

The form of ${\cal A}$ suggests that the proposition that
it satisfies Eq.~(\ref{eq:recrel}) is
well suited to an inductive proof.  We observe that
two kinds of terms appear on
the right hand side of Eq.~(\ref{eq:inteq2}). One is the
inhomogeneous term
$\psi^{\gamma}_{(\eta)i}({\bf{r}})$;
the other is the product of
$\overline{{\cal B}_{(\eta) i}^{\beta}}({\bf r})$
and ${\cal{M}}_{(\eta)}^{\vec{\alpha}}({\bf{r}})$.
$\overline{{\cal B}_{(\eta) i}^{\beta}}({\bf r})$ is a functional of
$\overline{{\textstyle{\cal A}_{i}^{\beta}}}({\bf r})$, and
${\cal{M}}_{(\eta)}^{\vec{\alpha}}({\bf{r}})$
is a multilinear functional of
$\overline{{\cal Y}^{\beta}}({\bf{r}})$, which is
given as a functional of
$\overline{{\textstyle{\cal A}_{i}^{\beta}}}({\bf r})$
in Eq.~(\ref{eq:defY}). It is useful to
examine the $r^{th}$ order components of
${\cal{M}}_{(\eta)}^{\vec{\alpha}}({\bf{r}})$ and
$\overline{{\cal B}_{(\eta) i}^{\beta}}({\bf r})$. These
are given, respectively, by
\begin{equation}
{\cal M}_{(\eta,r)}^{\vec{\alpha}}({\bf r})=
\Theta (r-\eta)\sum_{r[1],\cdots, r[\eta]}
\delta_{r[1]+\cdots +r[\eta]-r}\prod_{m=1}^{\eta}
{\cal Y}_{(r[m])}^{\alpha [m]}({\bf r})\;,
\label{eq:orderM}
\end{equation}
and
\begin{equation}
{\cal B}_{(\eta,r)i}^{\beta}({\bf r})=
\delta_{r}\,a_{i}^{\beta}({\bf r})
+(\,\delta_{ij}-{\textstyle\frac{\eta}{\eta+1}}\,
{\textstyle\frac{\partial_{i}\partial_{j}}{\partial^{2}}}\,)\,
{\cal A}_{(r)j}^{\beta}({\bf r})\;,
\label{eq:orderB}
\end{equation}
where the subscript $r$ is an integer-valued index that labels the
order in the expansion of
$\overline{{\textstyle{\cal A}_{i}^{\gamma}}}({\bf r}),$
and $\delta_{r}$ is the Kronecker `delta' that
vanishes unless $r=0$.
In Eqs.~(\ref{eq:defM}) and (\ref{eq:orderM}),
$\eta$ is a `multiplicity index' that defines
the multilinearity of
${\cal{M}}_{(\eta)}^{\vec{\alpha}}({\bf{r}})\,$ in
$\overline{{\cal Y}^{\beta}}({\bf{r}}).$
Eqs.~(\ref{eq:inteq2})-(\ref{eq:orderB}) demonstrate
that an ${\cal A}_{r}$ that appears on the l.h.s. of
Eq.~(\ref{eq:inteq2}) is given in terms of the $r^{th}$ order
inhomogeneous term $\psi_{(r)j}^{\gamma}({\bf r})\,
\Pi_{j}^{\gamma}({\bf r})$,
and ${\cal A}_{(r^{\prime }) j}^{\beta}$
terms on the r.h.s. of this equation in which  $r^\prime < r$.
To emphasize this very crucial observation,
we note that in addition to the $g^\eta$ that appears as an
overall factor in Eq.~(\ref{eq:inteq2}), each
$\overline{{\cal A}_{j}^{\beta}}({\bf r})$ in
${\cal{M}}_{(\eta)}^{\vec{\alpha}}({\bf{r}})$
and $\overline{{\cal B}_{(\eta) i}^{\beta}}$
carries its own complement of coupling
constants --- $g^r$ for each
order $r$. The $r^{th}$ order term on the
l.h.s. of Eq.~(\ref{eq:inteq2}) therefore
depends on r.h.s.
contributions from ${\cal{M}}_{(\eta)}^{\vec{\alpha}}({\bf{r}})$ and
 $\overline{{\cal B}_{(\eta) i}^{\beta}}\,({\bf{r}})$
whose orders do not add up to $r$,
but only to $r-\eta$. Since the summation in
Eq.~(\ref{eq:inteq2}) begins with $\eta=1,$ the
highest possible order of ${\cal A}^\gamma_{{(r^\prime)}j}$ that can
appear on the r.h.s. of Eq.~(\ref{eq:inteq2}),
when ${\cal A}_{r}$ is on the l.h.s., is
${\cal A}^\gamma_{{(r-1)}j}$  --- and that must stem from the
${\cal{M}}_{(\eta)}^{\vec{\alpha}}({\bf{r}})$ with
the multiplicity index $\eta=1$. Contributions
from ${\cal{M}}_{(\eta)}^{\vec{\alpha}}({\bf{r}})$
with higher multiplicity
indices are restricted to ${\cal A}^\gamma_{{(r^\prime)}j}$ with even
lower order $r^\prime$.  This feature of Eq.~(\ref{eq:inteq2})
naturally leads us to consider an
inductive proof --- one in which we assume Eq.~(\ref{eq:inteq2})
for all ${\cal A}_{r}$
with $r\leq N,$ and then use that assumption to
prove it for ${\cal A}_{r}$ with $r=N+1.$
\bigskip

The fact that Eq.~(\ref{eq:recrel}) is a `soft' equation, is
an impediment to an inductive proof of the proposition
that ${\cal A}_{n},$ defined by Eq.~(\ref{eq:inteq2}),
satisfies it.  In order to carry out the
needed inductive proof, we must infer correct `hard'
generalizations of both these
equations, in which ${\cal A}$ is replaced
by $i{\int}d{\bf{r}}\;
\overline{{\cal{A}}_{i}^{\gamma}}({\bf{r}})\;
V_i^{\gamma}({\bf{r}}),$
where $V_i^{\gamma}({\bf{r}})$ is
{\em any} field that transforms appropriately, and
$\partial_{i}V_i^{\gamma}({\bf{r}})$ is
not required to annihilate any states.
The generalization we seek is an exact equality between
operator-valued quantities --- one that is true in general, and
not only when both sides of the equation project on a specified
subset of states.  Such a generalization would, in
particular, allow us to use many different spatial vectors in
the role of  $V_i^{\gamma}({\bf{r}})$ in the
course of the inductive proof.
\bigskip

We have made the necessary generalization,
and have arrived at the defining equation for
the $n^{th}$ order term of
$i{\int}d{\bf{r}}\;
\overline{{\cal{A}}_{i}^{\gamma}}({\bf{r}})\;
V_i^{\gamma}({\bf{r}})$,
 that generalizes Eq.~(\ref{eq:inteq2}):
\begin{eqnarray}
& & ig^{n}\int d{\bf r}
{\cal A}_{(n)i}^{\gamma}({\bf r})\,V_{i}^{\gamma}({\bf r})
={\textstyle \frac{ig^{n}}{n!}}\int d{\bf r}\,
\psi_{(n)i}^{\gamma}({\bf r})\,V_{i}^{\gamma}({\bf r})
\nonumber \\
&  & \;\;\;\;+\sum_{\eta=1}{\textstyle\frac{ig^{\eta}}{\eta!} }\,
f^{\vec{\alpha}\beta\gamma}_{(\eta)} \,\sum_{u=0}\sum_{r=\eta}
\delta_{r+u+\eta-n}\int d{\bf r}\,
{\cal M}_{(\eta,r)}^{\vec{\alpha}}({\bf r})
\,{\cal B}_{(\eta,u)i}^{\beta}({\bf r})\,
V^{\gamma}_{i}({\bf r})\;.
\label{eq:LCA1}
\end{eqnarray}
The generalization of Eq.~(\ref{eq:recrel}) --- we make
use of the configuration-space representation
of the Gauss's law operator in this case, instead of its
Fourier transform --- is
\begin{eqnarray}
&& i  \int d{\bf r}^\prime\left[\,\partial_{i}\Pi_{i}^{a}({\bf r}),
{\cal A}_{(n)j}^{\gamma}({\bf r}^\prime)\,\right]\,
V_{j}^{\gamma}({\bf r}^\prime)
+  \delta_{n-1}f^{a\mu \gamma}\,A_{i}^{\mu}({\bf r})\,
V_{i}^{\gamma}({\bf r})
\nonumber \\
 & &\;\;\;\; -  \sum_{\eta=1}\sum_{r=\eta}\delta_{r+\eta-(n-1)}
{\textstyle\frac{B(\eta)}{\eta!}}\,f^{a\mu c}
f^{\vec{\alpha}c\gamma}_{(\eta)}
A_{i}^{\mu}({\bf r})\,{\textstyle \frac{\partial_{i}}{\partial^{2}}}
\left({\cal M}_{(\eta,r)}^{\vec{\alpha}}({\bf r})\,\partial_{j}
V_{j}^{\gamma}({\bf r})\right)
\nonumber \\
& &\;\;\;\; +
\sum_{\eta=0}\sum_{t=1}\sum_{r=\eta}\delta_{r+t+\eta-n}
(-1)^{t-1}{\textstyle \frac{B(\eta)}{\eta!(t-1)!(t+1)}}\,
f^{\vec{\mu} a
\lambda}_{(t)}f^{\vec{\alpha}\lambda\gamma}_{(\eta)}\,
{\cal R}_{(t)}^{\vec{\mu}}({\bf r})\,
{\cal M}_{(\eta,r)}^{\vec{\alpha}}({\bf r})\,
\partial_{i}V_{i}^{\gamma}({\bf r})
\nonumber \\
 & &\;\;\;\; +   f^{a\mu d}A_{i}^{\mu}({\bf r})
\sum_{\eta=0}\sum_{t=1}\sum_{r=\eta}\delta_{r+t+\eta-(n-1)}
(-1)^{t}{\textstyle\frac{B(\eta)}{\eta!(t+1)!}}\,
f^{\vec{\nu}d\lambda}_{(t)}f^{\vec{\alpha}\lambda\gamma}_{(\eta)}
{\textstyle\frac{\partial_{i}}{\partial^{2}}}
\left({\cal R}_{(t)}^{\vec{\nu}}({\bf r})\,
{\cal M}_{(\eta,r)}^{\vec{\alpha}}({\bf r})\,
\partial_{j}V_{j}^{\gamma}({\bf r})\right)
\nonumber \\
&&\;\;\;\;=   -if^{a\mu\sigma}A_{i}^{\mu}({\bf r})\;
\int d{\bf r}^\prime\left[\Pi_{i}^{\sigma}({\bf r}),\,
{\cal A}_{(n-1)j}^{\gamma}({\bf r}^\prime)\right]\,
V_{j}^{\gamma}({\bf r}^\prime)\;,
\label{eq:LCA2}
\end{eqnarray}
where $B(\eta)$ denotes the $\eta^{th}$ Bernoulli number.
Eq.~(\ref{eq:LCA2})
relates  ${\cal A}_{(n)j}^{\gamma}({\bf r})$ with $n\geq 1$,
on the l.h.s. of the equation, to
${\cal A}_{(n-1)j^\prime}^{\gamma^{\,\prime}}({\bf r}^\prime)$ on
the r.h.s.; ${\cal A}_{(n)j}^{\gamma}({\bf r})$
with $n=0$  is not required for the representation
of $\overline{{\cal{A}}_{i}^{\gamma}}({\bf{r}})$
given in Eq.~(\ref{eq:calAbar}),
and therefore does not have to be considered.
${\cal A}_{(n)j}^{\gamma}({\bf r})$ with
$n=1$ {\em is} required, but $\left[\partial_{i}\Pi_{i}^{a}({\bf r}),
{\cal A}_{(1)j}^{\gamma}({\bf r}^\prime)\right]$ can not be described
properly by Eq. (\ref{eq:LCA2}),
unless ${\cal A}_{(0)j}^{\gamma}({\bf r})$ on the r.h.s.
of Eq. (\ref{eq:LCA2}) is given an
appropriate definition.
The only equation like Eq. (\ref{eq:LCA2}), but with
$\int d{\bf r}^\prime\left[\partial_{i}\Pi_{i}^{a}({\bf r})
{\cal A}_{(1)j}^{\gamma}({\bf r}^\prime)\right]
V_{j}^{\gamma}({\bf r}^\prime)$  appearing on its l.h.s.,
is Eq.~(\ref{eq:thetabqcom}) with $\Pi_{i}^{\gamma}({\bf r})$
replaced by
$V_{i}^{\gamma}({\bf r})$.  We have
formulated Eq. (\ref{eq:LCA2}) so that it includes the case of
$\int d{\bf r}^\prime\left[\partial_{i}\Pi_{i}^{a}({\bf r}),
{\cal A}_{(1)j}^{\gamma}({\bf r}^\prime)\right]
V_{j}^{\gamma}({\bf r}^\prime)$
on the l.h.s., by including the r.h.s. of Eq.~(\ref{eq:thetabqcom})
for the $n=1$ case. To include that case
correctly, we define the degenerate
${\cal M}_{(\eta,r)}^{\vec{\alpha}}({\bf r})$ with  $\eta=r=0$
as ${\cal M}_{(0,0)}^{\vec{\alpha}}({\bf r})=1$, and the degenerate
${\cal A}_{(0)j}^{\gamma}({\bf r})=0$.  We will refer to
Eq.~(\ref{eq:LCA2}) as the
`fundamental theorem' for this construction of ${\Psi}$.
\bigskip

The general plan for the inductive proof of
Eq.~(\ref{eq:LCA2}) is as follows: We {\em assume}
Eq.~(\ref{eq:LCA2})
for all values of $n\leq N.$ We then observe that, in the $n=N+1$
case to be proven, the r.h.s. of Eq.~(\ref{eq:LCA2}) becomes
${\sf RHS}_{(N+1)}=-if^{a\mu\sigma}A_{i}^{\mu}({\bf r})\;
\int d{\bf r}^\prime\left[\Pi_{i}^{\sigma}({\bf r}),\,
{\cal A}_{(N)\,j}^{\gamma}({\bf r}^\prime)\right]\,
V_{j}^{\gamma}({\bf r}^\prime).$
We  use Eq.~(\ref{eq:LCA1}) to
substitute for the  ${\cal A}_{(N)j}^{\gamma}\,V_{j}^{\gamma}\,$
in ${\sf RHS}_{(N+1)}$, and
evaluate the resulting commutators
$\left[\,\Pi_{i}^{\sigma}({\bf r}),\,
\psi_{(N)j}^{\gamma}({\bf r}^\prime)\,\right]$,
$\left[\,\Pi_{i}^{\sigma}({\bf r}),\,
{\cal M}_{(\eta,r)}^{\vec{\alpha}}({\bf r}^\prime)\,\right]$, and
$\left[\,\Pi_{i}^{\sigma}({\bf r}),\,
{\cal B}_{(\eta,u)j}^{\beta}({\bf r}^\prime)\,\right]$.  Since
$\psi_{(N)j}^{\gamma}({\bf r}^\prime)$ is a
known inhomogeneity in Eq.~(\ref{eq:LCA1}),
$\left[\,\Pi_{i}^{\sigma}({\bf r}),\,
\psi_{(N)j}^{\gamma}({\bf r}^\prime)\,\right]$
can be explicitly evaluated.  In expanding the
$f^{\vec{\alpha}\beta\gamma}_{(\eta)}
\left[\,\Pi_{i}^{\sigma}({\bf r}),\,
{\cal M}_{(\eta,r)}^{\vec{\alpha}}({\bf r}^\prime)\,\right]$ that
result from the substitution of Eq.~(\ref{eq:LCA1}) into
${\sf RHS}_{(N+1)}$, we make use
of the identity
\begin{eqnarray}
& &  f^{\vec{\alpha}\delta\gamma}_{(\eta)}\sum_{r=\eta}
\delta_{r+\eta+u-N}\left[{\sf Q}({\bf r}),
{\cal M}_{(\eta,r)}^{\vec{\alpha}}({\bf r}^\prime)\right]
\nonumber \\
&&\;\;\;\;=- \left[{\cal P}^{(0)}_{(\alpha ,\beta [\eta-1])}
f^{\alpha\delta e}f^{\vec{\beta} e\gamma}_{(\eta-1)}\right]
\sum_{p=\eta-1}\sum_{r[\eta]=1}\delta_{p+r[\eta]+u+\eta-N}
\left[\,{\sf Q}({\bf r}),\,
{\cal Y}_{(r[\eta])}^{\alpha}({\bf r}^\prime)\,\right]
{\cal M}_{(\eta-1,p)}^{\vec{\beta}}({\bf r}^\prime)\;,
\label{eq:perm}
\end{eqnarray}
where ${\sf Q}({\bf r})$ is any arbitrary
operator; at times, the commutator $[{\sf Q}({\bf r})\,,
{\cal M}_{(\eta,r)}^{\vec{\alpha}}({\bf r}^\prime)\,]$ will represent
a partial derivative $\partial_{j}
{\cal M}_{(\eta,r)}^{\vec{\alpha}}({\bf r}^\prime)\,.$
 ${\cal P}^{(0)}_{(\alpha ,\beta [\eta-1])}$ represents
a sum over permutations over the indices labeling
the ${\cal Y}_{(r[\eta ])}^{\alpha[\eta]}({\bf r}^\prime)$ factors that
constitute ${\cal M}_{(\eta,r)}^{\vec{\alpha}}({\bf r}^\prime)$,
as shown in Eq.~(\ref{eq:defM}).
${\cal P}^{(0)}_{(\alpha ,\beta [\eta-1])}$
is defined by
\begin{equation}
\left[{\cal P}_{(e ,\beta [\eta-1])}^{(0)}f^{e\delta f}
f^{\vec{\beta} f\gamma}_{(\eta-1)}\right]
{\cal M}_{(\eta-1)}^{\vec{\beta}}=\sum_{s=0}^{\eta-1}
f^{\vec{\beta}\delta u}_{(s)}f^{ue v}
f^{\vec{\sigma}v\gamma}_{(\eta-s-1)}
{\cal M}_{(s)}^{\vec{\beta}}{\cal M}_{(\eta-s-1)}^{\vec{\sigma}}\;.
\label{eq:defperm}
\end{equation}
Eqs.~(\ref{eq:perm}) and (\ref{eq:defperm}) apply not only to
those specific cases, but
also to all other operators --- such as
${\cal R}_{(\eta)}^{\vec{\alpha}}({\bf r}^\prime)$ ---
that similarly are products of factors, identical
except for their Lie group indices
contracted over chains of structure functions.
\bigskip

With the substitution of of Eq.~(\ref{eq:LCA1}) into
${\sf RHS}_{(N+1)}$, and extensive integration by parts,
we have replaced the commutator
$[\,\Pi_{i}^{\sigma}({\bf r}),\,
{\cal A}_{(N)j}^{\gamma}({\bf r}^\prime)\,]$ which appears
in ${\sf RHS}_{(N+1)}$, with products of
chains of ${\cal A}_{(n^\prime)j}^{\beta^\prime}({\bf r}^\prime)\,$
and one commutator $[\,\Pi_{i}^{\sigma}({\bf r}),\,
{\cal A}_{(l)j}^{\gamma}({\bf r}^\prime)\,]$ with $l\leq N-1$.
Although the $[\,\Pi_{i}^{\sigma}({\bf r}),\,
{\cal A}_{(N)j}^{\gamma}({\bf r}^\prime)\,]$ in ${\sf RHS}_{(N+1)}$
is {\em not} covered by the inductive axiom --- it is
the r.h.s. of the equation
for the $n=N+1$ case --- the $[\,\Pi_{i}^{\sigma}({\bf r}),\,
{\cal A}_{(l)j}^{\gamma}({\bf r}^\prime)\,]$ with
$l\leq N-1$, which have been substituted into ${\sf RHS}_{(N+1)}$,
{\em are} covered by this axiom.
We can therefore use the inductive axiom to
replace all these latter commutators by their corresponding
left hand side equivalents from Eq.~(\ref{eq:LCA2}). After
extensive algebraic manipulations, we can demonstrate that
${\sf RHS}_{(N+1)}$ has been transformed into
the {\em left hand side}
of Eq.~(\ref{eq:LCA2}) for the case in which all $n$ have
been replaced by $n=N+1$. This, then, completes the
inductive proof of Eq.~(\ref{eq:LCA2}).
The details of the argument are given in two appendices. Appendix
A proves some necessary lemmas; Appendix B proves the
fundamental theorem.

\section{The inclusion of quarks}

In  Eq.~(\ref{eq:subcon}), we have implemented
the `pure glue' form of Gauss's law. The complete Gauss's law
operator, when the quarks are included as sources
for the chromoelectric field, takes the form
\begin{equation}
{\hat {\cal G}}^{a}({\bf r})=\partial_{i}
\Pi_{i}^{a}({\bf r})+gf^{abc}A_{i}^{b}({\bf r})
\Pi_{i}^{c}({\bf r})+j_{0}^{a}({\bf r})\;,
\label{eq:Ghat}
\end{equation}
where
\begin{equation}
j^a_{0}({\bf{r}})=
g\,\,\psi^\dagger({\bf{r}})\,
{\textstyle\frac{\lambda^a}{2}}\,\psi({\bf{r}})\;,
\label{eq:quark}
\end{equation}
and where the ${\lambda^a}$ represent the Gell-Mann
matrices. To implement the `complete' Gauss's law --- a form that
incorporates quark as well as gluon color
--- we must solve the equation
\begin{equation}
{\hat {\cal G}}^{a}({\bf r})\,{\hat {\Psi}}\,|{\phi}\rangle =0\;.
\label{eq:Ggqlaw}
\end{equation}
Our approach to this problem will be based
on the fact that ${\hat {\cal G}}^{a}({\bf r})$ and
${\cal G}^{a}({\bf r})$ are unitarily equivalent, so that
\begin{equation}
\hat{\cal{G}}^a({\bf{r}})={\cal{U}}_{\cal{C}}\,
{\cal{G}}^a({\bf{r}})\,{\cal{U}}^{-1}_{\cal{C}}\;,
\label{eq:GgqGg}
\end{equation}
where ${\cal{U}}_{\cal{C}}=e^{{\cal C}_{0}}
e^{\bar {\cal C}}$ and where
${{\cal C}_{0}}$ and ${\bar {\cal C}}$ are given by
\begin{equation}
{\cal C}_{0}=i\,\int d{\bf{r}}\,
{\textstyle {\cal X}^{\alpha}}({\bf r})\,j_{0}^{\alpha}({\bf r})\;,
\;\;\;\;\;\mbox{and}\;\;\;\;\;\;\;
{\bar {\cal C}}=i\,\int d{\bf{r}}\,
\overline{{\cal Y}^{\alpha}}({\bf r})\,j_{0}^{\alpha}({\bf r})\;.
\label{eq:CCbar}
\end{equation}
We can demonstrate this unitary equivalence by noting
that Eq~(\ref{eq:GgqGg}) can be rewritten as
\begin{equation}
e^{-{\cal C}_{0}}\,{\hat {\cal G}}^{a}({\bf{r}})\,e^{{\cal C}_{0}}=
e^{\bar {\cal C}}\,{\cal G}^{a}({\bf{r}})\,e^{-\bar {\cal C}}\;.
\label{eq:gghatp}
\end{equation}
In this form, the unitary equivalence can be shown
to be a direct consequence of the
fundamental theorem --- $i.\,e.$ Eq.~(\ref{eq:LCA2}).
We observe that the l.h.s. of
Eq.~(\ref{eq:gghatp}) can be expanded, using the
Baker-Hausdorff-Campbell (BHC) theorem, as
\begin{equation}
e^{-{\cal C}_{0}}\,{\hat {\cal G}}^{a}({\bf{r}})\,e^{{\cal C}_{0}}=
{\hat {\cal G}}^{a}({\bf{r}})+{\cal S}_{(1)}^a+
\cdots +{\cal S}_{(n)}^a+\cdots\;,
\label{eq:uniser}
\end{equation}
where ${\cal S}_{(1)}^a=-[\,{\cal C}_{0},\,\hat{\cal{G}}^a({\bf{r}})\,]$
and
 ${\cal S}_{(n)}^a=-(1/n)[\,{\cal C}_{0},\,{\cal S}_{(n-1)}^a\,]$.
We observe that
\begin{equation}
{\cal S}_{(1)}^a=-\left[\,\delta_{a,c}+gf^{abc}{\cal X}^b({\bf{r}})
+gf^{abc}A_i^b({\bf{r}})
{\textstyle\frac{\partial_{i}}{\partial^2}}\,\right]\,j_0^c({\bf{r}})\;,
\label{eq:Sone}
\end{equation}
and that
\begin{eqnarray}
{\cal S}_{(n)}^a={\textstyle\frac{{(-1)}^{n+1}}{n\,!}}
&&\left[\left(\,g^{n -1}f^{\vec{\alpha}a\gamma}_{(n-1)}\,
{\cal R}^{\vec{\alpha}}_{(n-1)}({\bf r})\,+
g^{n}\, f^{\vec{\alpha}a\gamma}_{(n)}\,
{\cal R}^{\vec{\alpha}}_{(n)}({\bf r})\,\right)\,
j_0^{\gamma}({\bf{r}}\,)\right.
\nonumber\\
&&+ \left .g^n f^{abc}\,f^{\vec{\alpha}c\gamma}_{(n-1)}\,
A_i^b({\bf{r}})\,
{\textstyle\frac{\partial_{i}}{\partial^2}}
\left(\,{\cal R}^{\vec{\alpha}}_{(n-1)}({\bf r})\,
j_0^{\gamma}({\bf{r}})\,\right)\, \right]\;.
\label{eq:Snth}
\end{eqnarray}
Eq~(\ref{eq:Snth}) shows that two $g^{n}
f^{\vec{\alpha}a\gamma}_{(n)}\,
{\cal R}^{\vec{\alpha}}_{(n)}({\bf r})\,j_0^{\gamma}({\bf{r}})$ terms
will appear in this series:
one in ${\cal S}_{(n)}^a$, and one in ${\cal S}_{(n+1)}^a$. The sum of
these terms will have the
coefficient $\left[\frac{1}{n!}-\frac{1}{(n+1)!}\right]=
\frac{1}{(n+1)(n-1)!}$. When
the BHC series is summed, we find that
\begin{eqnarray}
e^{-{\cal C}_{0}}\,{\hat {\cal G}}^{a}({\bf r})\,e^{{\cal C}_{0}}
&&={\hat {\cal G}}^{a}({\bf r})-j_{0}^{a}({\bf r})-gf^{abc}
A_{i}^{b}({\bf r})\,{\textstyle\frac{\partial_{i}}{\partial^{2}}}
j_{0}^{c}({\bf r})
\nonumber \\
&&-\sum_{n=1} (-1)^{n} g^{n}{\textstyle \frac{1}{(n-1)!(n+1)}}
f^{\vec{\alpha}a\gamma}_{(n)}\,
{\cal R}_{(n)}^{\vec{\alpha}}({\bf r})\,
j_{0}^{\gamma}({\bf r})
\nonumber \\
&&+gf^{abc}A_{i}^{b}({\bf r})\sum_{n=1} (-1)^{n}g^{n}
{\textstyle \frac{1}{(n+1)!}}f^{\vec{\alpha}c\gamma}_{(n)}\,
{\textstyle \frac{\partial_{i}}{\partial^{2}}}
\left(\,{\cal R}_{(n)}^{\vec{\alpha}}({\bf r})\,
j_{0}^{\gamma}({\bf r})\,\right)\;.
\label{eq:GGhatleft}
\end{eqnarray}
\bigskip

To prepare for the evaluation of
$e^{\bar {\cal C}}\,{\cal G}^{a}({\bf r})\,e^{-\bar {\cal C}}$,
the r.h.s. of Eq.~(\ref{eq:gghatp}), we multiply both sides of
Eq~(\ref{eq:LCA2}) for the
$n^{th}$ order term, ${\cal A}_{(n)i}^{\gamma}({\bf r})$, by $g^n$,
and then sum over the integer-valued indices
 $r$ and $n$ (in that order).  The result --- a formulation
of the fundamental theorem that
no longer applies to the individual orders,
${\cal A}_{(n)j}^{\gamma}({\bf r}),$ but to their sum,
$\overline{{\cal A}_{j}^{\gamma}}({\bf r})$ --- is
\begin{eqnarray}
i\int d{\bf r}^\prime&&[\,\partial_{i}\Pi_{i}^{a}({\bf r}),\,
\overline{{\cal A}_{j}^{\gamma}}({\bf r}^\prime)\,]\,
V_{j}^{\gamma}({\bf r}^\prime)
+ igf^{a\beta d}A_{i}^{\beta}({\bf r})\int d{\bf r}^\prime
[\,\Pi_{i}^{d}({\bf r}),\,
\overline{{\cal A}_{j}^{\gamma}}({\bf r}^\prime)\,]\,
V_{j}^{\gamma}({\bf r}^\prime)
\nonumber \\
&&= -gf^{a\mu d}\,A_{i}^{\mu}({\bf r})\,V_{i}^{d}({\bf r})
\nonumber \\
&&+\sum_{\eta=1}{\textstyle\frac{g^{\eta +1}B(\eta)}{\eta!}}\,
f^{a\beta c}f^{\vec{\alpha}c\gamma}_{(\eta)}\,A_{i}^{\beta}({\bf r})\,
{\textstyle \frac{\partial_{i}}{\partial^{2}}}\left(\,
{\cal M}_{(\eta)}^{\vec{\alpha}}({\bf r})\,
\partial_{j}V_{j}^{\gamma}({\bf r})\,\right)
\nonumber \\
&&-\sum_{\eta=0}\sum_{t=1} (-1)^{t-1}g^{t+\eta}\,
{\textstyle \frac{B(\eta)}{\eta!(t-1)!(t+1)}}\,
f^{\vec{\mu}a\lambda}_{(t)}f^{\vec{\alpha}\lambda\gamma}_{(\eta)}\,
{\cal R}_{(t)}^{\vec{\mu}}({\bf r})\,
{\cal M}_{(\eta)}^{\vec{\alpha}}({\bf r})\,
\partial_{i}V_{i}^{\gamma}({\bf r})
\nonumber \\
&&-gf^{a\beta d}A_{i}^{\beta}({\bf r})\,
\sum_{\eta=0}\sum_{t=1} (-1)^{t}g^{t+\eta}\,
{\textstyle\frac{B(\eta)}{\eta!(t+1)!}}
f^{\vec{\mu}d\lambda}_{(t)}
f^{\vec{\alpha}\lambda\gamma}_{(\eta)}
{\textstyle\frac{\partial_{i}}{\partial^{2}}}\,
\left(\,{\cal R}_{(t)}^{\vec{\mu}}({\bf r})\,
{\cal M}_{(\eta)}^{\vec{\alpha}}({\bf r})\,
\partial_{j}V_{j}^{\gamma}({\bf r})\,\right)\;.
\label{eq:Asum}
\end{eqnarray}
If we again use the BHC expansion, as in Eq~(\ref{eq:uniser}),
but this time to represent
\begin{equation}
e^{\bar {\cal C}}\,{\cal G}^{a}({\bf r})\,e^{-\bar {\cal C}}=
{\cal G}^{a}({\bf r})+{\bar{\cal S}}_{(1)}^a+\cdots
+{\bar{\cal S}}_{(n)}^a+\cdots,
\end{equation}
we find that the first order term, ${\bar{\cal S}}_{(1)}^a$,
can be obtained directly from Eq.~(\ref{eq:Asum}) and is
\begin{eqnarray}
{\bar{\cal S}}_{(1)}^a &&=
-gf^{a\mu\gamma}\,A_{i}^{\mu}({\bf r})\,
{\textstyle\frac{\partial_{i}}{\partial^2}}
j_0^{\gamma}({\bf{r}})
\nonumber \\
&&+\sum_{s=1}{\textstyle\frac{g^{s+1}B(s)}{s!}}\,
f^{a\beta c}f^{\vec{\alpha}c\gamma}_{(s)}\,A_{i}^{\beta}({\bf r})\,
{\textstyle \frac{\partial_{i}}{\partial^{2}}}\left(\,
{\cal M}_{(s)}^{\vec{\alpha}}({\bf r})\,
j_0^{\gamma}({\bf{r}})\,\right)
\nonumber \\
&&-\sum_{s=0}\sum_{t=1} (-1)^{t-1}g^{t+s}\,
{\textstyle \frac{B(s)}{s!(t-1)!(t+1)}}\,
f^{\vec{\mu}a\lambda}_{(t)}f^{\vec{\alpha}\lambda\gamma}_{(s)}\,
{\cal R}_{(t)}^{\vec{\mu}}({\bf r})\,
{\cal M}_{(s)}^{\vec{\alpha}}({\bf r})\,
j_0^{\gamma}({\bf{r}})
\nonumber \\
&&-gf^{a\beta d}\,A_{i}^{\beta}({\bf r})\,
\sum_{s=0}\sum_{t=1} (-1)^{t}g^{t+s}
{\textstyle\frac{B(s)}{s!(t+1)!}}\,
f^{\vec{\mu}d\lambda}_{(t)}
f^{\vec{\alpha}\lambda\gamma}_{(s)}\,
{\textstyle\frac{\partial_{i}}{\partial^{2}}}
\left(\,{\cal R}_{(t)}^{\vec{\mu}}({\bf r})\,
{\cal M}_{(s)}^{\vec{\alpha}}({\bf r})\,
j_0^{\gamma}({\bf{r}})\,\right)\;;
\label{eq:SRone}
\end{eqnarray}
the $k^{th}$ order term is
\begin{eqnarray}
{\bar{\cal S}}_{(k)}^a & = & {\textstyle\frac{g^{k}}{k\, !}}\,
f^{a\mu d}f^{\vec{\alpha}d\gamma}_{(k-1)}\,A_{i}^{\mu}({\bf r})\,
{\textstyle \frac{\partial_{i}}{\partial^{2}}}
\left(\,{\cal M}_{(k-1)}^{\vec{\alpha}}({\bf r})\,
j_0^{\gamma}({\bf{r}})\,\right)
\nonumber \\
&+& \sum_{s=1}{\textstyle\frac{g^{s+k}B(s)}{s!\,k!}}\,
f^{a\beta c}f^{\vec{\alpha}c\gamma}_{(s+k-1)}\,
A_{i}^{\beta}({\bf r})\,
{\textstyle \frac{\partial_{i}}{\partial^{2}}}
\left(\,{\cal M}_{(s+k-1)}^{\vec{\alpha}}({\bf r})\,
j_0^{\gamma}({\bf{r}})\,\right)
\nonumber \\
&-& \sum_{s=0}\sum_{t=1} (-1)^{t-1}g^{t+s+k-1}\,
{\textstyle \frac{B(s)}{s!k!(t-1)!(t+1)}}\,
f^{\vec{\mu}a\lambda}_{(t)}
f^{\vec{\alpha}\lambda\gamma}_{(s+k-1)}\,
{\cal R}_{(t)}^{\vec{\mu}}({\bf r})\,
{\cal M}_{(s+k-1)}^{\vec{\alpha}}({\bf r})\,
j_0^{\gamma}({\bf{r}})
\nonumber \\
&-& gf^{a\beta d}\,A_{i}^{\beta}({\bf r})\,
\sum_{s=0}\sum_{t=1} (-1)^{t}g^{t+s+k-1}\,
{\textstyle\frac{B(s)}{s!k!(t+1)!}}\,
f^{\vec{\mu}d\lambda}_{(t)}
f^{\vec{\alpha}\lambda\gamma}_{(s+k-1)}\,
{\textstyle\frac{\partial_{i}}{\partial^{2}}}
\left(\,{\cal R}_{(t)}^{\vec{\mu}}({\bf r})\,
{\cal M}_{(s+k-1)}^{\vec{\alpha}}({\bf r})\,
j_0^{\gamma}({\bf{r}})\,\right)\;.
\label{eq:SRkth}
\end{eqnarray}
When we sum over the entire series, we can change
variables in the integer-valued indices to $\eta=k+s-1$,
and perform the summation over
$\eta$ and $s$, with $k=\eta-s+1$. The summation over $s$ then
involves nothing but the Bernoulli numbers
and fractional coefficients, so that we obtain
\begin{eqnarray}
e^{\bar {\cal C}}\,{\cal G}^{a}({\bf{r}})\,e^{-\bar {\cal C}}& =
& {\cal G}^{a}({\bf r})-gf^{a\beta\gamma}\,
A_{i}^{\beta}({\bf r})
{\textstyle\frac{\partial_{i}}{\partial^{2}}}\,j_{0}^{\gamma}({\bf r})
\nonumber \\
&+& \sum_{\eta=1}g^{\eta +1}D_0^\eta (\eta)f^{a\beta c}
f^{\vec{\alpha}c\gamma}_{(\eta)}\,A_{i}^{\beta}({\bf r})\,
{\textstyle \frac{\partial_{i}}{\partial^{2}}}\left(\,
{\cal M}_{(\eta)}^{\vec{\alpha}}({\bf r})\,
\,j_{0}^{\gamma}({\bf r})\,\right)
\nonumber \\
&+& \sum_{\eta=0}
\sum_{t=1} (-1)^{t}g^{t+\eta}\,
{\textstyle \frac{D_0^\eta (\eta)}{(t-1)!(t+1)}}\,
f^{\vec{\mu}a\lambda}_{(t)}
f^{\vec{\alpha}\lambda\gamma}_{(\eta)}\,
{\cal R}_{(t)}^{\vec{\mu}}({\bf r})\,
{\cal M}_{(\eta )}^{\vec{\alpha}}({\bf r})\,
j_{0}^{\gamma}({\bf r})
\nonumber \\
&-& gf^{a\mu d}\,A_{i}^{\mu}({\bf r})\,\sum_{\eta=0}\sum_{t=1}
(-1)^{t}g^{t+\eta}
{\textstyle\frac{D_0^\eta (\eta)}{(t+1)!}}\,
f^{\vec{\mu}d\lambda}_{(t)}
f^{\vec{\alpha}\lambda\gamma}_{(\eta)}\,
{\textstyle\frac{\partial_{i}}{\partial^{2}}}
\left(\,{\cal R}_{(t)}^{\vec{\mu}}({\bf r})\,
{\cal M}_{(\eta )}^{\vec{\alpha}}({\bf r})\,
j_{0}^{\alpha}({\bf r})\,\right)\;,
\label{eq:SRsum}
\end{eqnarray}
where $D_0^\eta (\eta)$ is the sum over Bernoulli numbers
defined in Eq.~(\ref{bernoulli}).  $D_0^\eta (\eta)$ has the
values $D_0^\eta (\eta)=0$ for $\eta\neq 0$, and $D_0^0 (0)=1$.
Since $f^{\vec{\alpha}\lambda\gamma}_{(0)}=
-\delta_{\lambda ,\gamma}$,
we find that substitution of these values into
Eq.~(\ref{eq:SRsum}) reduces it identically to
Eq.~(\ref{eq:GGhatleft}) and thereby proves
Eqs.~(\ref{eq:GgqGg}) and (\ref{eq:gghatp}),
demonstrating the unitary equivalence of
${\hat {\cal G}}^{a}({\bf r})$ and ${\cal G}^{a}({\bf r})$.
\bigskip

The demonstration of unitary equivalence of
${\hat {\cal G}}^{a}({\bf r})$ and
${\cal G}^{a}({\bf r})$ enables us to assign two different roles to
${\cal G}^{a}({\bf r})$.  On the one
hand, ${\cal G}^{a}({\bf r})$ can be viewed as the Gauss's law
operator for `pure glue' QCD and ${\hat {\cal G}}^{a}({\bf r})$ as the
Gauss's law operator for the theory that includes  quarks as
well as gluons.   But ${\cal G}^{a}({\bf r})$
can also be viewed as the Gauss's law operator for QCD {\em with}
interacting quarks and gluons, in a
representation in which all operators and states have been
transformed with a similarity transformation
that transforms ${\hat {\cal G}}^{a}({\bf r})$ into
${\cal G}^{a}({\bf r})\,$ and that similarly transforms all other
operators and states as well,
but that leaves matrix elements unchanged.
We will designate the representation in which
${\hat {\cal G}}^{a}({\bf r})$ represents the
Gauss's law operator for QCD with quarks as well
as gluons, and in which ${\cal G}^{a}({\bf r})$
represents the `pure glue' Gauss's law operator,
 as the `common' or ${\cal C}$ representation.
The unitarily transformed representation,
in which  ${\cal G}^{a}({\bf r})$ represents the Gauss's
law operator for QCD with interacting
quarks and gluons, will be designated the ${\cal N}$ representation.
We can use the relationship between
these two representations to construct states
that implement the `complete' Gauss's law ---
Eq.~(\ref{eq:Ggqlaw}) --- from
\begin{equation}
{\cal G}^{a}({\bf r})\,{\Psi}\,|{\phi}\rangle =0,
\label{eq:Gglaw}
\end{equation}
which is the `pure glue' form of Gauss's law in the ${\cal C}$
representation.
We can simply view Eq.~(\ref{eq:Gglaw}) as the statement of
the complete Gauss's law --- the version
that includes interacting quarks and gluons ---
but in the ${\cal N}$ representation.
In order to transform Eq.~(\ref{eq:Gglaw}) ---
now representing Gauss's law with interacting quarks and gluons ---
from the
${\cal N}$ to the ${\cal C}$ representation, we
make use of the fact that
\begin{equation}
{\hat {\cal G}}^{a}({\bf r})\,{\hat {\Psi}}\,
|{\phi}\rangle={\cal{U}}_{\cal{C}}\,
{\cal G}^{a}({\bf r})\,{\cal{U}}^{-1}_{\cal{C}}\,
{\cal{U}}_{\cal{C}}\,{\Psi}\,|{\phi}\rangle =0\;,
\label{eq:Gtrans}
\end{equation}
identifying ${\hat {\Psi}}\,|{\phi}\rangle =
{\cal{U}}_{\cal{C}}\,{\Psi}\,|{\phi}\rangle$ as a state
that implements Gauss's law for a theory with
quarks and gluons, in the ${\cal C}$ representation.

\section{Gauge-invariant spinor and gauge fields}

We can apply the unitary equivalence
demonstrated in the preceding section to
the construction of gauge-invariant spinor
and gauge field operators. We observe that
Gauss's Law has a central role in generating
local gauge transformations, in which
the operator-valued gauge and spinor fields in a
gauge theory --- QCD in this case ---
are gauge-transformed by an arbitrary c-number
field $\omega^a({\bf{r}})$ consistent with the gauge condition that
underlies the canonical theory.  In this, the temporal gauge, such
gauge transformations are implemented by
\begin{equation}
{\cal{O}}({\bf{r}})\,\rightarrow\,
{\cal{O}}^\prime({\bf{r}})
=\,\exp\left(-\frac{i}{g}{\int}\hat{\cal{G}}^a({\bf{r}}^\prime)\,
\omega^a({\bf{r}}^\prime)\,d{\bf{r}}^\prime\,\right)\,
{\cal{O}}({\bf{r}})\,
\exp\left(\frac{i}{g}{\int}\hat{\cal{G}}^a({\bf{r}}^\prime)\,
\omega^a({\bf{r}}^\prime)\,d{\bf{r}}^\prime\,\right)\;,
\label{eq:gaugetrans}
\end{equation}
where $\omega^a({\bf{r}})$ is time-independent, and where
${\cal{O}}({\bf{r}})$ represents any of the
operator-valued fields of the gauge theory and
${\cal{O}}^\prime({\bf{r}})$ its gauge-transformed
form\cite{jacktop}. Eq.~(\ref{eq:gaugetrans}) applies
to QCD with quarks and gluons, and is expressed
in the ${\cal C}$ representation. It is obvious
that any operator-valued
field that commutes with ${\hat{\cal{G}}}^a({\bf{r}})$
is gauge-invariant.
\bigskip

We can also formulate the same gauge transformations
in the ${\cal N}$ representation, in which case they
take the form
\begin{equation}
{\cal{O}}_{\cal N}({\bf{r}})\,\rightarrow\,
{\cal{O}}_{\cal N}^\prime({\bf{r}})
=\,\exp\left(-\frac{i}{g}{\int}{\cal{G}}^a({\bf{r}}^\prime)\,
\omega^a({\bf{r}}^\prime)\,d{\bf{r}}^\prime\,\right)\,
{\cal{O}}_{\cal N}({\bf{r}})\,
\exp\left(\frac{i}{g}{\int}{\cal{G}}^a({\bf{r}}^\prime)\,
\omega^a({\bf{r}}^\prime)\,d{\bf{r}}^\prime\,\right)\;,
\label{eq:gaugetransN}
\end{equation}
where ${\cal{O}}_{\cal N}({\bf{r}})$ now represents a spinor
or gauge field in the ${\cal N}$ representation.
Eq.~(\ref{eq:gaugetransN}) has the same form as the equation  that
implements gauge-transformations for `pure glue' QCD in the ${\cal C}$
representation, but it has
a very different meaning.  In Eq.~(\ref{eq:gaugetransN}),
the operator-valued field ${\cal{O}}_{\cal N}({\bf{r}})$, and
${\cal{G}}^a({\bf{r}})$ which here
represents the {\em entire} Gauss's law --- quarks and
gluons included --- both are in the ${\cal N}$ representation.
\bigskip

It is easy to see that the spinor field $\psi({\bf{r}})$ is
a gauge-invariant spinor in the ${\cal N}$ representation, because
$\psi({\bf{r}})$ and ${\cal{G}}^a({\bf{r^\prime}})$ trivially
commute. To produce ${\psi}_{\sf GI}({\bf{r}}),$ this
gauge-invariant spinor
transposed into the ${\cal C}$ representation, we make use of
\begin{equation}
{\psi}_{\sf GI}({\bf{r}})={\cal{U}}_{\cal C}\,
\psi({\bf{r}})\,{\cal{U}}^{-1}_{\cal C}\;
\;\;\;\mbox{and}\;\;\;
{\psi}_{\sf GI}^\dagger({\bf{r}})={\cal{U}}_{\cal C}\,
\psi^\dagger({\bf{r}})\,{\cal{U}}^{-1}_{\cal C}\;.
\label{eq:psiqcd}
\end{equation}
We can easily carry out the unitary transformations in
Eq.~(\ref{eq:psiqcd})
to give
\begin{equation}
{\psi}_{\sf GI}({\bf{r}})=V_{\cal{C}}({\bf{r}})\,\psi ({\bf{r}})
\;\;\;\mbox{\small and}\;\;\;
{\psi}_{\sf GI}^\dagger({\bf{r}})=
\psi^\dagger({\bf{r}})\,V_{\cal{C}}^{-1}({\bf{r}})\;,
\label{eq:psiqcdg1}
\end{equation}
where
\begin{equation}
V_{\cal{C}}({\bf{r}})=
\exp\left(\,-ig{\overline{{\cal{Y}}^\alpha}}({\bf{r}})
{\textstyle\frac{\lambda^\alpha}{2}}\,\right)\,
\exp\left(-ig{\cal X}^\alpha({\bf{r}})
{\textstyle\frac{\lambda^\alpha}{2}}\right)\;,
\label{eq:el1}
\end{equation}
and
\begin{equation}
V_{\cal{C}}^{-1}({\bf{r}})=
\exp\left(ig{\cal X}^\alpha({\bf{r}})
{\textstyle\frac{\lambda^\alpha}{2}}\right)\,
\exp\left(\,ig{\overline{{\cal{Y}}^\alpha}}({\bf{r}})
{\textstyle\frac{\lambda^\alpha}{2}}\,\right)\;.
\label{eq:eldagq1}
\end{equation}
Because we have given an explicit expression for
$\overline{{\cal Y}^{\alpha}}({\bf{r}})$ in
Eqs.~(\ref{eq:defY}) and (\ref{eq:inteq2}),  Eq.~(\ref{eq:psiqcdg1})
represents complete, non-perturbative expressions for
gauge-invariant spinors in the
${\cal C}$ representation.  We can, if we choose,
expand  Eqs.~(\ref{eq:psiqcdg1})  to arbitrary order. We then
find that to $O(g^3)$, we agree with Refs.~\cite{lavelle2,lavelle5} in
which a perturbative construction of a
gauge-invariant spinor is carried out to $O(g^3).$ Furthermore, in the
${\cal C}$ representation,  ${\psi}({\bf{r}})$ gauge-transforms as
\begin{equation}
{\psi}({\bf{r}})\,\rightarrow\,
\psi^\prime({\bf{r}})=\,
\exp\left(i\omega^\alpha({\bf{r}})\,
{\textstyle\frac{\lambda^\alpha}{2}}\,\right)\,\psi({\bf{r}})\;.
\label{eq:psitransf}
\end{equation}
Since ${\psi}_{\sf GI}({\bf{r}})$ has been shown to be
gauge-invariant, it
immediately follows that $V_{\cal{C}}({\bf{r}})$
gauge-transforms as
\begin{equation}
V_{\cal{C}}({\bf{r}})\rightarrow
V_{\cal{C}}({\bf{r}})\exp\left(-i\omega^\alpha({\bf{r}})\,
{\textstyle\frac{\lambda^\alpha}{2}}\,\right)\;\;\;\;
\mbox{and}\;\;\;
V^{-1}_{\cal{C}}({\bf{r}})\rightarrow
\exp\left(i\omega^\alpha({\bf{r}})\,
{\textstyle\frac{\lambda^\alpha}{2}}\,\right)
V_{\cal{C}}^{-1}({\bf{r}})\;.
\end{equation}

The procedure we have used to construct gauge-invariant spinors
is not applicable to the construction of gauge-invariant gauge
fields, because we do not have ready access to a form of the gauge
field that is trivially gauge invariant in either the ${\cal C}$
or the ${\cal N}$ representation.  We will, however, discuss two
methods for constructing gauge-invariant gauge fields.
One method is based on the observation that the states $|{\phi}\rangle$
for which
\begin{equation}
{\hat {\cal G}}^a{\hat {\Psi}}\,|{\phi}\rangle=
{\hat {\cal G}}^a{\cal{U}}_{\cal{C}}\,{\Psi}\,|{\phi}\rangle=0
\label{eq:phiset}
\end{equation}
include any state $|\phi_{A_{T\,i}^{b}({\bf{r}})}\rangle$ in which
the transverse gauge field $A_{T\,i}^{b}({\bf{r}})$
acts on another $|{\phi}\rangle$ state.  This is an immediate
consequence of the fact that
${\hat {\cal G}}^a\,{\hat {\Psi}}={\hat {\Psi}}\,b_{Q}^{a}({\bf{k}})
+B_{Q}^{a}({\bf{k}}),$ and that $A_{T\,i}^{b}({\bf{r}})$
trivially commutes with $\partial_i\Pi_i^{a}({\bf{r}}^\prime\,).$
We use the commutator algebra
for the operator-valued fields to maneuver the transverse gauge field,
along with all further gauge field functionals generated in this
process, to the left of ${\cal{U}}_{\cal{C}}\,{\Psi}$ in
${\cal{U}}_{\cal{C}}\,{\Psi}\,A_{T\,i}^{b}({\bf{r}})\,
|{\phi}\rangle.$ We then obtain the result that
\begin{equation}
{\hat {\Psi}}\,A_{T\,i}^{b}({\bf{r}})\,|{\phi}\rangle=
A_{{\sf GI}\,i}^b({\bf{r}})\,{\hat {\Psi}}\,|{\phi}\rangle,
\label{eq:Agi}
\end{equation}
where $A_{{\sf GI}\,i}^b({\bf{r}})$ is a gauge-invariant gauge field
created in the process of commuting $A_{T\,i}^{b}({\bf{r}})$ past the
$\Psi$ to its left.
The gauge-invariance of $A_{{\sf GI}\,i}^b({\bf{r}})$ follows from
the fact that  the Gauss's law operator ${\hat {\cal G}}^a$ annihilates
both sides of Eq.~(\ref{eq:Agi}).
 Eqs.~(\ref{eq:phiset}) and (\ref{eq:Agi}) require that the
commutator $\left[{\hat {\cal G}}^a,\,
A_{{\sf GI}\,i}^b({\bf{r}})\, \right]=0,$ and it then follows
directly from Eq.~(\ref{eq:gaugetrans}) that
$A_{{\sf GI}\,i}^b({\bf{r}})$ is gauge-invariant.
It only remains for us to
find an explicit expression for $A_{{\sf GI}\,i}^b({\bf{r}}).$
We first observe from Eqs.~(\ref{eq:GgqGg}) and (\ref{eq:CCbar})
that the gauge field and all functionals of gauge fields commute with
${\cal{U}}_{\cal{C}}.$ We further see that
\begin{equation}
A_{{\sf GI}\,i}^b({\bf{r}})\,{\Psi}=\left[\Psi,\,
A_{T\,i}^{b}({\bf{r}})\, \right]
+A_{T\,i}^{b}({\bf{r}})\,\Psi.
\label{eq:Agieq}
\end{equation}
When we expand $\Psi$ as
\begin{eqnarray}
\Psi&&={\|}\,\exp({\cal{A}})\,{\|}={\|}\,
\exp\left(i{\int}\,d{\bf{r}}\,
\overline{{\cal{A}}^\gamma_{k}}({\bf{r}})\,
\Pi^\gamma_{k}({\bf{r}})\right){\|}
\nonumber\\
&&=1+i{\int}\,d{\bf{r}}_1)\,
\overline{{\cal{A}}^{\gamma}_{k}}({\bf{r}}_1)\,
\Pi^\gamma_{k}({\bf{r}}_1)+{\textstyle\frac{(i)^2}{2}}
{\int}\,d{\bf{r}}_1\,d{\bf{r}}_2\,
\overline{{\cal{A}}^{\gamma_1}_{k_1}}({\bf{r}}_1)\,
\overline{{\cal{A}}^{\gamma_2}_{k_2}}({\bf{r}}_2)\,
\Pi^{\gamma_1}_{k_1}({\bf{r}}_1)\,
\Pi^{\gamma_2}_{k_2}({\bf{r}}_2)+\cdots
\nonumber\\
&&+{\textstyle\frac{(i)^n}{n\,!}}{\int}\,
d{\bf{r}}_1\,d{\bf{r}}_2\,\cdots\,d{\bf{r}}_n\,
\overline{{\cal{A}}^{\gamma_1}_{k_1}}({\bf{r}}_1)\,
\overline{{\cal{A}}^{\gamma_2}_{k_2}}({\bf{r}}_2)\,\cdots\,
\overline{{\cal{A}}^{\gamma_n}_{k_n}}({\bf{r}}_n)\,
\Pi^{\gamma_1}_{k_1}({\bf{r}}_1)\,
\Pi^{\gamma_2}_{k_2}({\bf{r}}_2)\,\cdots\,
\Pi^{\gamma_n}_{k_n}({\bf{r}}_n)
\nonumber \\
&&+\,\cdots
 \label{eq:Psiexp}
\end{eqnarray}
it becomes evident that
\begin{eqnarray}
[\,{\Psi},\,A_{T\,i}^b({\bf{r}})\,]&=&
\,(\delta_{ij}-{\textstyle\frac{\partial_{i}\partial_j}
{\partial^2}})\overline{{\cal{A}}^b_{j}}({\bf{r}})\,
+(\delta_{ij}-{\textstyle\frac{\partial_{i}\partial_j}
{\partial^2}})\overline{{\cal{A}}^b_{j}}({\bf{r}})\,
i{\int}\,d{\bf{r}}_1\,
\overline{{\cal{A}}^{\gamma}_{k}}({\bf{r}}_1)\,
\Pi^{\gamma}_{k}({\bf{r}}_1)+\cdots
\nonumber\\
&&+(\delta_{ij}-{\textstyle\frac{\partial_{i}\partial_j}
{\partial^2}})\overline{{\cal{A}}^b_{j}}({\bf{r}})\,
{\textstyle\frac{(i)^{n-1}}{(n-1)\,!}}{\int}\,
d{\bf{r}}_1\,d{\bf{r}}_2\,\cdots\,d{\bf{r}}_{n-1}\,
\overline{{\cal{A}}^{\gamma_1}_{k_1}}({\bf{r}}_1)\,
\overline{{\cal{A}}^{\gamma_2}_{k_2}}({\bf{r}}_2)\,\cdots\,
\overline{{\cal{A}}^{\gamma_{n-1}}_{k_{n-1}}}({\bf{r}}_{n-1})\,
\nonumber\\
&&\;\;\;\;\;\;\;\;\;\;\times
\Pi^{\gamma_1}_{k_1}({\bf{r}}_1)\,
\Pi^{\gamma_2}_{k_2}({\bf{r}}_2)\,\cdots\,
\Pi^{\gamma_{n-1}}_{k_{n-1}}({\bf{r}}_{n-1})
+\cdots \nonumber \\
&=&(\delta_{ij}-{\textstyle\frac{\partial_{i}\partial_j}
{\partial^2}})\overline{{\cal{A}}^b_{j}}({\bf{r}})\,\Psi\;,
\label{eq:bob4a}
\end{eqnarray}
and therefore that the gauge-invariant gauge field is
\begin{equation}
A_{{\sf GI}\,i}^b({\bf{r}})=
A_{T\,i}^b({\bf{r}}) +
[\delta_{ij}-{\textstyle\frac{\partial_{i}\partial_j}
{\partial^2}}]\overline{{\cal{A}}^b_{j}}({\bf{r}})=
a_{i}^{b} ({\bf{r}})+\overline{{\cal{A}}^b_{i}}({\bf{r}})
-\partial_{i}\overline{{\cal Y}^{b}}({\bf{r}})\;.
\label{eq:Adressedthree1b}
\end{equation}

Confirmation of this result can be obtained from the fact that
$A_{{\sf GI}\,i}^b({\bf{r}})$ commutes with ${\cal G}^a$ --- and
therefore also with
${\hat {\cal G}}^a.$  We observe that
\begin{eqnarray}
\left[\,{\cal{G}}^a({\bf{r}}),\,
A_{{\sf GI}\,i}^b({\bf{r}}^\prime)\right]=
\left[\,{\cal{G}}^a({\bf{r}}),\,
\left(A_{i\,T}^{b}({\bf{r}}^\prime)
+(\delta_{ij}-{\textstyle\frac{\partial_{i}\partial_j}{\partial^2}})
\overline{{\cal{A}}^b_{j}}({\bf{r}}^\prime)\,\right)\,\right]&=&
\nonumber \\
{\int}\,d{\bf{y}}\,
\left\{\left[\,{\cal{G}}^a({\bf{r}}),\,
{A}^b_{j}({\bf{y}})\,\right]+\left[\,{\cal{G}}^a({\bf{r}}),\,
\overline{{\cal{A}}^b_{j}}({\bf{y}})\,\right]\,\right\}
V_{ij}({\bf{y}}-{\bf{r}}^\prime)&=&0\;,
\label{eq:dirgi}
\end{eqnarray}
where
\begin{equation}
V_{ij}({\bf{y}}-{\bf{r}}^\prime)=
(\delta_{ij}-{\textstyle\frac{\partial_{i}\partial_j}{\partial^2}})\,
\delta({\bf{y}}-{\bf{r}}^\prime)\;.
\label{eq:vjay}
\end{equation}
Eq.~(\ref{eq:dirgi}) follows directly from Eq.~(\ref{eq:Asum});
$\int d{\bf y}\,\left[\,{\cal{G}}^a({\bf{r}}),\,
\overline{{\cal{A}}^b_{j}}({\bf{y}})\,\right]\,
V_{ij}({\bf{y}}-{\bf{r}}^\prime)$ can be identified as the
first line of that equation, when the integration
over ${\bf{y}}$ in Eq.~(\ref{eq:dirgi})
is identified with
the integration over ${\bf{r}}^\prime$ in Eq.~(\ref{eq:Asum}), and
when the tensor element
$V_{ij}({\bf{y}}-{\bf{r}}^\prime),$ with ${\bf r}^\prime$ and
$i$ fixed, is substituted for the
vector component $V^\gamma_{j}$ in Eq.~(\ref{eq:Asum}). Similarly,
$\int d{\bf y}\left[\,{\cal{G}}^a({\bf{r}}),\,
{A}^b_{i}({\bf{y}})\,\right]\,
V_{ij}({\bf{y}}-{\bf{r}}^\prime)$ can be identified as the second
line of Eq.~(\ref{eq:Asum}).  The remaining three lines of
Eq.~(\ref{eq:Asum}) vanish because
$\partial_{j}V_{ij}({\bf{y}}-{\bf{r}}^\prime)=0$ is
an identity. In this way,  Eq.~(\ref{eq:Asum}) accounts for the
gauge-invariance of $A_{{\sf GI}\,i}^b({\bf{r}}).$
\bigskip

Another method for constructing a gauge-invariant gauge field is based
on the observation that
$V_{\cal{C}}({\bf{r}})$ can be written as an exponential function.
We can make use of the BHC theorem that
$e^{\sf u}e^{\sf v}=e^{\sf w}\,,$
where ${\sf w}$ is a series whose initial term is ${\sf u}+{\sf v},$
and whose higher order terms are multiples of
successive commutators of ${\sf u}$ and ${\sf v}$ with earlier
terms in that series.
Since the commutator algebra of the Gell-Mann
matrices ${\lambda}^\alpha$ is closed,
$V_{\cal{C}}({\bf{r}})$ must be of the form
$\exp[-ig{\cal Z}^\alpha({\lambda}^\alpha/2)],$ where
\begin{equation}
\exp\left[-ig{\cal Z}^\alpha
{\textstyle\frac{{\lambda}^\alpha}{2}}\right]=
\exp\left[-ig{\overline {\cal Y}^\alpha}
{\textstyle\frac{{\lambda}^\alpha}{2}}\right]\,
\exp\left[-ig{\cal X}^\alpha
{\textstyle\frac{{\lambda}^\alpha}{2}}\right]
\label{eq:Zxy}
\end{equation}
and ${\cal Z}^\alpha$ is a functional of gauge fields
(but not of their canonical momenta).
$V_{\cal{C}}({\bf{r}})$ therefore can be viewed as a
particular case of the operator $\exp\left[i\omega^\alpha({\bf{r}})
\left(\lambda^{\alpha}/2\right)\right]$ that
gauge-transforms the spinor field ${\psi}({\bf r})\,;$
$\omega^\alpha$ in this
case is ${\cal Z}^\alpha$ and therefore a functional of gauge fields
that commutes with all other functionals
of gauge and spinor fields.
Moreover, we can refer to the Euler-Lagrange
equation (in the $A_0=0$ gauge)
for the spinor field ${\psi}({\bf r}),$
\begin{equation}
\left[im+\gamma_j\left(\partial_{j}-ig\,A_{j}^\alpha({\bf{r}})
{\textstyle\frac{\lambda^\alpha}{2}}\right)+
\gamma_{0}\partial_{0}\right]\,{\psi}({\bf r})=0,
\label{eq:Diracspin}
\end{equation}
where we have used the same non-covariant notation for the
gauge fields as in
Ref.~\cite{khymtemp} ($i.e.$ $A_{j}^\alpha({\bf{r}})$
designates contravariant and
$\partial_{j}$ covariant quantities), and where $\gamma_0=\beta$ and
$\gamma_j={\beta}{\alpha_j}.$  Although the gauge fields are
operator-valued, they commute with all other operators in
Eq.~(\ref{eq:Diracspin}) --- with the exception of the
derivatives $\partial_j\,$ --- so that, when only time-independent
gauge-transformations are considered, $V_{\cal{C}}({\bf{r}}),$ acting
as an operator that gauge-transforms $\psi,$
behaves as though ${\cal Z}^\alpha$ were a c-number.
The gauge-transformed
gauge field, that corresponds to the gauge-transformed spinor
${\psi}_{{\sf GI}}({\bf{r}})=V_{\cal{C}}({\bf{r}})\,\psi ({\bf{r}})$,
therefore also is gauge-invariant; it is given by
\begin{equation}
[\,A_{{\sf GI}\,i}^{b}({\bf{r}})\,{\textstyle\frac{\lambda^b}{2}}\,]
=V_{\cal{C}}({\bf{r}})\,[\,A_{i}^b({\bf{r}})\,
{\textstyle\frac{\lambda^b}{2}}\,]\,
V_{\cal{C}}^{-1}({\bf{r}})
+{\textstyle\frac{i}{g}}\,V_{\cal{C}}({\bf{r}})\,
\partial_{i}V_{\cal{C}}^{-1}({\bf{r}})\;.
\label{eq:AdressedAxz}
\end{equation}
Since {\em further} gauge transformations must be carried
out simultaneously
on $\psi ({\bf{r}})$ and $V_{\cal{C}}({\bf{r}}),$ and must leave
${\psi}_{{\sf GI}}({\bf{r}})$ untransformed,
$A_{{\sf GI}\,i}^{b}({\bf{r}})$ must also
therefore remain untransformed by further gauge transformations.
$A_{{\sf GI}\,i}^{b}({\bf{r}})$ thus is identified as a
gauge-invariant gauge field.
\bigskip

To find an explicit form for $[\,A_{{\sf GI}\,i}^{b}({\bf{r}})\,
{\textstyle\frac{\lambda^b}{2}}\,]$ from the r.h.s. of
Eq.~(\ref{eq:AdressedAxz}),
we use Eq.~(\ref{eq:LCA1}), with $V_{j}^{\gamma}({\bf r})=
\delta_{ij}({\lambda}^{\gamma}
/2)\,,$ to obtain
\begin{equation}
\left[a_{i}^{\gamma} ({\bf{r}})+
{\overline{{\cal{A}}^{\gamma}_i}}({\bf{r}})-\sum_{\eta=1}^\infty
{\textstyle\frac{g^\eta}{\eta!}}\,f^{\vec{\alpha}\beta\gamma}_{(\eta)}\,
{\cal{M}}_{(\eta)}^{\vec{\alpha}}({\bf{r}})\,
\overline{{\cal{B}}_{(\eta) i}^{\beta}}({\bf{r}})\,\right]
{\textstyle\frac{\lambda^\gamma}{2}}=
\left[a_{i}^{\gamma} ({\bf{r}})+\sum_{\eta=1}^\infty
{\textstyle\frac{g^\eta}{\eta!}}
\,\psi^{\gamma}_{(\eta)i}({\bf{r}})\,\right]
{\textstyle\frac{{\lambda}^\gamma}{2}}.
\label{eq:inteq3}
\end{equation}
It is straightforward but tedious to show that
\begin{equation}
\left[a_{i}^{\gamma} ({\bf{r}})+\sum_{\eta=1}^\infty
{\textstyle\frac{g^\eta}{\eta!}}
\,\psi^{\gamma}_{(\eta)i}({\bf{r}})\,\right]
{\textstyle\frac{{\lambda}^\gamma}{2}}=
\exp\left(\,-ig\,{\cal X}^\alpha ({\bf{r}})
{\textstyle\frac{\lambda^\alpha}{2}}\,\right)\,
\left[A_{i}^{\gamma} ({\bf{r}})
{\textstyle\frac{{\lambda}^\gamma}{2}\,+\frac{i}{g}}\partial_i\,\right]
\exp\left(\,ig\,{\cal X}^\alpha ({\bf{r}})
{\textstyle\frac{\lambda^\alpha}{2}}\,\right)\,,
\label{eq:psident}
\end{equation}
\begin{equation}
\left[a_{i}^{\gamma} ({\bf{r}})-\sum_{\eta=1}^\infty
{\textstyle\frac{g^\eta}{\eta!}}\,
f^{\vec{\alpha}\beta\gamma}_{(\eta)}\,
{\cal{M}}_{(\eta)}^{\vec{\alpha}}({\bf{r}})\,
a_{i}^{\gamma} ({\bf{r}})\right]
{\textstyle\frac{{\lambda}^\gamma}{2}}=
\exp\left(\,ig\,{\overline{{\cal Y}^\alpha}}({\bf{r}})
{\textstyle\frac{\lambda^\alpha}{2}}\,\right)\,
\left[a_{i}^{\gamma} ({\bf{r}})
{\textstyle\frac{{\lambda}^\gamma}{2}}\right]
\exp\left(\,-ig\,{\overline{{\cal Y}^\alpha}}({\bf{r}})
{\textstyle\frac{\lambda^\alpha}{2}}\,\right)\,,
\label{eq:aident}
\end{equation}

\begin{equation}
\left[\partial_{i}{\overline{{\cal Y}^{\gamma}}} ({\bf{r}})
-\sum_{\eta=1}^\infty
{\textstyle\frac{g^\eta}{\eta!}}\,
f^{\vec{\alpha}\beta\gamma}_{(\eta)}\,
{\cal{M}}_{(\eta)}^{\vec{\alpha}}({\bf{r}})\,
\partial_{i}{\overline{{\cal Y}^{\gamma}}} ({\bf{r}})\right]
{\textstyle\frac{{\lambda}^\gamma}{2}}=
\exp\left(\,ig\,{\overline{{\cal Y}^\alpha}}({\bf{r}})
{\textstyle\frac{\lambda^\alpha}{2}}\,\right)\,
\left[\partial_{i}{\overline{{\cal Y}^\gamma}} ({\bf{r}})
{\textstyle\frac{{\lambda}^\gamma}{2}}\right]
\exp\left(\,-ig\,{\overline{{\cal Y}^\alpha}}({\bf{r}})
{\textstyle\frac{\lambda^\alpha}{2}}\,\right)\,,
\label{eq:dyident}
\end{equation}
and
\begin{eqnarray}
&&\left[\overline{{\cal A}_{i}^{\gamma}}+\partial_{i}
{\overline{{\cal Y}^{\gamma}}} ({\bf{r}})-\sum_{\eta=1}^\infty
{\textstyle\frac{g^\eta}{\eta!}}\,
f^{\vec{\alpha}\beta\gamma}_{(\eta)}\,
{\cal{M}}_{(\eta)}^{\vec{\alpha}}({\bf{r}})\,
\left(\overline{{\cal A}_{i}^{\gamma}} ({\bf{r}})
+{\textstyle\frac{1}{\eta +1}}\partial_{i}
{\overline{{\cal Y}^\gamma}} ({\bf{r}})\right)\right]
{\textstyle\frac{{\lambda}^\gamma}{2}}
\nonumber \\
&&\;\;\;\;\;\;\;\;\;\;\;\;\;\;\;\;=
\exp\left(\,ig\,{\overline{{\cal Y}^\alpha}}({\bf{r}})
{\textstyle\frac{\lambda^\alpha}{2}}\,\right)\,
\left[\overline{{\cal A}_{i}^{\gamma}} ({\bf{r}})
{\textstyle\frac{{\lambda}^\gamma}{2}+\frac{i}{g}}\partial_i\,\right]
\exp\left(\,-ig\,{\overline{{\cal Y}^\alpha}}({\bf{r}})
{\textstyle\frac{\lambda^\alpha}{2}}\,\right)\,.
\label{eq:calaident}
\end{eqnarray}
Eqs. (\ref{eq:inteq3})-(\ref{eq:calaident}) leads to

\begin{equation}
V_{\cal{C}}({\bf{r}})\,[\,A_{i}^b({\bf{r}})\,
{\textstyle\frac{\lambda^b}{2}}\,]\,
V_{\cal{C}}^{-1}({\bf{r}})
+{\textstyle\frac{i}{g}}\,V_{\cal{C}}({\bf{r}})\,
\partial_{i}V_{\cal{C}}^{-1}({\bf{r}})\,=
A_{T\,i}^b ({\bf{r}}){\textstyle\frac{{\lambda}^b}{2}} +
[\delta_{ij}-{\textstyle\frac{\partial_{i}\partial_j}
{\partial^2}}]\overline{{\cal A}_{i}^b} ({\bf{r}})
{\textstyle\frac{{\lambda}^b}{2}}\;,
\end{equation}
so that the identical gauge-invariant gauge field is given in
Eqs.~(\ref{eq:Adressedthree1b}) and (\ref{eq:AdressedAxz}).
In  the gauge-invariant gauge field, as in the earlier case
of the gauge-invariant spinor, we find that when we
expand  Eq.~(\ref{eq:Adressedthree1b}) --- this time to $O(g^2)$ ---
we agree with Refs.~\cite{lavelle2,lavelle5} in which a perturbative
construction of a gauge-invariant gauge field is
carried out to that order.

\section{The case of Yang-Mills Theory}
Because of the simplicity of the $SU(2)$ structure constants,
it is instructive to examine
$\overline{{\cal{A}}^a_{j}}({\bf{r}})$ --- its defining equation and
its role in the `fundamental theorem' --- for the
case of Yang-Mills theory. For that purpose, we
substitute $\epsilon^{abc}$ --- the structure constants of
$SU(2)$ --- for the $f^{abc}$ required for $SU(3),$ in the
equations that pertain to
$\overline{{\cal{A}}^a_{j}}({\bf{r}})$.
$\epsilon^{\vec{\alpha}\beta\gamma}_{(\eta)},$
the $SU(2)$ equivalent of the $f^{\vec{\alpha}\beta\gamma}_{(\eta)}$
that are important in the definition of
$\overline{{\cal{A}}^a_{j}}({\bf{r}}),$ is given by
\begin{equation}
\epsilon^{\vec{\alpha}\beta\gamma}_{(\eta)}=(-1)^{\frac{\eta}{2}-1}
\delta_{\alpha[1]\alpha[2]}\,\delta_{\alpha[3]\alpha[4]}\,
\cdots\,\delta_{\alpha[\eta-3]\alpha[\eta-2]}\,
\epsilon^{\alpha[\eta-1]\beta b}\,
\epsilon^{b\alpha[\eta]\gamma}\;
\label{eq:fproductN2}
\end{equation}
and
\begin{equation}
\epsilon^{\vec{\alpha}\beta\gamma}_{(\eta)}=(-1)^{\frac{\eta-1}{2}}
\delta_{\alpha[1]\alpha[2]}\,\delta_{\alpha[3]\alpha[4]}\,
\cdots\,\delta_{\alpha[\eta-2]\alpha[\eta-1]}\,
\epsilon^{\alpha[\eta]\beta \gamma}\;
\label{eq:fproductN3}
\end{equation}
for even and odd $\eta$ respectively.
We can use Eqs.~(\ref{eq:fproductN2}) and (\ref{eq:fproductN3})  to
write the $SU(2)$ version of Eq.~(\ref{eq:inteq2}) for
$\overline{{\cal{A}}_{i}^{\gamma}}({\bf{r}}),$ which
appears (implicitly) as the coefficient of the
$\Pi_i^{\gamma}({\bf r})$ on the l.h.s. of that equation.
In doing so, we separate
$\overline{{\cal{A}}_{i}^{\gamma}}({\bf{r}})$ into two
parts
\begin{equation}
\overline{{\cal{A}}_{i}^{\gamma}}({\bf{r}})=
\overline{{\cal{A}}_{i}^{\gamma}}({\bf{r}})_{\cal X}+
\overline{{\cal{A}}_{i}^{\gamma}}({\bf{r}})_{\overline{\cal Y}}\;,
\label{eq:azero}
\end{equation}
where $\overline{{\cal{A}}_{i}^{\gamma}}({\bf{r}})_{\cal X}$
represents the part of
$\overline{{\cal{A}}_{i}^{\gamma}}({\bf{r}})$ that depends only
on `known' quantities
 that stem from the $\psi^{\gamma}_{(n)i}({\bf{r}})$ and are
functionals of gauge fields;
$\overline{{\cal{A}}_{i}^{\gamma}}({\bf{r}})_{\overline{\cal Y}}$
represents the
part that implicitly contains the
$\overline{{\cal{A}}_{i}^{\gamma}}({\bf{r}})$ itself.
In Section~\ref{sec-Implementing}, we showed how the perturbative
expansion of
$\overline{{\cal{A}}_{i}^{\gamma}}({\bf{r}})$ proceeds with the
construction of the
$n^{th}$ order term, ${\cal{A}}_{(n)i}^{\gamma}({\bf{r}}),$ from the
$\psi^{\gamma}_{(n)i}({\bf{r}})$ of
the same order, and from
${\cal{A}}_{(n^\prime)i}^{\gamma}({\bf{r}})$ of lower
orders --- in the $SU(2)$ case, the latter originating from
$\overline{{\cal{A}}_{i}^{\gamma}}({\bf{r}})_{\overline{\cal Y}}.$
The explicit forms of
$\overline{{\cal{A}}_{i}^{\gamma}}({\bf{r}})_{\cal X}$ and
$\overline{{\cal{A}}_{i}^{\gamma}}({\bf{r}})_{\overline{\cal Y}}$ are
\begin{eqnarray}
\overline{{\cal{A}}_{i}^{\gamma}}({\bf{r}})_{\cal X}
&&\,=g\epsilon^{\alpha\beta\gamma}\,
{\cal{X}}^\alpha({\bf{r}})\,A_i^\beta({\bf{r}})\,
{\textstyle\frac{\sin({\cal{N}})}{{\cal{N}}}}
\nonumber\\
&&-g\epsilon^{\alpha\beta\gamma}\,
{\cal{X}}^\alpha({\bf{r}})\,\partial_i{\cal{X}}^\beta({\bf{r}})\,
{\textstyle\frac{1-\cos({\cal{N}})}{{\cal{N}}^2}}
\nonumber\\
&&-g^2\epsilon^{\alpha\beta b}
\epsilon^{b\mu\gamma}\,
{\cal{X}}^\mu({\bf{r}})\,{\cal{X}}^\alpha({\bf{r}})
\,A_i^\beta({\bf{r}})\,
{\textstyle\frac{1-\cos({\cal{N}})}{{\cal{N}}^2}}
\nonumber\\
&&+g^2\epsilon^{\alpha\beta b}
\epsilon^{b\mu\gamma}\,
{\cal{X}}^\mu({\bf{r}})\,{\cal{X}}^\alpha({\bf{r}})\,
\partial_i{\cal{X}}^\beta({\bf{r}})\,
[{\textstyle\frac{1}{{\cal{N}}^2}}
-{\textstyle\frac{\sin({\cal{N}})}{{\cal{N}}^3}}]
\label{eq:a2X}
\end{eqnarray}
and
\begin{eqnarray}
\overline{{\cal{A}}_{i}^{\gamma}}({\bf{r}})_{\overline{\cal Y}}
&&\;=g\epsilon^{\alpha\beta\gamma}\,
\overline{{\cal{Y}}^\alpha}({\bf{r}})
\,\left(\,A_{T\,i}^\beta({\bf{r}}) +
(\delta_{ij}-{\textstyle\frac{\partial_{i}\partial_j}
{\partial^2}})\overline{{\cal{A}}^\beta_{j}}({\bf{r}})\,\right)\,
{\textstyle\frac{\sin(\overline{\cal{N}})}{\overline{\cal{N}}}}
\nonumber\\
&&+g\epsilon^{\alpha\beta\gamma}\,
\overline{{\cal{Y}}^\alpha}({\bf{r}})
\,\partial_i\overline{{\cal{Y}}^\beta}({\bf{r}})\,
{\textstyle\frac{1-\cos(\overline{\cal{N}})}
{\overline{\cal{N}}^2}}
\nonumber\\
&&+g^2\epsilon^{\alpha\beta b}
\epsilon^{b\mu\gamma}\,
\overline{{\cal{Y}}^\mu}({\bf{r}})\,
\overline{{\cal{Y}}^\alpha}({\bf{r}})
\,\left(\,A_{T\,i}^\beta({\bf{r}}) +
(\delta_{ij}-{\textstyle\frac{\partial_{i}\partial_j}
{\partial^2}})\overline{{\cal{A}}^\beta_{j}}({\bf{r}})\,\right)\,
{\textstyle\frac{1-\cos(\overline{\cal{N}})}{\overline{\cal{N}}^2}}
\nonumber\\
&&+g^2\epsilon^{\alpha\beta b}
\epsilon^{b\mu\gamma}\,
\overline{{\cal{Y}}^\mu}({\bf{r}})\,
\overline{{\cal{Y}}^\alpha}({\bf{r}})\,
\partial_i\overline{{\cal{Y}}^\beta}({\bf{r}})\,
[{\textstyle\frac{1}{\overline{\cal{N}}^2}}-
{\textstyle\frac{\sin(\overline{\cal{N}})}
{\overline{\cal{N}}^3}}]\;,
\label{eq:a2Y}
\end{eqnarray}
where
\begin{equation}
{\cal{N}}({\bf{r}})\equiv{\cal{N}}=\left[g^2\,
{\cal{X}}^\delta({\bf{r}})\,
{\cal{X}}^\delta({\bf{r}})\,\right]^{\frac{1}{2}}\;,
\end{equation}
and
\begin{equation}
\overline{\cal{N}}({\bf{r}})\equiv\overline{\cal{N}}=
\left[g^2\,\overline{{\cal{Y}}^\delta}({\bf{r}})\,
\overline{{\cal{Y}}^\delta}({\bf{r}})\,\right]^{\frac{1}{2}}\;.
\end{equation}
There is a striking resemblance in the structure of
Eqs.~(\ref{eq:a2X}) and (\ref{eq:a2Y}) on the one
hand, and $\left(A^{\gamma}\,\right)^{\prime}_{i},$ the
gauge-transformed gauge field $A^{\gamma}_{i},$
where the gauge transformation is by a finite gauge
function $\omega^\gamma.$
$\left(A^{\gamma}\,\right)^{\prime}_{i}$ is given by
\begin{eqnarray}
&&\left(A^{\gamma}\,\right)^{\prime}_{i}=\,
(\,A^\gamma_{i}+{\textstyle\frac{1}{g}}\,
\partial_i\omega^\gamma\,)
\nonumber\\
&&-\,\epsilon^{\alpha\beta\gamma}
\left(\,\omega^\alpha\,A_{i}^\beta\,
{\textstyle\frac{\sin(|\omega|)}{|\omega|}}\,
+\,{\textstyle\frac{1}{g}}\,
\omega^\alpha\,\partial_i\omega^\beta\,
{\textstyle\frac{1-\cos(|\omega|)}{|\omega|^2}}\,\right)
\nonumber\\
&&-\,\epsilon^{\alpha\beta b}\epsilon^{b\mu\gamma}\,
\left(\,\omega^\mu\omega^\alpha\,A^\beta_{i}\,
\,{\textstyle\frac{1-\cos(|\omega|)}{|\omega|^2}}\,
+{\textstyle\frac{\omega^\mu
\omega^\alpha\partial_i\omega^\beta}{g}}\,\,
(\,{\textstyle\frac{1}{|\omega|^2}}
-{\textstyle\frac{\sin(|\omega|)}{|\omega|^3}})\,\right)\;.
\label{eq:atranssu2}
\end{eqnarray}
The $SU(2)$ version of Eq.~(\ref{eq:Asum}) --- our
so-called `fundamental theorem' --- can similarly be given.
In that case, the summations over order and multiplicity
indices can be absorbed into trigonometric functions, and
we obtain the much
simpler equation
\begin{eqnarray}
i\int d{\bf r}^\prime&&[\,\partial_{i}\Pi_{i}^{a}({\bf r}),\,
\overline{{\cal A}_{j}^{\gamma}}({\bf r}^\prime)\,]\,
V_{j}^{\gamma}({\bf r}^\prime)
+ ig\epsilon^{a\beta d}A_{i}^{\beta}({\bf r})\int d{\bf r}^\prime
[\,\Pi_{i}^{d}({\bf r}),\,\overline{{\cal A}_{j}^{\gamma}}({\bf
r}^\prime)\,]\,
V_{j}^{\gamma}({\bf r}^\prime)
\nonumber \\
&&= -g\epsilon^{a\mu d}\,A_{i}^{\mu}({\bf r})\,V_{i}^{d}({\bf r})
\nonumber \\
&&-{\textstyle\frac{g^{2}}{2}}\,
\epsilon^{a\beta c}\epsilon^{{\alpha}c\gamma}\,A_{i}^{\beta}({\bf r})\,
{\textstyle \frac{\partial_{i}}{\partial^{2}}}\left(\,
\overline{{\cal Y}^{{\alpha}}}({\bf r})\,
\partial_{j}V_{j}^{\gamma}({\bf r})\,\right)
\nonumber \\
&&-g^{3}
\epsilon^{a\beta c}\epsilon^{\vec{\alpha}c\gamma}_{(2)}\,
A_{i}^{\beta}({\bf r})\,
{\textstyle \frac{\partial_{i}}{\partial^{2}}}
\left({\cal M}_{(2)}^{\vec{\alpha}}({\bf r})
\left[{\textstyle\frac{1}{2{\overline{\cal{N}}}}}
\cot{\left(\textstyle\frac{{\overline{\cal{N}}}}{2}\right)}
-{\textstyle\frac{1}{{\overline{\cal{N}}^2}}}\right]\,
\partial_{j}V_{j}^{\gamma}({\bf r})\,\right)
\nonumber \\
&&+g\epsilon^{{\mu}a\gamma}
\,{\cal X}^{{\mu}}({\bf r})
\left[{\textstyle\frac{\sin({\cal{N}})}{{\cal{N}}}}
-{\textstyle\frac{1-\cos({\cal{N}})}{{\cal{N}}^2}}
\right]\partial_{i}V_{i}^{\gamma}({\bf r})
\nonumber \\
&&+g^{2}\,
\epsilon^{\vec{\mu}a\gamma}_{(2)}\,
{\cal R}_{(2)}^{\vec{\mu}}({\bf r})\,
\left[{\textstyle\frac{\cos({\cal{N}})}{{\cal{N}}^2}}
-{\textstyle\frac{\sin({\cal{N}})}{{\cal{N}}^{3}}}\right]
\partial_{i}V_{i}^{\gamma}({\bf r})
\nonumber \\
&&+{\textstyle\frac{g^2}{2}}
\epsilon^{{\mu}a\lambda}\epsilon^{\alpha\lambda\gamma}
\,{\cal X}^{{\mu}}({\bf r})\overline{{\cal Y}^{{\alpha}}}({\bf r})\,
\left[{\textstyle\frac{\sin({\cal{N}})}{{\cal{N}}}}
-{\textstyle\frac{1-\cos({\cal{N}})}{{\cal{N}}^2}}
\right]\partial_{i}V_{i}^{\gamma}({\bf r})
\nonumber \\
&&+{\textstyle\frac{g^3}{2}}
\epsilon^{\vec{\mu}a\gamma}_{(2)}\,\epsilon^{\alpha\lambda\gamma}
{\cal R}_{(2)}^{\vec{\mu}}({\bf r})\,
\overline{{\cal Y}^{{\alpha}}}({\bf r})\,
\left[{\textstyle\frac{\cos({\cal{N}})}{{\cal{N}}^2}}
-{\textstyle\frac{\sin({\cal{N}})}{{\cal{N}}^{3}}}\right]
\partial_{i}V_{i}^{\gamma}({\bf r})
\nonumber \\
&&+g^3
\epsilon^{{\mu}a\lambda}
\,\epsilon^{\vec{\alpha}\lambda\gamma}_{(2)}
{\cal X}^{{\mu}}({\bf r}){\cal M}_{(2)}^{\vec{\alpha}}({\bf r})\,
\left[{\textstyle\frac{\sin({\cal{N}})}{{\cal{N}}}}
-{\textstyle\frac{1-\cos({\cal{N}})}{{\cal{N}}^2}}
\right]\left({\textstyle\frac{1}{2{\overline{\cal{N}}}}}
\cot{\left(\textstyle\frac{{\overline{\cal{N}}}}{2}\right)}
-{\textstyle\frac{1}{{\overline{\cal{N}}^2}}}\right)
\partial_{i}V_{i}^{\gamma}({\bf r})
\nonumber \\
&&+g^{4}\,
\epsilon^{\vec{\mu}a\lambda}_{(2)}\,
\epsilon^{\vec{\alpha}\lambda\gamma}_{(2)}
{\cal R}_{(2)}^{\vec{\mu}}({\bf r})\,
{\cal M}_{(2)}^{\vec{\alpha}}({\bf r})\,
\left[{\textstyle\frac{\cos({\cal{N}})}{{\cal{N}}^2}}
-{\textstyle\frac{\sin({\cal{N}})}{{\cal{N}}^{3}}}\right]
\left({\textstyle\frac{1}{2{\overline{\cal{N}}}}}
\cot{\left(\textstyle\frac{{\overline{\cal{N}}}}{2}\right)}
-{\textstyle\frac{1}{{\overline{\cal{N}}^2}}}\right)
\partial_{i}V_{i}^{\gamma}({\bf r})
\nonumber \\
&&-g^2 \epsilon^{a\beta d}
\epsilon^{{\mu}d\gamma}
A_{i}^{\beta}({\bf r})\,{\textstyle\frac{\partial_{i}}{\partial^{2}}}\,
\left(\,{\cal X}^{{\mu}}({\bf r})\,
{\textstyle\frac{1-\cos({\cal{N}})}{{\cal{N}}^2}}
\partial_{j}V_{j}^{\gamma}({\bf r})\,\right)
\nonumber \\
&&-g^3 \epsilon^{a\beta d}
\epsilon^{\vec{\mu}d\gamma}_{(2)}
A_{i}^{\beta}({\bf r})\,{\textstyle\frac{\partial_{i}}{\partial^{2}}}\,
\left(\,{\cal R}_{(2)}^{\vec{\mu}}({\bf r})\,
{\textstyle\frac{\sin({\cal{N}})-{\cal{N}}}{{\cal{N}}^3}}
\partial_{j}V_{j}^{\gamma}({\bf r})\,\right)
\nonumber \\
&&-{\textstyle\frac{g^3}{2}} \epsilon^{a\beta d}
\epsilon^{{\mu}d\lambda}\epsilon^{\alpha\lambda\gamma}
A_{i}^{\beta}({\bf r})\,
{\textstyle\frac{\partial_{i}}{\partial^{2}}}\,
\left(\,{\cal X}^{{\mu}}({\bf r})\,
\overline{{\cal Y}^{{\alpha}}}({\bf r})\,
{\textstyle\frac{1-\cos({\cal{N}})}{{\cal{N}}^2}}
\partial_{j}V_{j}^{\gamma}({\bf r})\,\right)
\nonumber \\
&&-{\textstyle\frac{g^4}{2}} \epsilon^{a\beta d}
\epsilon^{\vec{\mu}d\lambda}_{(2)}
\epsilon^{\alpha\lambda\gamma}
A_{i}^{\beta}({\bf r})\,
{\textstyle\frac{\partial_{i}}{\partial^{2}}}\,
\left(\,{\cal R}_{(2)}^{\vec{\mu}}({\bf r})\,
\overline{{\cal Y}^{{\alpha}}}({\bf r})\,
{\textstyle\frac{\sin({\cal{N}})-{\cal{N}}}{{\cal{N}}^3}}
\partial_{j}V_{j}^{\gamma}({\bf r})\,\right)
\nonumber \\
&&-g^4 \epsilon^{a\beta d}
\epsilon^{{\mu}d\lambda}\epsilon^{\vec{\alpha}\lambda\gamma}_{(2)}
A_{i}^{\beta}({\bf r})\,
{\textstyle\frac{\partial_{i}}{\partial^{2}}}\,
\left(\,{\cal X}^{{\mu}}({\bf r})\,
{\cal M}_{(2)}^{\vec{\alpha}}({\bf r})\,
{\textstyle\frac{1-\cos({\cal{N}})}{{\cal{N}}^2}}
\left[{\textstyle\frac{1}{2{\overline{\cal{N}}}}}
\cot{\left(\textstyle\frac{{\overline{\cal{N}}}}{2}\right)}
-{\textstyle\frac{1}{{\overline{\cal{N}}^2}}}\right]
\partial_{j}V_{j}^{\gamma}({\bf r})\,\right)
\nonumber \\
&&-g^5 \epsilon^{a\beta d}
\epsilon^{\vec{\mu}d\lambda}_{(2)}
\epsilon^{\vec{\alpha}\lambda\gamma}_{(2)}
A_{i}^{\beta}({\bf r})\,
{\textstyle\frac{\partial_{i}}{\partial^{2}}}\,
\left(\,{\cal R}_{(2)}^{\vec{\mu}}({\bf r})\,
{\cal M}_{(2)}^{\vec{\alpha}}({\bf r})\,
{\textstyle\frac{\sin({\cal{N}})-{\cal{N}}}{{\cal{N}}^3}}
\left[{\textstyle\frac{1}{2{\overline{\cal{N}}}}}
\cot{\left(\textstyle\frac{{\overline{\cal{N}}}}{2}\right)}
-{\textstyle\frac{1}{{\overline{\cal{N}}^2}}}\right]
\partial_{j}V_{j}^{\gamma}({\bf r})\,\right)\;.
\label{eq:fundthesu2}
\end{eqnarray}

To account for the general structure of
Eqs.~(\ref{eq:a2X}) and (\ref{eq:a2Y}),
we observe from Eqs.~(\ref{eq:Zxy}) and (\ref{eq:Diracspin}) that the
unitary transformation that transforms the spinor field
to its gauge-invariant
form {\em is itself a gauge transformation}.
$V_{\cal{C}}({\bf{r}})$ therefore
is an operator that gauge-transforms the spinor ${\psi}({\bf r})$ to a
form {\em that is then invariant to any
further gauge transformations}.
And $A_{{\sf GI}\,i}^{b}({\bf{r}}),$ which
is the corresponding gauge transform of the
gauge field $A_{i}^{b}({\bf{r}}),$  is similarly
invariant to any further gauge
transformations.  Eq.~(\ref{eq:Adressedthree1b}) identifies
$\overline{{\cal{A}}_{i}^{b}}({\bf{r}})$ as an essential constituent of
$A_{{\sf GI}\,i}^{b}({\bf{r}}),$ and Eqs.~(\ref{eq:a2X})
and (\ref{eq:a2Y})
specialize $\overline{{\cal{A}}_{i}^{b}}({\bf{r}})$
to its $SU(2)$ structure.
It is therefore not surprising to find that the relation between
$\overline{{\cal{A}}_{i}^{b}}({\bf{r}})$ and
$A_{i}^{b}({\bf{r}})$ anticipates the
relation between $A_{{\sf GI}\,i}^{b}({\bf{r}})$ and
$A_{i}^{b}({\bf{r}})$ --- $i.e.$ that
$A_{{\sf GI}\,i}^{b}({\bf{r}})$ is the gauge-transform of
$A_{i}^{b}({\bf{r}})$ by the finite gauge function
${\cal Z}^{b}({\bf{r}}),$ defined in Eq.~(\ref{eq:Zxy}).

\section{Discussion}

This paper has addressed four main topics:  The first
has been a proof of a previously published conjecture that states,
constructed in an earlier work\cite{bellchenhall} and
given in Eqs.~(\ref{eq:subcon}), (\ref{eq:Apsi}), and (\ref{eq:inteq2}),
implement the `pure glue' form of Gauss's law for QCD. Another
has been the construction of a unitary transformation
that extends these states so that they implement Gauss's
law for QCD with quarks as well as gluons. The third topic
is the construction of gauge-invariant spinor
and gauge field operators. And the last topic is
the application of the formalism
to the $SU(2)$ Yang-Mills case.
\bigskip

Implementation of Gauss's law is always required in a gauge theory,
but in earlier work it
was shown that in QED and other Abelian gauge theories, the failure to
implement Gauss's law does not affect the theory's physical
consequences\cite{khqedtemp,khelqed}.
And, in fact, it is known that the renormalized S-matrix in
perturbative QED
is correct, in spite of the fact that incident and
scattered charged particles are detached from all
fields, including the ones required to implement Gauss's law. In contrast,
the validity of perturbative QCD is more limited.  It is not
applicable to low energy phenomena.  And, it is likely
that all perturbative results in QCD are obscured, in some measure,
 by long-range effects, so that the implications of QCD for
even high-energy phenomenology are still not fully known.
In particular, color confinement is not
well understood.  One possible avenue for exploring QCD dynamics
beyond the perturbative
regime is the use of gauge-invariant operators and states in formulating
QCD dynamics.  Although  dynamical equations for gauge-invariant
operator-valued
fields have not yet been developed, we believe that
the mathematical apparatus we have constructed in this
paper can serve as a basis
for reaching such an objective.
\bigskip

We also note a feature of this work that is most clearly evident in
the $SU(2)$ example. The recursive equation for
$\overline{{\cal{A}}_{i}^{b}}({\bf{r}})$ --- Eq.~(\ref{eq:inteq2})
in the $SU(3)$ case,
with an arbitrary $V_i^\gamma({\bf r})$ replacing
the $\Pi_i^\gamma({\bf r})$,
and  Eqs.~(\ref{eq:azero})--(\ref{eq:a2Y}) in the
$SU(2)$ Yang-Mills theory  --- have many of the features that
we associate with  finite gauge transformations applied to a
gauge field. This is particularly conspicuous for the parts of
$\overline{{\cal{A}}_{i}^{\gamma}}({\bf{r}})_{\cal X}$ and
$\overline{{\cal{A}}_{i}^{\gamma}}({\bf{r}})_{\overline{\cal Y}}$ that
correspond to the `pure gauge' components of
$\left(A^{\gamma}\,\right)^{\prime}_{i}$ displayed in
Eq.~(\ref{eq:atranssu2}).
These `pure gauge' parts are
$\overline{{\cal{A}}_{i}^{\gamma}}({\bf{r}})_{\cal X}^{(pg)}$ and
$\overline{{\cal{A}}_{i}^{\gamma}}({\bf{r}})_{\overline{\cal Y}}^{(pg)}$
respectively, and are given by
\begin{eqnarray}
\overline{{\cal{A}}_{i}^{\gamma}}({\bf{r}})_{\cal X}^{(pg)}\,
&&=-g\epsilon^{\alpha\beta\gamma}\,
{\cal{X}}^\alpha({\bf{r}})\,\partial_i{\cal{X}}^\beta({\bf{r}})\,
{\textstyle\frac{1-\cos({\cal{N}})}{{\cal{N}}^2}}
\nonumber\\
&&+g^2\epsilon^{\alpha\beta b}
\epsilon^{b\mu\gamma}\,
{\cal{X}}^\mu({\bf{r}})\,{\cal{X}}^\alpha({\bf{r}})\,
\partial_i{\cal{X}}^\beta({\bf{r}})\,
[{\textstyle\frac{1}{{\cal{N}}^2}}
-{\textstyle\frac{\sin({\cal{N}})}{{\cal{N}}^3}}]
\label{eq:a2Xpg}
\end{eqnarray}
and
\begin{eqnarray}
\overline{{\cal{A}}_{i}^{\gamma}}({\bf{r}})_{\overline{\cal Y}}^{(pg)}\,
&&=g\epsilon^{\alpha\beta\gamma}\,
\overline{{\cal{Y}}^\alpha}({\bf{r}})
\,\partial_i\overline{{\cal{Y}}^\beta}({\bf{r}})\,
{\textstyle\frac{1-\cos(\overline{\cal{N}})}
{\overline{\cal{N}}^2}}
\nonumber\\
&&+g^2\epsilon^{\alpha\beta b}
\epsilon^{b\mu\gamma}\,
\overline{{\cal{Y}}^\mu}({\bf{r}})\,
\overline{{\cal{Y}}^\alpha}({\bf{r}})\,
\partial_i\overline{{\cal{Y}}^\beta}({\bf{r}})\,
[{\textstyle\frac{1}{\overline{\cal{N}}^2}}-
{\textstyle\frac{\sin(\overline{\cal{N}})}
{\overline{\cal{N}}^3}}]\;.
\label{eq:a2Ypg}
\end{eqnarray}
The `pure gauge' parts of
$\overline{{\cal{A}}_{i}^{\gamma}}({\bf{r}})_{\cal X}$ and
$\overline{{\cal{A}}_{i}^{\gamma}}({\bf{r}})_{\overline{\cal Y}}$
correspond to the pure gauge part of
$\left(A^{\gamma}\,\right)^{\prime}_{i}$, with
$-g{\cal{X}}^\gamma({\bf{r}})$ and
$g\overline{{\cal{Y}}^\gamma}({\bf{r}})$
corresponding to the gauge function
$\omega^\gamma({\bf{r}}), $ and ${\cal{N}}$ and
$\overline{\cal{N}}$ corresponding to $|\omega|$ respectively.
This correspondence suggests that, in addition to the iterative
solution of  Eq.~(\ref{eq:inteq2}), which we have discussed
extensively in this work, there may be non-perturbative solutions
that can not be represented as an iterated
series and that are related to the non-trivial topological sectors of
non-Abelian gauge fields\cite{topsect}.

\section{acknowledgements}
This research was supported by the Department of Energy under Grant
No.~DE-FG02-92ER40716.00.

\appendix
\section{Some necessary lemmas}

In this appendix we will prove a number of lemmas required for the
inductive proof of Eq. (\ref{eq:LCA2}) ---
the fundamental identity that enables us to construct states that
implement the non-Abelian
Gauss's law. The first group of lemmas pertains to the sums over
permutations of structure
constants that arise when $\Pi_i^\sigma ({\bf r})$ and $\partial_i
\Pi_i^a ({\bf r})$ are
commuted with $\overline{{\cal A}_j^\gamma}({\bf r}^\prime)$. The
first of
these identities is
\begin{equation}
\left[{\cal P}_{(e ,\beta [m-1])}^{(0)}f^{e\delta f}
f^{\vec{\beta}f\gamma}_{(m-1)}\right]
{\cal U}^{\vec{\beta}}_{(m-1)}  =
\sum_{s=0}^{m-1}{\textstyle\frac{m!}
{(m-s-1)!(s+1)!}}f^{\vec{\beta}\delta g}_{(m-s-1)}
f^{g f \gamma}f^{\vec{\sigma}e f}_{(s)}\,
{\cal U}^{\vec{\beta}}_{(m-s-1)}\,{\cal U}^{\vec{\sigma}}_{(s)}\;.
\label{eqpo}
\end{equation}
${\cal U}_{(\eta)}^{\vec{\alpha}}=
\prod_{m=1}^{\eta}{\cal U}^{\alpha[m]}({\bf r})$
is a product of operator-valued functions
${\cal U}^{\alpha[m]}({\bf r})$ that differ only in the index
$\alpha[m]$ that refers to the adjoint representation of
the Lie group to which the gauge fields
belong, and for which $[\,{\cal U}^{\gamma}({\bf r}),\,
{\cal U}^{\lambda}({\bf r}^\prime)\,]=0$.
To prove Eq.~(\ref{eqpo}) we generalize Eq.~(\ref{eq:defperm})
by defining the more general permutation
operator
\begin{equation}
\left[{\cal P}_{(e ,\beta [m-j-1])}^{(j)}f^{e\delta f}f^{\vec{\beta}
f\gamma}_{(m-j-1)}\right]
{\cal U}_{(m-j-1)}^{\vec{\beta}}=
\sum_{s=0}^{m-j-1}{\textstyle\frac{(s+j)!}{s!j!}}
f^{\vec{\beta}\delta u}_{(s)}f^{u e v}
f^{\vec{\sigma}v\gamma}_{(m-s-j-1)}\,
{\cal U}_{(s)}^{\vec{\beta}}\,{\cal U}_{(m-s-j-1)}^{\vec{\sigma}}\;.
\label{eqde}
\end{equation}

We can designate the individual permutations that appear in
Eq.~(\ref{eqde}) as
\begin{equation}
{\sf p}_{e,{\beta}[m-j-1]}(s)=f^{\vec{\beta}\delta u}_{(s)}f^{u e v}
f^{\vec{\tau} v \gamma}_{(m-j-s-1)}{\cal U}^{\vec{\beta}}_{(s)}
{\cal U}_{(m-j-s-1)}^{\vec{\tau}}
\end{equation}
for $s=0,1,2,\cdots ,m-j-1$, so that Eq.~(\ref{eqde})
can be expressed as
\begin{equation}
 \left[{\cal P}_{(e ,\beta [m-j-1])}^{(j)}f^{e\delta f}f^{\vec{\beta}
f\gamma}_{(m-j-1)}\right]
{\cal U}_{(m-j-1)}^{\vec{\beta}}=\sum_{s=0}^{m-j-1}
{\textstyle\frac{(s+j)!}{s!j!}}
\,{\sf p}_{e,{\beta}[m-j-1]}(s)\;.
\label{eqdef}
\end{equation}
\bigskip

We can transform ${\sf p}_{e,{\beta}[m-j-1]}(s)$ by using the Jacobi
identity
\begin{equation}
f^{ceb[1]}f^{{b[1]\beta b[2]}}=f^{cb[2]b[1]}f^{{\beta eb[1]}}+
f^{{c\beta b[1]}}f^{eb[2]b[1]}\;.
\end{equation}
As we use the Jacobi identity to transform each permutation in Eq.
(\ref{eqdef}), turn by turn, each such transformation augments the
coefficient of the immediately following  permutation
by the accumulated sum of all preceding permutations
($i.e.$ applying the Jacobi identity to
${\sf p}_{e,{\beta}[m-j-1]}(s)$ contributes an additional
${\sf p}_{e,{\beta}[m-j-1]}(s+1)$ term).
Since $\sum_{n=0}^{s} \frac{(n+j)!}{n!j!}=\frac{(s+j+1)!}{s!(j+1)!}$,
we find, after the Jacobi identity has been applied to the last possible
set of permutations on the r.h.s. of Eq. (\ref{eqdef}), that we obtain
\begin{eqnarray}
&&\left[{\cal P}_{(e ,\beta [m-j-1])}^{(j)}f^{e\delta f}
f^{\vec{\beta}f\gamma}_{(m-j-1)}\right]\,
{\cal U}^{\vec{\beta}}_{(m-j-1)}
\nonumber \\
&&\;\;\;\;\;\;\;\;=
\left[{\cal P}_{(f ,\beta [m-j-2])}^{(j+1)}f^{f\delta g}
f^{\vec{\beta}g\gamma}_{(m-j-2)}\right]f^{e\sigma f}
{\cal U}^{\vec{\beta}}_{(m-j-2)}{\cal U}^{\sigma}   +
{\textstyle\frac{m!}{(m-j-1)!(j+1)!}}
f^{\vec{\beta}\delta f}_{(m-j-1)}
f^{ef \gamma}\,{\cal U}^{\vec{\beta}}_{(m-j-1)}\;.
\label{eqpj}
\end{eqnarray}
\bigskip

The last term on the r.h.s.
of Eq.~(\ref{eqpj}) is the last permutation in Eq.~(\ref{eqdef}), whose
coefficient has now been increased to
${(s+j+1)!}/{s!(j+1)!}$ with $s=m-j-1$ by the application of the Jacobi
identity to all the earlier permutations.
In this permutation, the structure constant that contains  $e$ is
already on the extreme right of all other structure constants, so that
the Jacobi identity can no longer be applied
to its product with the structure constant on its right.
For that reason, the sum over permutations on the
r.h.s. of Eq. (\ref{eqpj}) contains one
fewer elements than the sum over permutations on the l.h.s. of that
equation.
\bigskip

Applying Eq.~(\ref{eqpj}) sequentially to
\begin{equation}
\left[\,{\cal P}_{(e ,\beta [m-s-1])}^{(s)}\,f^{e\delta u}
f^{\vec{\beta}u\gamma}_{(m-s-1)}\,\right]\,
{\cal U}^{\vec{\beta}}_{(m-s-1)}\;,
\end{equation}
for $s=j,j+1,j+2,\cdots ,m-2$, thus decreasing the number of
terms in the sum over permutations by one with each operation
until the last permutation has vanished, leads to
\begin{eqnarray}
\left[\,{\cal P}_{(e,\beta [m-j-1])}^{(j)}f^{e\delta u}
f^{\vec{\beta}u\gamma}_{(m-j-1)}\,\right]\,
&&{\cal U}^{\vec{\beta}}_{(m-j-1)}
\nonumber\\
&& =\sum_{s=j}^{m-1}{\textstyle\frac{m!}{(m-s-1)!(s+1)!}}
f^{\vec{\beta}\delta v}_{(m-s-1)}f^{v u \gamma}
f^{\vec{\sigma}e u}_{(s-j)}\,{\cal U}^{\vec{\beta}}_{(m-s-1)}\,
{\cal U}^{\vec{\sigma}}_{(s-j)}\;,
\label{eqpjf}
\end{eqnarray}
one of the important lemmas established in this appendix.
For $j=0$, Eq.~(\ref{eqpjf}) becomes Eq.~(\ref{eqpo}). Eq.~(\ref{eqpjf})
with different values for $j$ can be combined to obtain other useful
identities. By combining the $j=0$ and $j=1$ versions of
Eq.~(\ref{eqpjf}), we obtain
\begin{eqnarray}
\left[\,{\cal P}_{(e ,\beta [m-1])}^{(0)}\,f^{e\delta u}
f^{\vec{\beta}u\gamma}_{(m-1)}\,\right]\,
{\cal U}^{\vec{\beta}}_{(m-1)}
=&&mf^{e\delta u}f^{\vec{\beta} u\gamma}_{(m-1)}\,
{\cal U}^{\vec{\beta}}_{(m-1)}
\nonumber\\
&&-\sum_{t=0}^{m-2}
{\textstyle\frac{m!(m-t-1)}{(m-t)!t!}}\,
f^{\vec{\beta}\delta v}_{(t)}f^{v u\gamma}
f^{\vec{\sigma}eu}_{(m-t-1)}{\cal U}^{\vec{\beta}}_{(t)}\,
{\cal U}^{\vec{\sigma}}_{(m-t-1)}\;,
\label{eqpotwo}
\end{eqnarray}
and
\begin{eqnarray}
\left[\,{\cal P}_{(e ,\beta [m-1])}^{(0)}\,f^{e\delta u}
f^{\vec{\beta}u\gamma}_{(m-1)}\,\right]\,
{\cal U}^{\vec{\beta}}_{(m-1)}\,
=&&f^{e\delta u}f^{\vec{\beta} u\gamma}_{(m-1)}\,
{\cal U}^{\vec{\beta}}_{(m-1)}
\nonumber\\
&&+\sum_{s=0}^{m-2}
{\textstyle\frac{(m-1)!}{(s+1)!(m-s-2)!}}\,
f^{\vec{\beta}\delta v}_{(m-s-1)}f^{v u\gamma}
f^{\vec{\sigma}e u}_{(s)}\,{\cal U}^{\vec{\beta}}_{(m-s-1)}\,
{\cal U}^{\vec{\sigma}}_{(s)}\;.
\end{eqnarray}
Our next objective is to evaluate
the contribution to Eq.~(\ref{eq:LCA2}) from
$\frac{i}{n!}\int d{\bf r}^\prime\,
\psi_{(n)j}^{\gamma}({\bf r}^\prime)\,
V_{j}^{\gamma}({\bf r}^\prime)$,
the inhomogeneous term in the
recursive equation for $i\int d{\bf r}^\prime\,
\overline{{\cal A}_{j}^{\gamma}}({\bf r}^\prime)\,
V_{j}^{\gamma}({\bf r}^\prime)$.
From Eq.~(\ref{eq:psindef2}) we observe that
\begin{eqnarray}
{\textstyle \frac{i}{n!}}\int d{\bf r}^\prime
\left[\,\Pi_{i}^{d}({\bf r}),\,
\psi_{(n)j}^{\gamma}({\bf r}^\prime)\,\right]&&
V_{j}^{\gamma}({\bf r}^\prime)
={\textstyle\frac{i}{n!}}(-1)^{n-1}
f^{\vec{\alpha}\beta\gamma}_{(n)}
\int d{\bf r}^\prime\,
{\cal R}^{\vec{\alpha}}_{(n)}({\bf r}^\prime)\,
\left[\Pi_{i}^{d}({\bf r}),\,
{\cal Q}_{(n)j}^{\beta}({\bf r}^\prime)
\right]V_{j}^{\gamma}({\bf r}^\prime)
\nonumber \\
&&+{\textstyle\frac{i}{n!}}(-1)^{n-1}
f^{\vec{\alpha}\beta\gamma}_{(n)}
\int d{\bf r}^\prime\,\left[\,\Pi_{i}^{d}({\bf r}),\,
{\cal R}^{\vec{\alpha}}_{(n)}({\bf r}^\prime)\,\right]\,
{\cal Q}_{(n)j}^{\beta}({\bf r}^\prime)\,
V_{j}^{\gamma}({\bf r}^\prime)\;.
\end{eqnarray}
We use integration by parts and the identity
\begin{eqnarray}
f^{\vec{\alpha}\delta \gamma}_{(m)}\,[\,{\sf Q},\,
{\cal U}_{(m)}^{\vec{\alpha}}\,]
=  -\left[\,{\cal P}^{(0)}_{(\alpha ,\beta [m-1])}\,
f^{\alpha\delta e}
f^{\vec{\beta} e\gamma}_{(m-1)}\,\right][\,{\sf Q},\,
{\cal U}^{\alpha}({\bf r})\,]\,
{\cal U}_{(m-1)}^{\vec{\beta}}\;,
\label{qu}
\end{eqnarray}
where ${\sf Q}$ represents an arbitrary operator for which
$[\,{\sf Q},\,{\cal U}^{\gamma}({\bf r})\,]$ commutes with
${\cal U}^{\lambda}({\bf r}^\prime)$, to obtain
\begin{eqnarray}
{\textstyle \frac{i}{n!}}\int d{\bf r}^\prime\,
\left[\,\Pi_{i}^{d}({\bf r}),\,
\psi_{(n)j}^{\gamma}({\bf r}^\prime)\,\right]\,
V_{j}^{\gamma}({\bf r}^\prime)
&&={\textstyle \frac{1}{n!}}(-1)^{n-1}
f^{\vec{\alpha} d\gamma}_{(n)}\,
{\cal R}^{\vec{\alpha}}_{(n)}({\bf r})\,V_{i}^{\gamma}({\bf r})
\nonumber \\
&&\!\!\!\!\!\!\!\!+{\textstyle \frac{1}{(n+1)!}(-1)^{n}
f^{\vec{\alpha}d\gamma}_{(n)}\,
\frac{\partial_{i}}{\partial^{2}}}
\left(\,{\cal R}^{\vec{\alpha}}_{(n)}({\bf r})\,
\partial_{j}V_{j}^{\gamma}({\bf r})\,\right)
\nonumber \\
&&\!\!\!\!\!\!\!\!\!\!\!\!\!\!\!\!\!
\!\!\!\!\!\!\!\!\!\!\!\!\!\!\!\!\!\!\!\!\!\!
+{\textstyle \frac{1}{(n+1)!}(-1)^{n+1}\,
\left[\,{\cal P}^{(0)}_{(\alpha,\sigma [n-1])}
f^{\alpha d e}f^{\vec{\sigma}e\gamma}_{(n-1)}\,\right]\,
\frac{\partial_{i}}{\partial^{2}}}
\left(\,x_{j}^{\alpha}({\bf r})\,
{\cal R}_{(n-1)}^{\vec{\sigma}}({\bf r})\,
V_{j}^{\gamma}({\bf r})\,\right)
\nonumber \\
&&\!\!\!\!\!\!\!\!\!\!\!\!\!\!\!\!\!\!
\!\!\!\!\!\!\!\!\!\!\!\!\!\!\!\!\!\!
+{\textstyle \frac{1}{n!}(-1)^{n-1}
\,\left[\,{\cal P}^{(0)}_{(d,\sigma[n-1])}f^{d\beta e}
f^{\vec{\sigma}e\gamma}_{(n-1)}\,\right]\,
\frac{\partial_{i}}{\partial^{2}}}
\left(\,{\cal R}^{\vec{\sigma}}_{(n-1)}({\bf r})\,
{\cal Q}_{(n)j}^{\beta}({\bf r})\,
V_{j}^{\gamma}({\bf r})\,\right)\;.
\label{pipsi}
\end{eqnarray}
Eq.~(\ref{eqpo}) enables us to rewrite Eq.~(\ref{pipsi}) as
\begin{eqnarray}
& &{\textstyle \frac{i}{n!}}
\int d{\bf r}^\prime\,\left[\,\Pi_{i}^{d}({\bf r}),\,
\psi_{(n)j}^{\gamma}({\bf r}^\prime)\,\right]\,
V_{j}^{\gamma}({\bf r}^\prime)
={\textstyle \frac{1}{n!}}(-1)^{n-1}
f^{\vec{\alpha} d\gamma}_{(n)}\,
{\cal R}^{\vec{\alpha}}_{(n)}({\bf r})\,
V_{i}^{\gamma}({\bf r})
\nonumber\\
&&\;\;\;\;\;\;\;\;\;\;\;\;\;\;\;\;\;\;
\;\;\;\;\;\;\;\;\;\;\;\;\;\;\;\;\;\;\;
\;\;\;\;\;\;\;\;\;\;\;\;\;\;\;\;\;
+{\textstyle \frac{1}{(n+1)!}(-1)^{n}
f^{\vec{\alpha}d\gamma}_{(n)}\frac{\partial_{i}}{\partial^{2}}}
\left(\,{\cal R}^{\vec{\alpha}}_{(n)}({\bf r})\,\partial_{j}
V_{j}^{\gamma}({\bf r})\,\right)
\nonumber \\
&&-\sum_{s=0}^{n-1}{\textstyle\frac{1}{(n+1)(s+1)!(n-s-1)!}
f^{\vec{\tau}\beta e}_{(s)}f^{e u\gamma}
f^{\vec{\sigma} d u}_{(n-s-1)}
(-1)^{n+1}\frac{\partial_{i}}{\partial^{2}}}
\left(\,{\cal R}_{(s)}^{\vec{\tau}}({\bf r})\,
{\cal R}_{(n-s-1)}^{\vec{\sigma}}({\bf r})\,
x_{j}^{\beta}({\bf r})\,V_{j}^{\gamma}({\bf r})\,\right)
\nonumber \\
&&+\sum_{s=0}^{n-1}{\textstyle\frac{1}{s!(n-s)!}
f^{\vec{\sigma} d u}_{(n-s-1)}f^{e u\gamma}
f^{\vec{\tau}\beta e}_{(s)}(-1)^{n+1}\,
\frac{\partial_{i}}{\partial^{2}}}
\left({\cal R}_{(n-s-1)}^{\vec{\sigma}}({\bf r})\,
{\cal R}_{(s)}^{\vec{\tau}}({\bf r})\,
{\cal Q}_{(n)j}^{\beta}({\bf r})\,
V_{j}^{\gamma}({\bf r})\,\right)\;,
\end{eqnarray}
and the identity
\begin{equation}
\frac{1}{s!(n-s)!}\,{\cal Q}^{\alpha}_{(s)j}({\bf r})
=\frac{1}{s!(n-s-1)!}\left[\frac{1}{(n-s)}\,
{\cal Q}^{\alpha}_{(n)j}({\bf r})
-\frac{1}{(n+1)(s+1)}\,x^{\alpha}_{j}({\bf r})\,\right]\;,
\end{equation}
finally leads to
\begin{eqnarray}
{\textstyle \frac{i}{n!}}
\int d{\bf r}^\prime\,\left[\,\Pi_{i}^{d}({\bf r}),\,
\psi_{(n)j}^{\gamma}({\bf r}^\prime)\,\right]&&\,
V_{j}^{\gamma}({\bf r}^\prime)
={\textstyle \frac{1}{n!}}(-1)^{n-1}
f^{\vec{\alpha}d\gamma}_{(n)}\,
{\cal R}^{\vec{\alpha}}_{(n)}({\bf r})\,V_{i}^{\gamma}({\bf r})
\nonumber \\
&&\;\;\;\;\;+{\textstyle \frac{1}{(n+1)!}(-1)^{n}
f^{\vec{\alpha}d\gamma}_{(n)}
\frac{\partial_{i}}{\partial^{2}}}
\left(\,{\cal R}^{\vec{\alpha}}_{(n)}({\bf r})\,\partial_{j}
V_{j}^{\gamma}({\bf r})\right)\nonumber \\
&&\!\!\!\!\!\!\!\!\!\!\!\!\!\!\!\!\!\!
+{\textstyle\frac{1}{n!}
f^{\vec{\sigma} d u}_{(n-1)}
f^{e u\gamma}(-1)^{n}\frac{\partial_{i}}{\partial^{2}}}
\left(\,{\cal R}_{(n-1)}^{\vec{\sigma}}({\bf r})\,
a_{j}^{e}({\bf r})\,V_{j}^{\gamma}({\bf r})\,\right)
\nonumber \\
&&\!\!\!\!\!\!\!\!\!\!\!\!\!\!\!\!\!\!
+\sum_{s=1}^{n-1}{\textstyle\frac{1}{s!(n-s)!}
f^{\vec{\sigma} d u}_{(n-s-1)}
f^{e u\gamma}(-1)^{n-s}\frac{\partial_{i}}{\partial^{2}}}
\left(\,{\cal R}_{(n-s-1)}^{\vec{\sigma}}({\bf r})\,
\psi_{(s)j}^{e}({\bf r})\,V_{j}^{\gamma}({\bf r})\,\right)\;.
\label{pi}
\end{eqnarray}
Similarly,  the same algebraic identities
used to obtain Eq.~(\ref{pi}), can be used to transform
\begin{eqnarray}
{\textstyle \frac{i}{n!}}\int d{\bf r}^\prime\,
\left[\,\partial_{i}\Pi_{i}^{b}({\bf r}),\,
\psi_{(n)j}^{\gamma}({\bf r}^\prime)\,\right]&&\,
V_{j}^{\gamma}({\bf r}^\prime)
\nonumber \\
&&\!\!\!\!\!\!\!\!\!\!\!\!\!
={\textstyle\frac{1}{n!}}(-1)^{n-1}
f^{\vec{\alpha}\beta\gamma}_{(n)}
\int d{\bf r}^\prime\,{\cal R}^{\vec{\alpha}}_{(n)}({\bf r}^\prime)\,
\left[\,\partial_{i}\Pi_{i}^{b}({\bf r}),\,
{\cal Q}_{(n)j}^{\beta}({\bf r}^\prime)\,
\right]\,V_{j}^{\gamma}({\bf r}^\prime)
\nonumber \\
&&\!\!\!\!\!\!\!\!\!\!\!
+{\textstyle\frac{i}{n!}}(-1)^{n-1}f^{\vec{\alpha}\beta\gamma}_{(n)}
\int d{\bf r}^\prime\,
\left[\,\partial_{i}\Pi_{i}^{b}({\bf r}),\,
{\cal R}^{\vec{\alpha}}_{(n)}({\bf r}^\prime)\,\right]\,
{\cal Q}_{(n)j}^{\beta}({\bf r}^\prime)\,
V_{j}^{\gamma}({\bf r}^\prime)\;,
\end{eqnarray}
to
\begin{eqnarray}
{\textstyle \frac{i}{n!}}\int d{\bf r}^\prime
\left[\,\partial_{i}\Pi_{i}^{b}({\bf r}),\,
\psi_{(n)j}^{\gamma}({\bf r}^\prime)\,\right]&&\,
V_{j}^{\gamma}({\bf r}^\prime)
={\textstyle \frac{1}{(n-1)!(n+1)}}
(-1)^{n-1}f^{\vec{\alpha}b\gamma}_{(n)}\,
{\cal R}^{\vec{\alpha}}_{(n)}({\bf r})\,
\partial_{j}V_{j}^{\gamma}({\bf r})
\nonumber \\
&&\;\;\;\;\;\;\;+{\textstyle \frac{1}{(n-1)!}}(-1)^{n-1}f^{b\beta e}
f^{\vec{\sigma}e\gamma}_{(n-1)}\,A_{j}^{\beta}({\bf r})\,
{\cal R}^{\vec{\sigma}}_{(n-1)}({\bf r})\,V_{j}^{\gamma}({\bf r})
\nonumber \\
&&+{\textstyle \frac{n-1}{n!}}(-1)^{n-1}
f^{e d\gamma}f^{\vec{\tau} b d}_{(n-1)}\,
{\cal R}^{\vec{\tau}}_{(n-1)}({\bf r})\,a_{j}^{e}({\bf r})\,
V_{j}^{\gamma}({\bf r})
\nonumber \\
&&+\sum_{s=1}^{n-2}{\textstyle \frac{n-s-1}{s!(n-s)!}}(-1)^{n-s-1}
f^{e d\gamma}f^{\vec{\tau} b d}_{(n-s-1)}\,
{\cal R}^{\vec{\tau}}_{(n-s-1)}({\bf r})\,\psi_{(s)j}^{e}({\bf r})\,
V_{j}^{\gamma}({\bf r})\;.
\label{ppi}
\end{eqnarray}
\bigskip

We can combine Eqs.~(\ref{pi}) and (\ref{ppi}) to obtain
\begin{eqnarray}
&&-f^{b\mu d}\,A_{i}^{\mu}({\bf r}){\textstyle \frac{i}{(n-1)!}}\int
d{\bf r}^\prime\,\left[\,\Pi_{i}^{d}({\bf r}),\,
\psi_{(n-1)j}^{\gamma}({\bf r}^\prime)\,\right]\,
V_{j}^{\gamma}({\bf r}^\prime)
={\textstyle \frac{i}{n!}}\int d{\bf r}^\prime\,
\left[\,\partial_{i}\Pi_{i}^{b}({\bf r}),\,
\psi_{(n)j}^{\gamma}({\bf r}^\prime)\,\right]\,
V_{j}^{\gamma}({\bf r}^\prime)
\nonumber \\
&&\;\;\;\;\;\;\;\;\;\;\;\;\;\;\;\;\;\;\;\;\;\;\;\;\;\;
\;\;\;\;\;\;\;\;\;\;\;\;\;\;\;\;\;\;\;\;\;\;\;\;\;\;
\;\;\;\;\;\;\;\;\;\;\;\;\;\;\;\;\;\;\;
-{\textstyle \frac{1}{(n-1)!(n+1)}}(-1)^{n-1}
f^{\vec{\alpha}b\gamma}_{(n)}\,
{\cal R}^{\vec{\alpha}}_{(n)}({\bf r})\,
\partial_{j}V_{j}^{\gamma}({\bf r})\nonumber \\
&&\;\;\;\;\;\;\;\;\;\;\;\;\;\;\;\;\;\;\;\;\;\;\;\;\;\;
\;\;\;\;\;\;\;\;\;\;\;\;\;\;\;\;\;\;\;
-{\textstyle \frac{n-1}{n!}}
(-1)^{n-1}f^{e d\gamma}f^{\vec{\tau} b d}_{(n-1)}\,
{\cal R}^{\vec{\tau}}_{(n-1)}({\bf r})\,
a_{j}^{e}({\bf r})\,V_{j}^{\gamma}({\bf r})
\nonumber \\
&&\;\;\;\;\;\;\;\;\;\;\;\;\;\;\;\;\;\;\;\;\;\;\;\;\;\;
\;\;\;\;\;\;\;\;\;\;\;\;\;\;\;\;\;\;\;
-\sum_{s=1}^{n-2}{\textstyle \frac{n-s-1}{s!(n-s)!}}
(-1)^{n-s-1}f^{e d\gamma}f^{\vec{\tau} b d}_{(n-s-1)}\,
{\cal R}^{\vec{\tau}}_{(n-s-1)}({\bf r})\,
\psi_{(s)j}^{e}({\bf r})\,V_{j}^{\gamma}({\bf r})
\nonumber \\
&&\;\;\;\;\;\;\;\;\;\;\;\;\;\;\;\;\;\;\;\;\;\;\;\;\;\;
\;\;\;\;\;\;\;\;\;\;\;\;\;\;\;\;\;\;
-f^{b\mu d}A_{i}^{\mu}({\bf r})
{\textstyle\frac{1}{n!}(-1)^{n-1}f^{\vec{\alpha}d\gamma}_{(n-1)}
\,\frac{\partial_{i}}{\partial^{2}}}
\left(\,{\cal R}^{\vec{\alpha}}_{(n-1)}({\bf r})\,\partial_{j}
V_{j}^{\gamma}({\bf r})\,\right)\nonumber \\
&&-f^{b\mu d}A_{i}^{\mu}({\bf r})
{\textstyle\frac{1}{(n-1)!}
f^{\vec{\sigma} d h}_{(n-2)}f^{e h\gamma}(-1)^{n-1}\,
\frac{\partial_{i}}{\partial^{2}}}
\left(\,{\cal R}_{(n-2)}^{\vec{\sigma}}({\bf r})\,
a_{j}^{e}({\bf r})\,V_{j}^{\gamma}({\bf r})\,\right)
\nonumber \\
&&-f^{b\mu d}A_{i}^{\mu}({\bf r})
\sum_{s=1}^{n-2}{\textstyle\frac{1}{s!(n-s-1)!}
f^{\vec{\sigma} d h}_{(n-s-2)}f^{e h\gamma}(-1)^{n-s-1}\,
\frac{\partial_{i}}{\partial^{2}}}
\left(\,{\cal R}_{(n-s-2)}^{\vec{\sigma}}({\bf r})\,
\psi_{(s)j}^{e}({\bf r})\,V_{j}^{\gamma}({\bf r})\,\right)\;.
\label{gpsi}
\end{eqnarray}

\section{Proof of fundamental theorem}

In this section we will prove Eq.~(\ref{eq:LCA2}) by an inductive
argument that assumes that Eq.~(\ref{eq:LCA2}) holds for all $n\leq
N$,
and then demonstrates that it must also hold for $n=(N+1)$. The
theorem is trivial for $n=2$ and $n=1$, in the latter case with the
previously established convention that
${\cal A}_{(0)j}^\alpha({\bf r})=0$.
The structure of
Eq.~(\ref{eq:LCA1}), which defines ${\cal A}_{(n)j}^\gamma({\bf r})
V_{j}^{\gamma}({\bf r})$
recursively in terms of the inhomogeneous term
$\left({ig^n}/{n!}\right)
\int d{\bf r}\psi_{(n)j}^{\gamma}({\bf r})
V_{j}^{\gamma}({\bf r}) $ as well as other
${\cal A}_{(n^\prime )j^\prime}^\gamma({\bf r})
V_{j^\prime}^{\gamma}({\bf r})$ with
$n^\prime < n$, is ideally suited to an inductive argument.
\bigskip

We will transform
$-igf^{b\mu d}A_{i}^{\mu}({\bf r})\; \int d{\bf r}^\prime
[\Pi_{i}^{d}({\bf r}),\,
{\cal A}_{(N)j}^{\alpha}({\bf r}^\prime)]V_{j}^{\alpha}({\bf
r}^\prime)$ --- the r.h.s of Eq.~(\ref{eq:LCA2})
for $n=N+1$ --- into the corresponding l.h.s. of that equation,
using Eq.~(\ref{eq:LCA2}) as an inductive axiom
only for those ${\cal A}_{(n)j}^{\alpha}({\bf r}^\prime)$
that have $n<N$. We set
\begin{equation}
-igf^{b\mu d}A_{i}^{\mu}({\bf r})\; \int
d{\bf r}^\prime\,[\,\Pi_{i}^{d}({\bf r}),\,
{\cal A}_{(N)j}^{\alpha}({\bf r}^\prime)\,]\,
V_{j}^{\alpha}({\bf r}^\prime)
={\sf A}+{\sf B}+{\sf C}\;,
\end{equation}
where
\begin{equation}
{\sf A}= - gf^{b\mu d}\,A_{i}^{\mu}({\bf r})\,
{\textstyle\frac{ig^{N}}{N!}}\,\int d{\bf r}^\prime
\,\left[\,\Pi_{i}^{d}({\bf r}),\,
\psi_{(N)j}^{\gamma}({\bf r}^\prime)\,\right]\,
V_{j}^{\gamma}({\bf r}^\prime)\;,
\end{equation}
\begin{equation}
{\sf B}=-gf^{b\mu d}\,A_{i}^{\mu}({\bf r})\,
\sum_{m=1}{\textstyle\frac{ig^{N}}{m!} }\,
f^{\vec{\alpha}\delta\gamma}_{(m)}\,
\sum_{u=0}\sum_{r=m}\,\delta_{r+u+m-N}\,\int d{\bf r}^\prime\,
\left[\,\Pi_{i}^{d}({\bf r}),\,
{\cal M}_{(m,r)}^{\vec{\alpha}}({\bf r}^\prime)\,\right]\,
{\cal B}_{(m,u)j}^{\delta}({\bf r}^\prime)\,
V^{\gamma}_{j}({\bf r}^\prime)\;,
\end{equation} and
\begin{equation}
{\sf C}=-gf^{b\mu d}\,A_{i}^{\mu}({\bf r})\,
\sum_{m=1}{\textstyle\frac{ig^{N}}{m!} }\,
f^{\vec{\alpha}\delta\gamma}_{(m)}
\sum_{u=0}\sum_{r=m}\,\delta_{r+m+u-N}\,
\int d{\bf r}^\prime\,
{\cal M}_{(m,r)}^{\vec{\alpha}}({\bf r}^\prime)\,
\left[\,\Pi_{i}^{d}({\bf r}),\,
{\cal B}_{(m,u)j}^{\delta}({\bf r}^\prime)\,\right]\,
V^{\gamma}_{j}({\bf r}^\prime)\;.
\end{equation}
We will represent ${\sf A}$, ${\sf B}$, and ${\sf C}$
by dividing each of them into parts as shown by
\begin{equation}
{\sf A}={\sf A}_{(1)}+{\sf A}_{(2)}+{\sf A}_{(3)}\;,
\end{equation}
so that the subscript (1) designates those terms
that contain commutators
with $\partial_{i}\Pi_{i}^{b}({\bf r})$,
the subscript (3) designates terms that contain
$\partial_{i}V_{i}^{\alpha}({\bf r})$,
and the subscript (2) labels residues, most of which cancel as the
proof proceeds to its conclusion.
The representation of {\sf A} as a sum of its properly
subscripted constituents is easily obtained from  Eq.~(\ref{gpsi}).
We use Eq.~(\ref{qu}) to represent {\sf B} as
\begin{eqnarray}
{\sf B}=-gf^{b\mu d}\,&&A_{i}^{\mu}({\bf r})\,\sum_{m=1}
{\textstyle\frac{ig^{N}}{m!}}\,\left[\,
{\cal P}^{(0)}_{(\alpha ,\beta [m-1])}
\,f^{\alpha\delta e}
f^{\vec{\beta} e\gamma}_{(m-1)}\,\right]\,\sum_{u=0}
\sum_{p=m-1}\sum_{r[m]=1}\delta_{p+r[m]+u+m-N}
\nonumber \\
&&\;\;\;\;\times \int d{\bf r^{\prime}}\,
\left[\,\Pi_{i}^{d}({\bf r}),\,
{\cal A}_{(r[m])j}^{\alpha}({\bf r}^\prime)\,\right]\,
{\textstyle\frac{\partial_{j}}{\partial^{2}}}\,\left(\,
{\cal M}_{(m-1,p)}^{\vec{\beta}}({\bf r}^\prime)\,
{\cal B}_{(m,u)k}^{\delta}({\bf r}^\prime)\,
V^{\gamma}_{k}({\bf r}^\prime)\,\right)\;,
\label{two}
\end{eqnarray}
and invoke the inductive axiom to represent
$gf^{b\mu d}A_{i}^{\mu}({\bf r})\,i\int d{\bf r}^\prime\,
[\,\Pi_{i}^{d}({\bf r}),\,
{\cal A}_{(r[m])j}^{\alpha}({\bf r}^\prime)\,]\,
V_{j}^{\alpha}({\bf r}^\prime)$ in terms
of the l.h.s. of Eq.~(\ref{eq:LCA2}) for all values
of $r[m]<N.$ When we equate the operator-valued vector quantity
${\textstyle\frac{g^{N}}{m!}}
\left[\,{\cal P}^{(0)}_{(\alpha ,\beta [m-1])}
f^{\alpha\delta e}f^{\vec{\beta}e\gamma}_{(m-1)}\,\right]\,
{\textstyle\frac{\partial_{j}}{\partial^{2}}}
\left(\,{\cal M}_{(m-1,p)}^{\vec{\beta}}({\bf r}^\prime)
\,{\cal B}_{(m,u)k}^{\delta}({\bf r}^\prime)\,
V^{\gamma}_{k}({\bf r}^\prime)\,\right)$ in Eq.~(\ref{two}) to the
arbitrary vector field $V_{j}^{\alpha}({\bf r}^\prime)$ in
Eq.~(\ref{eq:LCA2}), we obtain
\begin{eqnarray}
{\sf B}_{(1)}=  &&\sum_{m=1}{\textstyle\frac{ig^{N+1}}{m!}}
\left[{\cal P}^{(0)}_{(\alpha,\beta [m-1])}
f^{\alpha\delta e}
f^{\vec{\beta} e\gamma}_{(m-1)}\right]
\sum_{u=0}\sum_{p=m-1}\sum_{r[m]=0}\delta_{p+r[m]+m+u-N}
\nonumber \\
&&\;\;\;\;\times \int d{\bf r}^\prime\,\left[\,\partial_{i}
\Pi_{i}^{b}({\bf r}),\,
{\cal A}_{(r[m]+1)j}^{\alpha}({\bf r}^\prime)\,\right]\,
{\textstyle\frac{\partial_{j}}{\partial^{2}}}
\left(\,{\cal M}_{(m-1,p)}^{\vec{\beta}}({\bf r}^\prime)\,
{\cal B}_{(m,u)k}^{\delta}({\bf r}^\prime)\,
V^{\gamma}_{k}({\bf r}^\prime)\,\right)\;,
\label{twoone}
\end{eqnarray}
and, after summing over $r[m]$ and
representing the sums over permutations by using the
lemma given in Eq.~(\ref{eqpo}),
\begin{eqnarray}
{\sf B}_{(2)}&& = f^{b\mu \alpha}A_{i}^{\mu}({\bf r})
\sum_{m=1}\sum_{u=0}\sum_{v=0}^{m-1}\sum_{r=m-v-1}
\sum_{q=v}\delta_{r+q+m+u-N}
{\textstyle\frac{g^{N+1}}{(v+1)!(m-v-1)!} }
f^{\vec{\beta}\delta g}_{(m-v-1)}f^{gf\gamma}
f^{\vec{\sigma}\alpha f}_{(v)}
\nonumber \\
& &\;\;\;\;\;\;\;\;\;\;\;\;\;\;\;\;
\;\;\;\;\;\;\;\;\;\;\;\;\;\;\;\;\;\;\;\;\;\;\;\;\;\;\;\;\;\;\;\;
\times {\textstyle\frac{\partial_{i}}{\partial^{2}}}
\left({\cal M}_{(m-v-1,r)}^{\vec{\beta}}({\bf r})
{\cal M}_{(v,q)}^{\vec{\sigma}}({\bf r})
{\cal B}_{(m,u)j}^{\delta}({\bf r})
V^{\gamma}_{j}({\bf r})\right)
\nonumber \\
&& - \sum_{m=1}\sum_{u=0}\sum_{n=1}
\sum_{v=0}^{m-1}\sum_{r=m-v-1}\sum_{q=n+v}
\delta_{r+q+m+n+u-N}f^{\vec{\beta}\delta h}_{(m-v-1)}
f^{\alpha h\gamma}f^{\vec{\sigma} c\alpha}_{(n+v)}
f^{b\mu c}
{\textstyle\frac{g^{N+1}B(n)}{n!(v+1)!(m-v-1)!}}
\nonumber \\
&&\;\;\;\;\;\;\;\;\;\;\;\;\;\;\;\;
\;\;\;\;\;\;\;\;\;\;\;\;\;\;\;\;\;\;\;\;\;\;\;
\times A_{i}^{\mu}({\bf r})\,
{\textstyle \frac{\partial_{i}}{\partial^{2}}}
\left(\,{\cal M}_{(n+v,q)}^{\vec{\sigma}}({\bf r})\,
{\cal M}_{(m-v-1,r)}^{\vec{\beta}}({\bf r})\,
{\cal B}_{(m,u)j}^{\delta}({\bf r})\,
V^{\gamma}_{j}({\bf r})\,\right)
\nonumber \\
&&+ \sum_{m=1}\sum_{u=0}\sum_{n=0}\sum_{t=1}
\sum_{v=0}^{m-1}\sum_{r=m-v-1}
\sum_{q=n+v}\delta_{r+q+n+t+u+m-(N+1)}
(-1)^{t-1}\,
{\textstyle\frac{g^{N+1}B(n)}{n!(t-1)!(t+1)(v+1)!(m-v-1)!}}
\nonumber \\
&&\;\;\;\;\;\;\;\;\;\;\;\;\;\;\;\;
\times f^{\vec{\beta} \delta h}_{(m-v-1)}
f^{\alpha h\gamma}f^{\vec{\sigma}\lambda\alpha}_{(n+v)}
f^{\vec{\mu} b\lambda}_{(t)}\,
{\cal R}_{(t)}^{\vec{\mu}}({\bf r})\,
{\cal M}_{(n+v,q)}^{\vec{\sigma}}({\bf r})\,
{\cal M}_{(m-v-1,r)}^{\vec{\beta}}({\bf r})\,
{\cal B}_{(m,u)i}^{\delta}({\bf r})\,
 V^{\gamma}_{i}({\bf r})
\nonumber \\
 &&+\sum_{m=1}\sum_{u=0}\sum_{n=0}\sum_{t=1}
\sum_{v=0}^{m-1}\sum_{r=m-v-1}\sum_{q=n+v}
\delta_{r+q+n+t+m+u-N} (-1)^{t}\,
{\textstyle\frac{g^{N+1}B(n)}{n!(t+1)!(v+1)!(m-v-1)!} }\,
f^{\vec{\beta} \delta h}_{(m-v-1)}
\nonumber \\
&&\;\;\;\;\;\;
\times f^{\alpha h\gamma}f^{\vec{\sigma}\lambda\alpha}_{(n+v)}
f^{b\mu d}f^{\vec{\nu} d\lambda}_{(t)}\,A_{i}^{\mu}({\bf r})\,
{\textstyle\frac{\partial_{i}}{\partial^{2}}}
\left({\cal R}_{(t)}^{\vec{\nu}}({\bf r})\,
{\cal M}_{(n+v,q)}^{\vec{\sigma}}({\bf r})\,
{\cal M}_{(m-v-1,r)}^{\vec{\beta}}({\bf r})\,
{\cal B}_{(m,u)j}^{\delta}({\bf r})\,
V^{\gamma}_{j}({\bf r})\right)\;.
\label{twotwo}
\end{eqnarray}
We will represent ${\sf C}$ as
\begin{equation}
{\sf C}={\sf C}(a)+{\sf C}({\cal A})\;,
\end{equation}
where ${\sf C}(a)$ includes the commutator of
$\Pi_i^\alpha({\bf r})$ with
the $a_j^\delta ({\bf r}^\prime)$ part of
${\cal B}_{(m,u)j}^{\delta}({\bf r}^\prime)$
and ${\sf C}({\cal A})$ includes the commutator of
$\Pi_i^\alpha({\bf r})$ with the
${\cal A}_{(u)j}^\delta ({\bf r}^\prime)$ part of
${\cal B}_{(m,u)j}^{\delta}({\bf r}^\prime)$.
${\sf C}(a)$ is given by
\begin{equation}
{\sf C}(a)= -gf^{b\mu\delta}\,A_{i}^{\mu}({\bf r})\,
\sum_{m=1}\sum_{r=m}\,\delta_{r+m-N}
{\textstyle\frac{g^{N}}{m!} }\,
f^{\vec{\alpha}\delta\gamma}_{(m)}\,
(\,\delta_{ij}-{\textstyle
\frac{\partial_{i}\partial_{j}}{\partial^{2}}}\,)
\left({\cal M}_{(m,r)}^{\vec{\alpha}}({\bf r})\,
V^{\gamma}_{j}({\bf r})\right)\;,
\label{cla}
\end{equation}
and ${\sf C}({\cal A})$ is given by
\begin{eqnarray}
{\sf C}({\cal A})&&=- gf^{b\mu d}\,A_{i}^{\mu}({\bf r})\,
\sum_{m=1}{\textstyle\frac{ig^{N}}{m!} }\,
f^{\vec{\alpha}\delta\gamma}_{(m)}\,
\sum_{u=1}\sum_{r=m}\delta_{r+m+u-N}
\nonumber\\
&&\;\;\;\;\;\;\;\;\;\;\;\;\;\;\;\;
\times\int d{\bf r}^\prime\,
\left[\,\Pi_{i}^{d}({\bf r}),\,
{\cal A}_{(u)j}^{\delta}({\bf r}^\prime)\,\right]
(\,\delta_{jk}
-{\textstyle\frac{m}{m+1}
\frac{\partial_{j}\partial_{k}}{\partial^{2}}}\,)\,
\left(\,{\cal M}_{(m,r)}^{\vec{\alpha}}({\bf r}^\prime)\,
V^{\gamma}_{k}({\bf r}^\prime)\,\right)\;.
\label{threeb}
\end{eqnarray}
We again invoke the inductive axiom to transform ${\sf C}({\cal A})$
by representing
$ig\,f^{b\mu d}\,A_{i}^{\mu}({\bf r})\,\int\,
d{\bf r}^\prime\,[\,\Pi_{i}^{d}({\bf r}),\,
{\cal A}_{(u)j}^{\delta}({\bf r}^\prime)\,]\,
V_{j}^{\delta}({\bf r}^\prime)$
in terms of the l.h.s. of Eq.~(\ref{eq:LCA2}) for values
of $u<N$, and identify  ${\textstyle\frac{g^{N}}{m!}
f^{\vec{\alpha}\delta \gamma}_{(m)}(\delta_{jk}-\frac{m}{m+1}
\frac{\partial_{j}\partial_{k}}{\partial^{2}}})\,\left(\,
{\cal M}_{(m,r)}^{\vec{\alpha}}({\bf r}^\prime)\,
V^{\gamma}_{k}({\bf r}^\prime)\,\right)$ as the vector
field $V_{j}^{\alpha}({\bf r}^\prime)$ in
Eq.~(\ref{eq:LCA2}); after summing over the common integer-valued
variable in the two Kronecker $\delta$-functions in the resulting
expression, we obtain
\begin{equation}
{\sf C}({\cal A})={\sf C}({\cal A})_{(1)}+{\sf C}({\cal A})_{(2)}
+{\sf C}({\cal A})_{(3)}\;,
\end{equation}
with
\begin{eqnarray}
{\sf C}({\cal A})_{(1)}=&&
\sum_{m=1}{\textstyle\frac{ig^{N+1}}{m!}}
f^{\vec{\alpha}\delta\gamma}_{(m)}
\sum_{u=0}\sum_{r=m}\delta_{r+m+u-N}
\nonumber \\
& &\;\;\;\;\times\,
\int d{\bf r}^\prime\,
\left[\,\partial_{i}\Pi_{i}^{b}({\bf r}),\,
{\cal A}_{(u+1)j}^{\delta}({\bf r}^\prime)\,\right]\,
(\,\delta_{jk}
-{\textstyle\frac{m}{m+1}
\frac{\partial_{j}\partial_{k}}{\partial^{2}}}\,)
\left(\,{\cal M}_{(m,r)}^{\vec{\alpha}}({\bf r}^\prime)\,
V^{\gamma}_{k}({\bf r}^\prime)\,\right)\;,
\label{threebone}
\end{eqnarray}
\begin{eqnarray}
{\sf C}({\cal A})_{(2)}&&= f^{b\mu \delta}\,
A_{i}^{\mu}({\bf r})\,\sum_{m=1}{\textstyle\frac{g^{N+1}}{m!} }\,
f^{\vec{\alpha}\delta\gamma}_{(m)} \sum_{r=m}
\delta_{r+m-N}\,(\,\delta_{ij}-{\textstyle
\frac{m}{m+1}\frac{\partial_{i}\partial_{j}}{\partial^{2}}}\,)\,
\left(\,{\cal M}_{(m,r)}^{\vec{\alpha}}({\bf r})\,
V^{\gamma}_{j}({\bf r})\,\right)
\nonumber \\
&&- \sum_{m=1}\sum_{r=m}
{\textstyle\frac{g^{N+1}}{(m+1)!}}\,
f^{\vec{\alpha}\delta\gamma}_{(m)}
\sum_{n=1}\sum_{s=n}\delta_{s+n+r+m-N}
{\textstyle\frac{B(n)}{n!}}\,
f^{b\mu c}f^{\vec{\sigma}c\delta}_{(n)}\,
A_{i}^{\mu}({\bf r})
\nonumber \\
&& \;\;\;\;\;\;\;\;\;\;\;\;\;\;\;\;\;\;\;\;
\;\;\;\;\;\;\;\;\;\;\;\;\;\;\;\;\;\;\;\;
\;\;\;\;\;\;\;\times{\textstyle\frac{\partial_{i}}{\partial^{2}}}
\left(\,{\cal M}_{(n,s)}^{\vec{\sigma}}({\bf r})\,
\partial_{j}{\cal M}_{(m,r)}^{\vec{\alpha}}({\bf r})\,
V_{j}^{\gamma}({\bf r})\,\right)
\nonumber \\
&&+ \sum_{m=1}\sum_{r=m}
{\textstyle\frac{g^{N+1}}{(m+1)!}}\,
f^{\vec{\alpha}\delta\gamma}_{(m)}
\sum_{n=0}\sum_{t=1}\sum_{s=n}
\delta_{r+m+s+n+t-(N+1)}\,
(-1)^{t-1}{\textstyle \frac{B(n)}{n!(t-1)!(t+1)}}
f^{\vec{\mu} b\lambda}_{(t)}
f^{\vec{\sigma}\lambda\delta}_{(n)}
 \nonumber \\
& &\;\;\;\;\;\;\;\;\;\;\;\;\;\;\;\;
\;\;\;\;\;\;\;\;\;\;\;\;\;\;\;\;\;\;\;\;
\;\;\;\;\;\times {\cal R}_{(t)}^{\vec{\mu}}({\bf r})\,
{\cal M}_{(n,s)}^{\vec{\sigma}}({\bf r})\,
\partial_{i}{\cal M}_{(m,r)}^{\vec{\alpha}}({\bf r})\,
V_{i}^{\gamma}({\bf r})
\nonumber \\
&&+\sum_{m=1}\sum_{r=m}{\textstyle\frac{g^{N+1}}{(m+1)!}}
f^{\vec{\alpha}\delta \gamma}_{(m)} f^{b\mu d}\,
A_{i}^{\mu}({\bf r})\sum_{n=0}
\sum_{t=1}\sum_{s=n}\delta_{r+m+s+n+t-N} (-1)^{t}
{\textstyle\frac{B(n)}{n!(t+1)!}}
f^{\vec{\nu}d\lambda}_{(t)}f^{\vec{\sigma}\lambda\delta}_{(n)}
\nonumber \\
&&\;\;\;\;\;\;\;\;\;\;\;\;\;\;\;\;\;\;\;
\;\;\;\;\;\;\;\;\;\;\;\;\;\;\;\;
\times {\textstyle\frac{\partial_{i}}{\partial^{2}}}
\left(\,{\cal R}_{(t)}^{\vec{\nu}}({\bf r})\,
{\cal M}_{(n,s)}^{\vec{\sigma}}({\bf r})\,
\partial_{j}{\cal M}_{(m,r)}^{\vec{\alpha}}({\bf r})\,
V_{j}^{\gamma}({\bf r})\,\right)\;,
\label{eqddapart}
\end{eqnarray}
and
\begin{eqnarray}
{\sf C}({\cal A})_{(3)}&&= - \sum_{m=1}
\sum_{r=m}{\textstyle\frac{g^{N+1}}{(m+1)!}}
f^{\vec{\alpha}\delta\gamma}_{(m)}\sum_{n=1}
\sum_{s=n}\delta_{r+m+s+n-N}
{\textstyle\frac{B(n)}{n!}}f^{b\mu c}
f^{\vec{\sigma} c\delta}_{(n)}\,A_{i}^{\mu}({\bf r})
\nonumber \\
&&\;\;\;\;\;\;\;\;\;\;\;\;\;\;\;\;\;\;\;
\;\;\;\;\;\;\;\;\;\;\;\;\;\;\;\;\;\;\;
\times{\textstyle \frac{\partial_{i}}{\partial^{2}}}
\left(\,{\cal M}_{(n,s)}^{\vec{\sigma}}({\bf r})\,
{\cal M}_{(m,r)}^{\vec{\alpha}}({\bf r})\,
\partial_{j}V_{j}^{\gamma}({\bf r})\,\right)
\nonumber \\
& &+ \sum_{m=1} \sum_{r=m}
{\textstyle\frac{g^{N+1}}{(m+1)!}}
f^{\vec{\alpha}\delta \gamma}_{(m)}
\sum_{n=0}\sum_{t=1}\sum_{s=n}
\delta_{r+m+s+n+t-(N+1)}
(-1)^{t-1}{\textstyle \frac{B(n)}{n!(t-1)!(t+1)}}
\nonumber \\
&&\;\;\;\;\;\;\;\;\;\;\;\;\;\;\;\;\;\;\;
\;\;\;\;\;\;\;\;\;\;\;\;\;\times f^{\vec{\mu} b\lambda}_{(t)}
f^{\vec{\sigma}\lambda\delta}_{(n)}\,
{\cal R}_{(t)}^{\vec{\mu}}({\bf r})\,
{\cal M}_{(n,s)}^{\vec{\sigma}}({\bf r})\,
{\cal M}_{(m,r)}^{\vec{\alpha}}({\bf r})\,
\partial_{i}V_{i}^{\gamma}({\bf r})
\nonumber \\
&&+ \sum_{m=1}\sum_{r=m}{\textstyle\frac{g^{N+1}}{(m+1)!}}
f^{\vec{\alpha}\delta \gamma}_{(m)} f^{b\mu d}\,
A_{i}^{\mu}({\bf r})\sum_{n=0}\sum_{t=1}
\sum_{s=n}\delta_{r+m+s+n+t-N} (-
1)^{t}{\textstyle\frac{B(n)}{n!(t+1)!}}
\nonumber \\
& &\;\;\;\;\;\;\;\;\;\;\;\;
\;\;\;\;\;\;\;\;\;\;\;\;\;\;
\times f^{\vec{\nu} d\lambda}_{(t)}
f^{\vec{\sigma}\lambda\delta}_{(n)}
{\textstyle\frac{\partial_{i}}{\partial^{2}}}
\left(\,{\cal R}_{(t)}^{\vec{\nu}}({\bf r})\,
{\cal M}_{(n,s)}^{\vec{\sigma}}({\bf r})\,
{\cal M}_{(m,r)}^{\vec{\alpha}}({\bf r})\,
\partial_{j}V_{j}^{\gamma}({\bf r})\,\right)\;.
\label{threebthree}
\end{eqnarray}
When we transform
$\partial_i{\cal M}_{(m,r)}^{\vec{\alpha}}({\bf r})$
and $\partial_j{\cal M}_{(m,r)}^{\vec{\alpha}}({\bf r})$ in
${\sf C}({\cal A})_{(2)}$ by using Eq.~(\ref{qu}) and
transform the resulting expression by
applying  Eq.~(\ref{eqpo}), we
obtain an equation that so resembles
Eq.~(\ref{twotwo}) in structure,
that it becomes very natural to add
${\sf B}_{(2)}$ and ${\sf C}({\cal A})_{(2)}$.
In carrying out this addition, we note that
\begin{equation}
{\cal B}_{(m,u)j}^{\delta}({\bf r})
+{\textstyle\frac{v+1}{(m+1)(m-v)}}
\partial_{j}{\cal Y}_{(u)}^{\delta}({\bf r})
={\cal B}_{(m-v-1,u)j}^{\delta}({\bf r})\;,
\label{byb}
\end{equation}
and obtain
\begin{eqnarray}
{\sf B}_{(2)}&&+{\sf C}({\cal A})_{(2)}=
f^{b\mu\alpha}A_{i}^{\mu}({\bf r})
\sum_{m=1}\sum_{u=0}\sum_{v=0}^{m-1}
\sum_{r=m-v-1}\sum_{q=v}\delta_{r+q+m+u-N}
{\textstyle\frac{g^{N+1}}{(v+1)!(m-v-1)!} }
f^{\vec{\beta}\delta g}_{(m-v-1)}
f^{gf\gamma}f^{\vec{\sigma}\alpha f}_{(v)}
\nonumber\\
& &\;\;\;\;\;\;\;\;\;\;\;\;\;\;\;\;\;\;\;\;
\;\;\;\;\;\;\;\;\;\;\;\;\;\;\;\;\;\;\;\;
\times {\textstyle\frac{\partial_{i}}{\partial^{2}}}
\left(\,{\cal M}_{(m-v-1,r)}^{\vec{\beta}}({\bf r})\,
{\cal M}_{(v,q)}^{\vec{\sigma}}({\bf r})\,
{\cal B}_{(m,u)j}^{\delta}({\bf r})\,
V^{\gamma}_{j}({\bf r})\,\right)
\nonumber \\
&&+ f^{b\mu \delta}\,A_{i}^{\mu}({\bf r})\,
\sum_{m=1}{\textstyle\frac{g^{N+1}}{m!} }
f^{\vec{\alpha}\delta\gamma}_{(m)} \sum_{r=m}
\delta_{r+m-N}(\,\delta_{ij}-{\textstyle\frac{m}{m+1}
\frac{\partial_{i}\partial_{j}}{\partial^{2}}}\,)\,
\left(\,{\cal M}_{(m,r)}^{\vec{\alpha}}({\bf r})\,
V^{\gamma}_{j}({\bf r})\,\right)
\nonumber \\
&&- \sum_{m=1}\sum_{u=0}
\sum_{n=1}\sum_{v=0}^{m-1}
\sum_{r=m-v-1}\sum_{q=n+v}\delta_{r+q+m+n+u-N}
f^{\vec{\beta}\delta h}_{(m-v-1)}f^{\alpha h\gamma}
f^{\vec{\sigma} c\alpha}_{(n+v)}f^{b\mu c}
\nonumber \\
&&\;\;\;\;\;\;\;\;\;\;\;
\times{\textstyle\frac{g^{N+1}B(n)}{n!(v+1)!(m-v-1)!}}\,
A_{i}^{\mu}({\bf r})\,
{\textstyle \frac{\partial_{i}}{\partial^{2}}}
\left(\,{\cal M}_{(n+v,q)}^{\vec{\sigma}}({\bf r})\,
{\cal M}_{(m-v-1,r)}^{\vec{\beta}}({\bf r})\,
{\cal B}_{(m-v-1,u)j}^{\delta}({\bf r})\,
V^{\gamma}_{j}({\bf r})\,\right)
\nonumber \\
&&+\sum_{m=1}\sum_{u=0}\sum_{n=0}
\sum_{t=1}\sum_{v=0}^{m-1}\sum_{r=m-v-1}
\sum_{q=n+v}\delta_{r+q+m+n+u+t-(N+1)}
(-1)^{t-1}
{\textstyle\frac{g^{N+1}B(n)}{n!(t-1)!(t+1)(v+1)!(m-v-1)!} }
\nonumber \\
& &\;\;\;\;\;\;\;\times
f^{\vec{\beta} \delta h}_{(m-v-1)}
f^{\alpha h\gamma}f^{\vec{\sigma}\lambda\alpha}_{(n+v)}
f^{\vec{\mu} b\lambda}_{(t)}\,
{\cal R}_{(t)}^{\vec{\mu}}({\bf r})\,
{\cal M}_{(n+v,q)}^{\vec{\sigma}}({\bf r})\,
{\cal M}_{(m-v-1,r)}^{\vec{\beta}}({\bf r})\,
{\cal B}_{(m-v-1,u)i}^{\delta}({\bf r})\,
V^{\gamma}_{i}({\bf r})
\nonumber \\
 &&+\sum_{m=1}\sum_{u=0}\sum_{n=0}\sum_{t=1}
\sum_{v=0}^{m-1}\sum_{r=m-v-1}\sum_{q=n+v}
\delta_{r+q+m+n+u+t-N} (-1)^{t}
{\textstyle\frac{g^{N+1}B(n)}{n!(t+1)!(v+1)!(m-v-1)!} }
f^{\vec{\beta} \delta h}_{(m-v-1)}f^{\alpha h\gamma}
\nonumber \\
& &\;\;\;\;\;\;\;\;\;
\times f^{\vec{\sigma}\lambda\alpha}_{(n+v)}
f^{b\mu d}f^{\vec{\nu} d\lambda}_{(t)}\,
A_{i}^{\mu}({\bf r})\,{\textstyle\frac{\partial_{i}}{\partial^{2}}}
\left(\,{\cal R}_{(t)}^{\vec{\nu}}({\bf r})\,
{\cal M}_{(n+v,q)}^{\vec{\sigma}}({\bf r})\,
{\cal M}_{(m-v-1,r)}^{\vec{\beta}}({\bf r})\,
{\cal B}_{(m-v-1,u)j}^{\delta}({\bf r})
V^{\gamma}_{j}({\bf r})\,\right)\;.
\label{twoplustwo}
\end{eqnarray}
We change the integer-valued variables in the summations of
the 3$^{rd}$  4$^{th}$ and 5$^{th}$ terms in
Eq.~(\ref{twoplustwo}) to $k=m+n$ and $\ell =v+n,$ and carry out
the summation over $k$, $\ell$, and $n$.  We then observe that
combining ${\sf B_{(2)}}$ and ${\sf C}({\cal A})_{(2)}$
and applying Eq.~(\ref{byb}) has left us with an expression
in which the only $n$-dependence is in the Bernoulli numbers, and
in fractional coefficients. The indices $p$ and $q$ in the
operator-valued functions
${\cal M}_{(p,q)}^{\vec{\sigma}}({\bf r})$, and
${\cal B}_{(p,q)i}^{\delta}({\bf r})$ all are either $k$ or $\ell$,
and have no further dependence on the integer-valued summation
index $n$.
We therefore can make use of the identity
\begin{equation}
D_s^\ell (\ell ) =0\;\;\;\;\mbox{for}\;\;\; s=0 \;\;
\mbox{and} \;\;\ell >0\; , \;\;\;\;\;\; \mbox{where}\;\;
D_s^\kappa (\ell ) = \sum_{n=s}^{\kappa}
{\textstyle\frac{B(n)}{n!({\ell}-n+1)!}}\;,
\label{bernoulli}
\end{equation}
and observe that the only surviving contributions to
Eq.~(\ref{twoplustwo}) from sums over Bernoulli numbers are
$D_1^\ell (\ell )=-\frac{1}{(\ell +1)!}$  and
$D_0^0 (0) =1$.  We represent ${\sf B}_{(2)}+{\sf C}({\cal A})_{(2)}$
as $\left[\,{\sf B}_{(2)}+{\sf C}({\cal A})_{(2)}\,\right]_{(a)}+
\left[\,{\sf B}_{(2)}+{\sf C}({\cal A})_{(2)}\,\right]_{(b)}+
\left[\,{\sf B}_{(2)}+{\sf C}({\cal A})_{(2)}\,\right]_{(c)}$, where
\begin{eqnarray}
&&\left[\,{\sf B}_{(2)}+{\sf C}({\cal A})_{(2)}\,\right]_{(a)}=
f^{b\mu \delta}A_{i}^{\mu}({\bf r})
\sum_{m=1}{\textstyle\frac{g^{N+1}}{m!} }
f^{\vec{\alpha}\delta\gamma}_{(m)}
\sum_{r=m}\delta_{r+m-N}\,
(\,\delta_{ij}
-{\textstyle\frac{m}{m+1}
\frac{\partial_{i}\partial_{j}}{\partial^{2}}}\,)
\left(\,{\cal M}_{(m,r)}^{\vec{\alpha}}({\bf r})\,
V^{\gamma}_{j}({\bf r})\,\right)\;,
\nonumber \\
\label{tpta}
\end{eqnarray}

\begin{eqnarray}
\left[\,{\sf B}_{(2)}+{\sf C}({\cal A})_{(2)}\,\right]_{(b)}&&=
f^{b\mu \alpha}A_{i}^{\mu}({\bf r})
\sum_{m=1}\sum_{u=0}\sum_{v=0}^{m-1}
\sum_{r=m-v-1}\sum_{q=v}\delta_{r+q+m+u-N}
{\textstyle\frac{g^{N+1}}{(v+1)!(m-v-1)!} }
\nonumber\\
&&\;\;\;\;\;\;\;
\times f^{\vec{\beta}\delta g}_{(m-v-1)}f^{gf\gamma}
f^{\vec{\sigma}\alpha f}_{(v)}
{\textstyle\frac{\partial_{i}}{\partial^{2}}}
\left(\,{\cal M}_{(m-v-1,r)}^{\vec{\beta}}({\bf r})\,
{\cal M}_{(v,q)}^{\vec{\sigma}}({\bf r})\,
{\cal B}_{(m,u)j}^{\delta}({\bf r})\,
V^{\gamma}_{j}({\bf r})\,\right)
\nonumber \\
&&+\sum_{k=2}
\sum_{\ell =1}^{k-1}\sum_{u=0}\sum_{r=k-\ell -1}
\sum_{q=\ell}\delta_{r+q+k+u-N}
f^{\vec{\beta} \delta h}_{(k-\ell -1)}
f^{\alpha h\gamma}f^{\vec{\sigma}\nu\alpha}_{(\ell )}
f^{b\mu \nu}  A_{i}^{\mu}({\bf r})
{\textstyle\frac{g^{N+1}}{(\ell +1)!(k-\ell -1)!}}
\nonumber \\
&&\;\;\;\;\;\;\;\;\;\;\;\;\;\;\;\;\;\;\;\;
\times {\textstyle \frac{\partial_{i}}{\partial^{2}}}
\left({\cal M}_{(\ell ,q)}^{\vec{\sigma}}({\bf r})\,
{\cal M}_{(k-\ell -1,r)}^{\vec{\beta}}({\bf r})\,
{\cal B}_{(k-\ell -1,u)j}^{\delta}({\bf r})\,
V^{\gamma}_{j}({\bf r})\,\right)\;,
\label{tptb}
\end{eqnarray}
 and
\begin{eqnarray}
\left[\,{\sf B}_{(2)}+{\sf C}({\cal A})_{(2)}\,\right]_{(c)}&&=
\sum_{k=1}\sum_{u=0}\sum_{r=k-1}\sum_{t=1}\delta_{r+k+u+t-(N+1)}
(-1)^{t-1}{\textstyle\frac{g^{N+1}}{(t-1)!(t+1)(k-1)!} }
f^{\vec{\beta} \delta h}_{(k-1)}f^{h\lambda\gamma}
f^{\vec{\mu} b\lambda}_{(t)}
\nonumber \\
&&\;\;\;\;\;\;\;\;\;\;\;\;\;\;\;\;\;\;\;\;\;\;\;\;\;\;\;\;\;\;\;
\times\,{\cal R}_{(t)}^{\vec{\mu}}({\bf r})
{\cal M}_{(k-1,r)}^{\vec{\beta}}({\bf r})\,
{\cal B}_{(k-1,u)i}^{\delta}({\bf r})\,
V^{\gamma}_{i}({\bf r})
\nonumber \\
&&+ \sum_{k=1}\sum_{u=0}\sum_{r=k-1}
\sum_{t=1}\delta_{r+k+u+t-N}
(-1)^{t}{\textstyle\frac{g^{N+1}}{(t+1)!(k-1)!} }\,
f^{\vec{\beta} \delta h}_{(k-1)}f^{h\lambda\gamma}
f^{b\mu d}f^{\vec{\nu} d\lambda}_{(t)}\,A_{i}^{\mu}({\bf r})
\nonumber \\
&&\;\;\;\;\;\;\;\;\;\;\;\;\;\;\;\;\;\;\;\;\;\;\;\;\;\;\;\;\;\;\;
\times{\textstyle\frac{\partial_{i}}{\partial^{2}}}
\left(\,{\cal R}_{(t)}^{\vec{\nu}}({\bf r})\,
{\cal M}_{(k-1,r)}^{\vec{\beta}}({\bf r})\,
{\cal B}_{(k-1,u)j}^{\delta}({\bf r})\,
V^{\gamma}_{j}({\bf r})\,\right)\;.
\label{tptc}
\end{eqnarray}
\bigskip

We then note that
the second term on r.h.s. of Eq.~(\ref{tptb}) is an expression
that has the form $\sum_{k=2}\sum_{\ell=1}^{k-1}{\varphi}(k,\ell )$,
and that this sum can be expressed as
$\sum_{k=2}\sum_{\ell=1}^{k-1}{\varphi}(k,\ell )
=\sum_{k=1}\sum_{\ell=0}^{k-1}{\varphi}(k,\ell )-
\sum_{k=2}{\varphi}(k,0 )-{\varphi}(1,0 )$; we further
observe that a number of the summations in the parts of
Eq.~(\ref{tptb}) that we have included in
$\sum_{k=2}{\varphi}(k,0 )$ and ${\varphi}(1,0 )$
can be eliminated because they become
degenerate, enabling us to make use of Eq.~(\ref{eq:LCA1})
to transform them.  We  use Eq.~(\ref{byb}) to combine the
$\sum_{k=1}\sum_{\ell=0}^{k-1}{\varphi}(k,\ell )$
part of this second term on the r.h.s. of Eq.~(\ref{tptb})
with the first term in that equation,  so that the two
${\cal B}_{(\eta,u)j}^{\delta}({\bf r})\,$ terms are combined into
a multiple of $\partial_{j}{\cal Y}_{(u)}^{\delta}({\bf r}).$
Finally, we use Eq.~(\ref{eq:perm})
with $[{\sf Q},~]=\partial_j,$ to obtain
\begin{eqnarray}
\left[\,{\sf B}_{(2)}+{\sf C}({\cal A})_{(2)}\,\right]_{(b)}&&=
-\sum_{m=1}\sum_{r=m}
\delta_{r+m-N}{\textstyle\frac{g^{N+1}}{(m+1)!}}
f^{\vec{\alpha}\delta\gamma}_{(m)}
f^{b\mu\delta}\,A_{i}^{\mu}({\bf r})\,
{\textstyle\frac{\partial_{i}}{\partial^{2}}}
\left(\,\partial_{j}{\cal M}_{(m,r)}^{\vec{\alpha}}({\bf r})\,
V^{\gamma}_{j}({\bf r})\,\right)
\nonumber \\
&&\;\;\;\; - {\textstyle\frac{g^{N+1}}{(N-1)!}}
f^{\delta h\gamma}f^{b\mu\delta}\,
A_{i}^{\mu}({\bf r})\,
{\textstyle\frac{\partial_{i}}{\partial^{2}}}
\left(\,\psi_{(N-1)j}^{h}({\bf r})\,
V^{\gamma}_{j}({\bf r})\,\right)\;.
\label{tptaf}
\end{eqnarray}
We also note that the r.h.s. of Eq.~(\ref{tptc}) contains an expression
of the form $\sum_{k=1}\vartheta (k)$, which can be expressed as
$\sum_{k=1}\vartheta (k)=\sum_{k=2}\vartheta (k)+\vartheta (1)$.
As in  the case of $\varphi (0)$ above, a number of the summations in
$\vartheta (1)$ can be eliminated; we make use of Eq.~(\ref{eq:LCA1})
to transform $\sum_{k=2}\vartheta (k)$, and then obtain
\begin{eqnarray}
&&\left[\,{\sf B}_{(2)}+{\sf C}({\cal A})_{(2)}\,\right]_{(c)}
= (-1)^{N-1}{\textstyle\frac{g^{N+1}}{(N-1)!(N+1)}}
f^{\lambda\delta\gamma}f^{\vec{\mu} b\lambda}_{(N)}\,
{\cal R}_{(N)}^{\vec{\mu}}({\bf r})\,
a_{i}^{\delta}({\bf r})\,
V^{\gamma}_{i}({\bf r})
\nonumber \\
&& \;\;\;\;\;\;\;\;\;\;\;\;+\sum_{t=1}^{N-1} (-1)^{t-1}
{\textstyle\frac{g^{N+1}}{(t-1)!(t+1)(N-t)!}}
f^{\lambda\delta\gamma}f^{\vec{\mu} b\lambda}_{(t)}\,
{\cal R}_{(t)}^{\vec{\mu}}({\bf r})\,
\psi_{(N-t)i}^{\delta}({\bf r})\,
V^{\gamma}_{i}({\bf r})\nonumber \\
&& \;\;\;\;\;\;\;\;\;\;\;\;+ (-1)^{N-1}
{\textstyle\frac{g^{N+1}}{N!} }
f^{\lambda\delta\gamma} f^{b\mu d}f^{\vec{\nu} d\lambda}_{(N-1)}\,
A_{i}^{\mu}({\bf r})\,{\textstyle\frac{\partial_{i}}{\partial^{2}}}
\left(\,{\cal R}_{(N-1)}^{\vec{\nu}}({\bf x})\,
a_{j}^{\delta}({\bf r})\,
V^{\gamma}_{j}({\bf r})\,\right)
\nonumber \\
&& \;\;\;\;\;\;\;\;\;\;\;\;+\sum_{t=1}^{N-2} (-1)^{t}
{\textstyle\frac{g^{N+1}}{(t+1)!(N-t-1)!} }
f^{\lambda\delta\gamma} f^{b\mu d}f^{\vec{\nu} d\lambda}_{(t)}\,
A_{i}^{\mu}({\bf r})\,{\textstyle\frac{\partial_{i}}{\partial^{2}}}
\left(\,{\cal R}_{(t)}^{\vec{\nu}}({\bf x})\,
\psi_{(N-t-1)j}^{\delta}({\bf r})\,
V^{\gamma}_{j}({\bf r})\,\right)\;.
\label{tptbf}
\end{eqnarray}
\bigskip

Making use of  Eq.~(\ref{gpsi}), we observe that
$\left[\,{\sf B}_{(2)}+{\sf C}({\cal A})_{(2)}\,\right]_{(c)}$
in Eq.~(\ref{tptbf}) has the same form as ${\sf A}_{(2)}$;
and since  Eq.~(\ref{tpta}) has the same structure as  ${\sf C}(a)$ in
Eq.~(\ref{cla}), it is natural to combine these terms to obtain:
\begin{eqnarray}
{\sf A}_{(2)}+{\sf B}_{(2)}+{\sf C}(a)&&
+{\sf C}({\cal A})_{(2)}
\nonumber \\
&& \;\;\;\;=  \sum_{m=1}\sum_{r=m}
\delta_{r+m-N}{\textstyle\frac{g^{N+1}}{(m+1)!}} f^{b\mu c}
f^{\vec{\sigma} c\gamma}_{(m)}\,A_{i}^{\mu}({\bf r})\,
{\textstyle \frac{\partial_{i}}{\partial^{2}}}
\left(\,{\cal M}_{(m,r)}^{\vec{\sigma}}({\bf r})\,
\partial_{j}V_{j}^{\gamma}({\bf r})\,\right)\;.
\label{twof}
\end{eqnarray}
We combine all the terms with subscript (1), use Eq.~(\ref{qu})
to eliminate permutations of structure constants, and note that
$\partial_{i}\Pi_{i}^{b}({\bf r})$ commutes with
$a_{j}^{\alpha}({\bf r}^\prime).$ We
then observe that
\begin{eqnarray}
&&{\sf A}_{(1)}+{\sf B}_{(1)}+{\sf C}({\cal A})_{(1)}
={\textstyle\frac{ig^{N+1}}{(N+1)!}}\int d{\bf r}^\prime\,
\left[\,\partial_{i}\Pi_{i}^{b}({\bf r}),\,
\psi_{(N+1)j}^{\gamma}({\bf r}^\prime)\,\right]\,
V_{j}^{\gamma}({\bf r}^\prime)
\nonumber \\
 & &\;\;\;\; + \sum_{m=1}{\textstyle\frac{ig^{N+1}}{m!}}
f^{\vec{\alpha}\delta \gamma}_{(m)}\sum_{u=0}
\sum_{r=m}\delta_{r+m+u-(N+1)}\int d{\bf r}^\prime\,
\left[\,\partial_{i}\Pi_{i}^{b}({\bf r}),\,
{\cal M}_{(m,r)}^{\vec{\alpha}}({\bf r}^\prime)\,\right]\,
{\cal B}_{(m,u)j}^{\delta}({\bf r}^\prime)\,
V^{\gamma}_{j}({\bf r}^\prime)
\nonumber \\
 & &\;\;\;\;+\sum_{m=1}{\textstyle\frac{ig^{N+1}}{m!} }
f^{\vec{\alpha}\delta\gamma}_{(m)}
\sum_{u=0}\sum_{r=m}\delta_{r+m+u-(N+1)}\int d{\bf r}^\prime\,
\left[\,\partial_{i}\Pi_{i}^{b}({\bf r}),\,
{\cal B}_{(m,u)j}^{\delta}({\bf r}^\prime)\,\right]\,
{\cal M}_{(m,r)}^{\vec{\alpha}}({\bf r}^\prime)\,
V^{\gamma}_{j}({\bf r}^\prime)\;.
\label{eqlast}
\end{eqnarray}
If we then use Eq. (\ref{eq:LCA1}) to transform
${\cal A}_{(N+1)j}^{\gamma}$,  Eq.~(\ref{eqlast}) can be written as
\begin{equation}
{\sf A}_{(1)}+{\sf B}_{(1)}+{\sf C}\,({\cal A})_{(1)}=i
\int d{\bf r}^\prime\,\left[\,\partial_{i}\Pi_{i}^{b}({\bf r}),\,
{\cal A}_{(N+1)j}^{\alpha}({\bf r}^\prime)\,\right]\,
V_{j}^{\alpha}({\bf r}^\prime) \; .
\label{gan}
\end{equation}
\bigskip

We change the integer-valued variables in the
summation in Eq.~(\ref{threebthree})
to $\ell =m+n$, and carry out the summation over $\ell$ and $n$; we
then obtain an expression in which the only $n$-dependence is in the
Bernoulli numbers and in fractional coefficients. We make use of
Eq.~(\ref{bernoulli}) and observe that the only surviving
contributions to
Eq.~(\ref{threebthree}) from sums over Bernoulli numbers are
$D_0^{\ell -1}(\ell )=-\frac{B(\ell)}{\ell !}$ and
$D_1^{\ell -1}(\ell )=-\frac{B(\ell)}{\ell !}
-\frac{1}{(\ell +1)!}$; we then obtain
\begin{eqnarray}
{\sf C}({\cal A})_{(3)}&&=-\sum_{\ell =2}\sum_{p=\ell}
\delta_{p+\ell -N} \; g^{N+1}\,
\left(\,{\textstyle\frac{B(\ell )}{\ell !}}
+{\textstyle\frac{1}{(\ell +1)!}}\,\right)
f^{b\mu c}f^{\vec{\sigma} c\gamma}_{(\ell )}\,
A_{i}^{\mu}({\bf r})\,{\textstyle
\frac{\partial_{i}}{\partial^{2}}}
\left(\,{\cal M}_{(\ell ,p)}^{\vec{\sigma}}({\bf r})\,
\partial_{j}V_{j}^{\gamma}({\bf r})\,\right)
\nonumber \\
& &\;\;\;\;+ \sum_{\ell =1} \sum_{t=1}
\sum_{p=\ell }\,\delta_{p+\ell +t-(N+1)}
(-1)^{t-1}{\textstyle\frac{g^{N+1}B(\ell )}{\ell !(t-1)!(t+1)}}
f^{\vec{\mu} b \lambda}_{(t)}
f^{\vec{\sigma}\lambda\gamma}_{(\ell )}\,
{\cal R}_{(t)}^{\vec{\mu}}({\bf r})\,
{\cal M}_{(\ell ,p)}^{\vec{\sigma}}({\bf r})\,
\partial_{i}V_{i}^{\gamma}({\bf r})
\nonumber \\
&&\!\!\!\!\!\!\!\!\!\!\!\!\!\!\!
+\sum_{\ell =1}\sum_{t=1}\sum_{p=\ell }
\delta_{p+\ell +t-N}(-1)^{t} {\textstyle\frac{g^{N+1}B(\ell )}
{\ell !(t+1)!}} f^{b\mu d}f^{\vec{\nu} d\lambda}_{(t)}
f^{\vec{\sigma}\lambda\gamma}_{(\ell )}\,
A_{i}^{\mu}({\bf r})\,
{\textstyle\frac{\partial_{i}}{\partial^{2}}}
\left(\,{\cal R}_{(t)}^{\vec{\nu}}({\bf r})\,
{\cal M}_{(\ell ,p)}^{\vec{\sigma}}({\bf r})\,
\partial_{j}V_{j}^{\gamma}({\bf r})\,\right)\;.
\label{threefinal}
\end{eqnarray}
\bigskip

Finally, we combine  Eqs.~(\ref{gpsi}), (\ref{twof}),  (\ref{gan}), and
(\ref{threefinal}), to obtain
\begin{eqnarray}
{\sf A}+{\sf B}+{\sf C}&&= i
\int d{\bf r}^\prime\,\left[\,\partial_{i}\Pi_{i}^{b}({\bf r}),\,
{\cal A}_{(N+1)j}^{\alpha}({\bf r}^\prime)\,\right]\,
V_{j}^{\alpha}({\bf r}^\prime)+ \delta_{N}
f^{b\mu \gamma}\,A_{i}^{\mu}({\bf r})\,
V_{i}^{\gamma}({\bf r})
\nonumber \\
&&\;\;\;\;-\sum_{m=1}\sum_{r=m}\delta_{r+m-N}
{\textstyle\frac{B(m)}{m!}}f^{b\mu c}
f^{\vec{\alpha} c\gamma}_{(m)}\,A_{i}^{\mu}({\bf r})\,
{\textstyle \frac{\partial_{i}}{\partial^{2}}}
\left(\,{\cal M}_{(m,r)}^{\vec{\alpha}}({\bf r})\,
\partial_{j}V_{j}^{\gamma}({\bf r})\,\right)
\nonumber \\
&&+\sum_{m=0}\sum_{t=1}\sum_{r=m}\delta_{r+m+t-(N+1)}
(-1)^{t-1}{\textstyle \frac{B(m)}{m!(t-1)!(t+1)}}
f^{\vec{\mu} b
\lambda}_{(t)}f^{\vec{\alpha}\lambda\gamma}_{(m)}\,
{\cal R}_{(t)}^{\vec{\mu}}({\bf r})\,
{\cal M}_{(m,r)}^{\vec{\alpha}}({\bf r})\,
\partial_{i}V_{i}^{\gamma}({\bf r})
\nonumber \\
&&\!\!\!\!\!\!\!\!\!\!\!\!\!\!\!\!\!\!\!\!\!\!\!
+ f^{b\mu d}A_{i}^{\mu}({\bf r})
\sum_{m=0}\sum_{t=1}\sum_{r=m}\delta_{r+m+t-N}
(-1)^{t}{\textstyle\frac{B(m)}{m!(t+1)!}}
f^{\vec{\nu} d\lambda}_{(t)}
f^{\vec{\alpha}\lambda\gamma}_{(m)}\,
{\textstyle\frac{\partial_{i}}{\partial^{2}}}
\left(\,{\cal R}_{(t)}^{\vec{\nu}}({\bf r})\,
{\cal M}_{(m,r)}^{\vec{\alpha}}({\bf r})\,
\partial_{j}V_{j}^{\gamma}({\bf r})\,\right)\;,
\label{eqqed}
\end{eqnarray}
where we have added a $\delta_{N}$ term that vanishes except
for the $N=0$ case; the need for this term in the $N=0$ case
has been discussed in  Section~\ref{sec-Implementing}.
The r.h.s. of Eq.~(\ref{eqqed}) is identical to the l.h.s.
of Eq. (\ref{eq:LCA2}) for the value $n=N+1,$ and therefore completes
the proof of the `fundamental theorem' for the
construction of $\Psi$.


\begin{references}
\bibitem{goldjack}J. Goldstone and R. Jackiw, Phys.\ Lett.\ B 74
(1978) 81.
\bibitem{jackiw}R. Jackiw, Rev.\ Mod.\ Phys. 52 (1980) 661.
\bibitem{khymtemp}K. Haller, Phys.\ Rev. D 36 (1987) 1839;
Phys. Lett. B 251 (1990) 575.
\bibitem{lands}K. A. James and P. V. Landshoff, Phys.\ Lett.\ B 251
(1990) 167.
\bibitem{bellchenhall}M. Belloni, L. Chen, and K. Haller,
Phys.\ Lett.\ B 373 (1996) 185.
\bibitem{jacktop}We will restrict our discussion in this section to
homotopically trivial gauge transformations, which are the only ones
that can be implemented in this way. See, for example, R.
Jackiw,``Topological Investigations of Quantized Gauge Theories''
(Section 3.3) in {\em Current Algebras and Anomalies,} S. B.
Treiman $et.al.$ editors, (World Scientific, 1985).
\bibitem{lavelle2}M. Lavelle and D. McMullan, Phys.\ Lett. B 329
(1994) 68.
\bibitem{lavelle5}M. Lavelle and D. McMullan,
Constituent Quarks From QCD, hep-ph/9509344 UAB-FT-369,
PLY-MS-95-03.
\bibitem{khqedtemp}K. Haller, Phys.\ Rev. D 36  (1987) 1830.
\bibitem{khelqed}K. Haller and E. Lim-Lombridas,
Found. of Phys. 24 (1994) 217.
\bibitem{topsect}See, for example, R. Jackiw,
{\em op. cit.;} R. Rajaraman, {\em Solitons and Instantons,}
(North Holland, 1982); S. Pokorski, {\em Gauge Field Theories,}
(Cambridge, 1987).
\end{references}
\end{document}